\documentclass[webpdf,contemporary,large]{oup-authoring-template}%
\setcitestyle{numbers,square,comma}
\onecolumn % for one column layouts

\graphicspath{
  {Fig/}
  {Fig/plot_2latent/}
  {Fig/plot_2obs/}
  {Fig/plot_individual/}
  {Fig/plot_noheter/}
  {Fig/plot_sc/}
}
\usepackage{longtable}
\usepackage{booktabs}
\usepackage{pdflscape}
\usepackage{threeparttable} % for table notes
\usepackage{rotating}
\usepackage{float}
\usepackage{placeins}
\usepackage{adjustbox}
\usepackage{placeins}
\usepackage{makecell}  
\usepackage{array}  
\usepackage{graphicx}
\usepackage{tikz}
\usetikzlibrary{arrows.meta, positioning}

\theoremstyle{thmstyleone}%
\theoremstyle{thmstyletwo}%
\theoremstyle{thmstylethree}%

\begin{document}

\journaltitle{Journal Title Here}
\DOI{DOI added during production}
\copyrightyear{YEAR}
\pubyear{YEAR}
\vol{XX}
\issue{x}
\access{Published: Date added during production}
\appnotes{Paper}

\firstpage{1}

%\subtitle{Subject Section}

\title[Multi‐marker tests for panel count data]{Multi‐marker genetic association tests for panel count data}

\author[1]{Kun Xia}
\author[2]{Jianrui Zhang}
\author[3]{Qing Lu}
\author[1$\ast$]{Chenxi Li\ORCID{0000-0003-0701-0757}}

\address[1]{\orgdiv{Department of Epidemiology and Biostatistics}, \orgname{Michigan State University}, \orgaddress{\street{909 Wilson Road}, \postcode{48824}, \state{MI}, \country{United States}}}
\address[2]{\orgdiv{Department of Statistics and Probability}, \orgname{Michigan State University}, \orgaddress{\street{619 Red Cedar Road}, \postcode{48824}, \state{MI}, \country{United States}}}
\address[3]{\orgdiv{Department of Biostatistics}, \orgname{University of Florida}, \orgaddress{\street{2004 Mowry Road}, \postcode{48824}, \state{FL}, \country{United States}}}

\corresp[$\ast$]{Corresponding author.  Department of Epidemiology and Biostatistics, Michigan State University, 909 Wilson Road, 48824 MI, United States. \href{email:email-id.com}{cli@msu.edu}}

\received{Date}{0}{Year}
\revised{Date}{0}{Year}
\accepted{Date}{0}{Year}

%\editor{Associate Editor: Name}

%\abstract{
%\textbf{Motivation:} .\\
%\textbf{Results:} .\\
%\textbf{Availability:} .\\
%\textbf{Contact:} \href{name@email.com}{name@email.com}\\
%\textbf{Supplementary information:} Supplementary data are available at \textit{Journal Name}
%online.}

\abstract{The existing multi-marker survival tests focus on time-to-event outcomes. However, recurrent events are common in real-world clinical and biomedical studies, especially in the research of chronic and recurrent diseases. In this paper, we develop a suite of set‐based genetic association tests for panel count outcomes under a unified weighted‐V‐statistic framework. These tests can effectively account for genetic effect heterogeneity in panel count data. Additionally, we develop small-sample corrections to the tests to enhance the accuracy of the tests under small samples. Simulation studies show that the proposed tests perform well in terms of size and power across various scenarios and have a bigger power than the existing set-based tests for interval-censored outcomes. A dental caries GWAS data set is analyzed to illustrate the utility of the tests.}


\maketitle

\section{Introduction}
Genome-Wide Association Studies (GWAS) have revolutionized our understanding of the genetic basis of complex traits and chronic diseases. Kernel Association Tests use kernel machine regression to evaluate the joint effects of multiple genetic variants within a predefined region, such as a gene or pathway. By aggregating genetic information, these tests improve power for association discovery and reduce the multiple testing burden compared with single-marker tests. 

Although recurrent events are common in biomedical studies, set-based genetic association tests for recurrent-event outcomes remain less developed. Existing multi-marker tests have been extensively studied for quantitative and binary traits, e.g., Wu et al. \cite{wu2011skat}, Lee et al. \cite{lee2012skatO}, and Ionita-Laza et al. \cite{ionita2013skato}, but these methods cannot be directly applied to recurrent-event outcomes because they do not account for event-time or follow-up information.

Many existing multi-marker tests with survival outcomes, as highlighted in Li et al. \cite{li2021weightedV}, are inaccurate in estimating the null distribution of p-values, particularly in small or moderate sample sizes, such as Goeman et al. \cite{goeman2005pathway}, Sinnott and Cai \cite{sinnott2013aftKernel}, and Chen et al. \cite{chen2014skatSurv}. Li et al. \cite{li2021weightedV} developed a multi‐marker test based on a weighted V statistic to address these challenges. The weighted V statistic can effectively capture complex effects of a SNP set or gene set on survival outcomes while accounting for genetic heterogeneity, and it can also provide accurate inference in studies with moderate sample sizes.

Almost all of the existing multi-marker survival tests are applied to time-to-event outcomes. But recurrent events are common in real-world clinical and biomedical studies. Because it's usually impractical to continuously monitor the subjects, such as in clinical trial, only the number of events between two consecutive observation times are recorded. For instance, in the Early Childhood Caries study or migraines, the subjects may experience multiple recurrences of the same event. The exact timing of these events is unknown, the number of events can only be recorded during the hospital visit. Wu et al. \cite{wu2021interval} developed a multi-marker test for interval-censored survival outcomes based on the weighted V-statistic framework for single-event outcomes. Choi et al. \cite{choi2024interval} developed tests for genetic association with multiple interval-censored outcomes. These methods are designed for single-event interval-censored outcomes or multiple interval-censored outcomes, rather than recurrent panel count outcomes in which only cumulative event counts are observed at discrete examination times. Many existing multi-marker tests, including those proposed by Goeman et al. \cite{goeman2005pathway}, Sinnott and Cai \cite{sinnott2013aftKernel}, and Chen et al. \cite{chen2014skatSurv}, may provide inaccurate approximations to the null distribution of the test statistic under small or even moderate sample sizes, as discussed by Li et al.    \cite{li2021weightedV}. This issue is particularly relevant for genetic association studies with complex event outcomes, where the number of adjustment covariates can be large relative to the sample size.

In this article, we develop a set of multi-marker genetic association tests for panel count data based on a proportional means model. Panel count data arise from counting processes observed only at discrete examination times. The baseline mean function is estimated using the semiparametric mean regression model for panel count data proposed by Wang et al. \cite{wang2013aeex}. The tests are built based on the weighted V-statistic framework as introduced by Li et al. \cite{li2021weightedV} and thus are referred to as the weighted V tests in the sequel. In Section 2, we present the data structure, notation and semiparametric mean regression model and weighted V statistic. In Section 3, we study the finite-sample performance of the methods via simulation. In Section 4, we apply the proposed tests to a dbGaP dental caries data set, the ZOE 2.0 study \cite{divaris2020zoe2}. The paper concludes with some discussion on future research directions in Section 5.

% the overview framework
% the overview framework
\begin{figure}[!t]
\centering

\begin{tikzpicture}[
    box/.style={
        draw,
        rounded corners=3pt,
        minimum width=0.95\linewidth,
        minimum height=0.9cm,
        align=center,
        inner xsep=5pt,
        inner ysep=5pt,
        font=\scriptsize
    },
    arrow/.style={
        -{Latex[length=2.0mm]},
        line width=0.35pt
    },
    node distance=0.75cm
]

\node[box] (input) {
    \textbf{Inputs}\\
    Panel count data; covariates $\mathbf{Z}$; marker set $\mathbf{G}$; optional heterogeneity factors $\mathbf{X}$
};

\node[box, below=of input] (fit) {
    \textbf{Fit the null model}\\
    Estimate $\Lambda_0(t)$ and $\boldsymbol{\gamma}$ under\\
$\mathrm{E}\{N_i(t)\mid \mathbf{Z}_i\}
= \Lambda_0(t)\exp(\mathbf{Z}_i^\top\boldsymbol{\gamma})$\\
using AEEX estimation
};

\node[box, below=of fit] (residual) {
    \textbf{Compute residuals}\\
Covariate-adjusted terminal residuals\\
$\widehat{M}_{Z,i}
= N_i(T_{iK_i})-\widehat{\Lambda}_0(T_{iK_i})
\exp(\mathbf{Z}_i^\top\widehat{\boldsymbol{\gamma}})$
};

\node[box, below=of residual] (matrix) {
    \textbf{Construct similarity matrices}\\
Genetic similarity matrix $\mathbf{F}=\{f(\mathbf{G}_i,\mathbf{G}_j)\}$\\
Covariate projection matrix $\mathbf{H}$\\
Heterogeneity similarity matrix $\mathbf{K}=\{\kappa_{ij}\}$\\
and $\mathbf{W}=(\mathbf{1}+\mathbf{K})\odot\mathbf{F}$
};

\node[box, below=of matrix] (stat) {
    \textbf{Compute test statistics}\\
    WV-PCD and HWV-PCD \\
    Small-sample corrected versions:\\
    $V^c_{Z,\mathrm{PCD}}$ and $V^{H,c}_{Z,\mathrm{PCD}}$
};

\node[box, below=of stat] (infer) {
    \textbf{Inference}\\
    Approximate the null distribution and compute association $p$-values
};

\draw[arrow] (input) -- (fit);
\draw[arrow] (fit) -- (residual);
\draw[arrow] (residual) -- (matrix);
\draw[arrow] (matrix) -- (stat);
\draw[arrow] (stat) -- (infer);

\end{tikzpicture}

\caption{Overview of the proposed weighted V-statistic framework for set-based genetic association testing with panel count data.}
\label{fig:method-flowchart}
\end{figure}
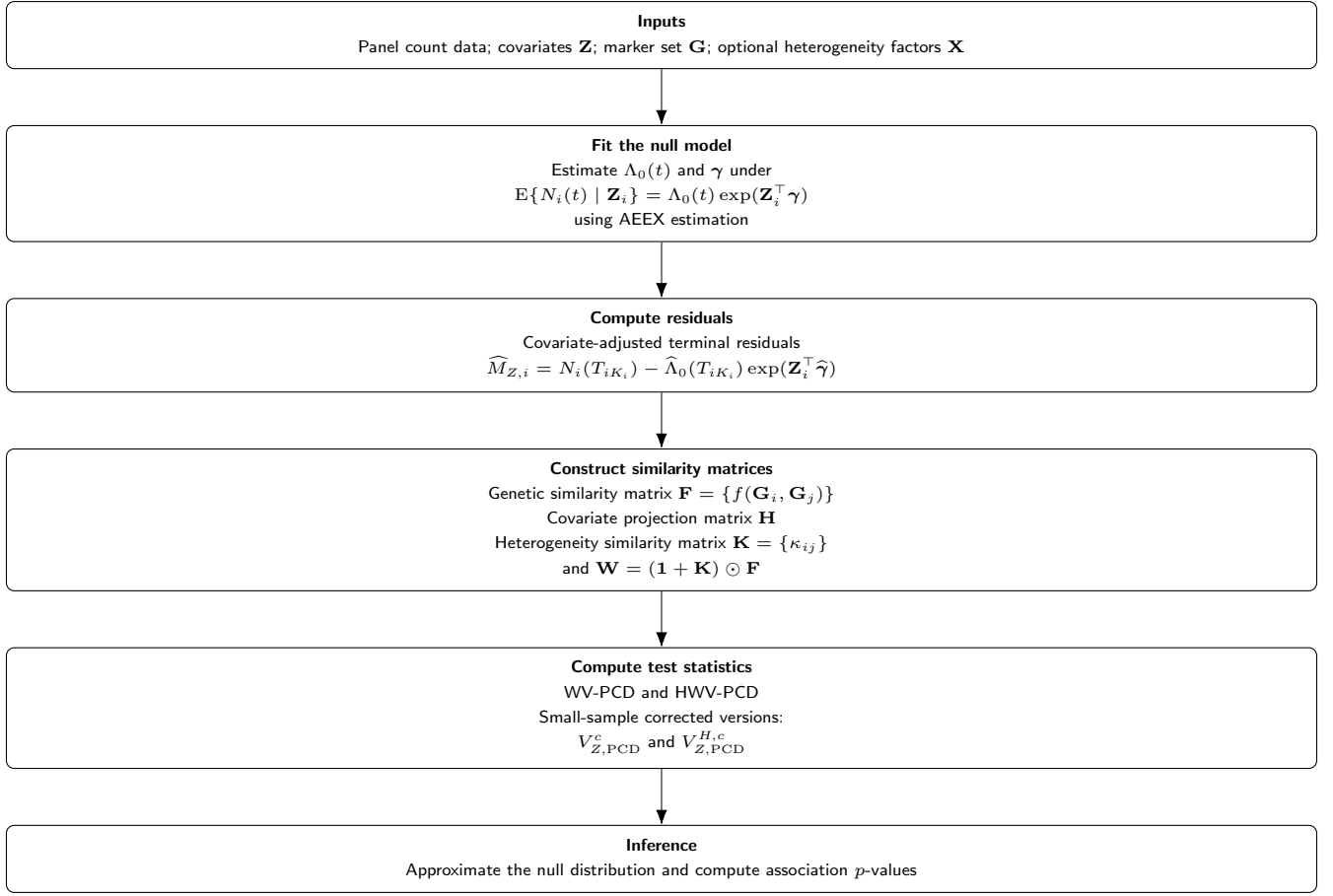

\section{Methods}\label{sec2}

\subsection{Data setup}\label{subsec2.1}

Suppose that the study consists of \(n\) subjects. Let \(N_i(t)\) denote the cumulative number of events experienced by subject \(i\) up to time \(t\). The counting process is observed only at \(K_i\) observation times, \(0<T_{i1}<\cdots<T_{iK_i}\) yielding observations
\( N_i(T_{i1}),\ldots,N_i(T_{iK_i}) \).
We are interested in testing the joint effect of a set of \(p\) genetic markers. Let
\(G_i=(G_{i1},\ldots,G_{ip})^T\)
denote the genetic marker vector for subject \(i\), where the marker set may consist of SNPs, gene expression measurements, or other genomic features. Marker sets can be defined according to genes, pathways, LD blocks, recombination hotspots, or sliding genomic windows.
Let
\(Z_i=(Z_{i1},\ldots,Z_{iq})^T\)
denote a \(q\)-dimensional covariate vector.
The observed data for subject \(i\) are
$$
D_i=\{K_i,(T_{ij},N_i(T_{ij}) )_{j=1}^{K_i} ,Z_i,G_i \},
$$
and the observed panel count dataset consists of \(n\) independent and identically distributed (i.i.d.) copies \(D_1,\ldots,D_n\).

\subsection{Weighted V tests for genetic association without considering genetic heterogeneity}\label{subsec2.2}

Under the null hypothesis of the covariate-adjusted association test, we assume that \(G\) has no effect on the event process \(N_i(t)\) given \(Z_i\). Under the null hypothesis, \(N_{i}(t)\) given
covariate \(Z_i\) follows a proportional means model for panel count data \cite{wang2013aeex},

\[E(N_{i}(t)|Z_{i}) = \Lambda_{0}(t) \exp(Z_{i}^T\gamma),\]

where \(\Lambda_0(t)\) is an unspecified baseline mean function and \(\gamma\) is a \(q\times1\) vector of regression coefficients. Define
\[
M_{Z,i}=N_i(T_{iK_i})-\Lambda_0(T_{iK_i})\exp(Z_i^T\gamma).
\]

Here, \(M_{Z,i}\) is a covariate-adjusted residual from the proportional mean model and measures the difference between the observed and expected cumulative numbers of events given \(Z_i\).

Following the weighted V-statistic framework of Li et al. \cite{li2021weightedV}, the proposed test, referred to as WV-PCD, is based on a weighted V statistic,

\begin{equation}
V^*_{Z,PCD}=n^{-2}\sum_{i=1}^{n}\sum_{j=1}^{n}\tilde f_Z(G_i,G_j)M_{Z,i}^{T}M_{Z,j},
\label{formula1}\tag{1}
\end{equation}

The quantity \(M_{Z,i}^TM_{Z,j} \) can be viewed as a measure of phenotypic concordance between subjects i and j after adjusting for covariate effects under the null model. \(\tilde{f}_{Z}(G_{i},G_{j})\) is considered as the weight function and serves as a
centered covariate-adjusted genetic similarity measure, which is defined
by

\begin{align*}
\tilde{f}_{Z}(G_{i}, G_{j}) 
&= f(G_{i}, G_{j}) 
    - E\big[ f(G_{i}, G_{k}) u(Z_{k}, Z_{j}) \mid G_{i}, Z_{j} \big] \\
&\quad - E\big[ u(Z_{i}, Z_{k}) f(G_{k}, G_{j}) \mid Z_{i}, G_{j} \big] \\
&\quad + E\big[ u(Z_{i}, Z_{m}) f(G_{m}, G_{k}) u(Z_{k}, Z_{j}) \mid Z_{i}, Z_{j} \big]
\end{align*},

where \( f(G_{i},G_{j})\) is a genetic similarity function and
\(u(Z_i,Z_j)=(1,Z_i^T)\Big[E\{(1,Z^T)^T(1,Z^T)\}\Big]^{-1}(1,Z_j^T)^T.\)
It can be shown that \(E[\tilde{f}_{Z}(G_{i},G_{j})(1,Z_{i}^T)|G_{i},Z_{i}] = 0\). Therefore, \(\tilde f_Z(G_i,G_j)\) is orthogonal to the covariates and represents a covariate-adjusted genetic similarity measure.

The choice of \(f(G_i,G_j)\) depends on the type of \(G\) and the expected form of the genetic effect. If the effect of \(G\) is expected to be linear, we use the linear kernel \(f(G_i,G_j)=G_i^TG_j\). For SNP data, if the genetic effect is not expected to be linear, we use the identity-by-state (IBS) kernel \(f(G_i,G_j)=\sum_{k=1}^p(2-|G_{i,k}-G_{j,k}|)/(2p)\). For gene expression data, polynomial or Gaussian kernels can be used when nonlinear or interactive effects are expected, with \(f(G_i,G_j)=(\rho+G_i^TG_j)^d\) and \(f(G_i,G_j)=\exp\{-\rho\|G_i-G_j\|^2\}\), respectively. Following Wei and Lu \cite{wei2017u}, we can also use a unified Laplacian-kernel-based genetic similarity, \(f(G_i,G_j)=\exp\{-\sum_{k=1}^p w_k|G_{i,k}-G_{j,k}|/\gamma\}\), where \(G_{i,k}\) can be discrete or continuous, \(\gamma=\sum_{k=1}^p w_k\), and \(w_k\) is a function of the variance \(\sigma_k^2\) of \(G_k\), defined by \(w_k=1/\sigma_k\). This kernel is particularly useful for association analyses with sequencing data, where a large proportion of genetic variants are rare.

Under the null hypothesis of no genetic effect on the event process \(N_i(t)\) while adjusting for the covariates \(Z\),
\( M_{Z,i}\) has mean zero given \(Z_{i}\), which is
\(E(M_{Z,i}|Z_{i}) =0\). When \(G\) is independent of \(Z\) (or
satisfies a linear model with \(Z\) as covariates) , following Li et
al.\cite{li2021weightedV}, the null limiting distribution of
\(V^*_{Z,PCD} \equiv n^{-2}\sum_{i=1}^{n} \sum_{j=1}^{n} \tilde f_Z(G_i,G_j)M_{Z,i}^TM_{Z,j} \)
satisfies

\begin{equation}
nV^*_{Z,PCD} \rightarrow \xi \sum_{t=1}^{\infty}\nu_{t}\chi_{1t}^2
\label{formula2}\tag{2}
\end{equation}

with \(\xi = E(M_{Z,i}^2)\), \(\chi_{1t}^2\)\textquotesingle s are
independent chi-square variates with degree \(1\), and \(\nu_t\)'s are the eigenvalues of \(\tilde{f_{Z}}(g_{i},g_{j})\), \( i.e., \tilde{f_{Z}}(g_{i},g_{j})= \sum_{t=1}^\infty \nu_{t}\phi_{t}(g_{i},z_{i})\phi_{t}(g_{j},z_{j})  \) ),
with \(E(\phi_{s}(G,Z)\phi_{t}(G,Z))=I(s=t))\). 

When Z has effects both on G and the event process, the test might be slightly conservative based on the discussion of Li et al. \cite{li2021weightedV}. However, the test is
still asymptotically correct under linear confounding when the linear
kernel is used for \(f(G_{i},G_{j})\), i.e. ,\(G = a+B^TZ+e\) , where
\(a\) and \(B\) are a constant vector and a constant matrix
respectively, and \(e\) is a zero-mean random error that is independent of \(Z\). Following the proof of Theorem 2 in Li et al. \cite{li2021weightedV}, it can be shown that the null limiting distribution of \(nV^*_{Z,PCD} \) with the linear kernel is still given by \((\ref{formula2})\) under linear confounding. 

Under the alternative hypothesis that \(G\) has effects on the event process adjusting for \(Z\), the covariate-adjusted phenotype similarity \(M_{Z,i}^TM_{Z,j} \) aligns with the covariate-centered genetic similarity \(\tilde{f_{Z}}(G_{i},G_{j})\). That is to say, the larger phenotype
similarity is weighted heavier, leading to a large value of
\(V^*_{Z-PCD}\) and the smaller phenotype similarity is given less weight accordingly.

However, \(M_{Z,i}\) involves unknown parameters \((\Lambda_{0}, \gamma)\) and \(\tilde f_Z(G_i,G_j)\) also depends on unknown conditional expectations in the population version. In applications, we replace \((\Lambda_{0}, \gamma)\) by \((\hat{\Lambda}_{0}, \hat{\gamma})\), and \(\hat{\Lambda}_{0}\) and \(\hat{\gamma}\) are the estimators based on the augmented estimating equations under conditional independent censoring \cite{wang2013aeex}. Wang et al. \cite{wang2013aeex} proved that \(\hat{\Lambda}\) and \(\hat{\gamma}\) are consistent under both of the observation schemes considered. And we also replace the population
version \(M_{Z,i}\) by \[
\hat M_{Z,i}=N_i(T_{iK_i})-\hat\Lambda_0(T_{iK_i})\exp(Z_i^T\hat\gamma).
\]
\(\tilde f_Z(G_i,G_j)\) is replaced by the corresponding sample averages. The resulting estimated \(\tilde f_Z(G_i,G_j)\) equals the
\((i,j)\)-th element of the matrix \((I-H)^TF(I-H)\), where
\(F= \{f(G_{i},G_{j})\}_{n\times n}\), \(I\) is an \(n \times n\)
identity matrix, and
\(H= \tilde{Z}(\tilde{Z}^T\tilde{Z})^{-1}\tilde{Z}^T\) with \(\tilde Z\) being an \(n\times (q+ 1)\) matrix whose \(i\)-th row is
\((1,Z_{i}^T)\) \((i=1,...,n)\), the corresponding sample version of the weighted V statistic is
\[
\bf{V_{Z,PCD}=\hat M_Z^T(I-H)^TF(I-H)\hat M_Z } ,
\]
where \(\hat M_Z=(\hat M_{Z,1},\ldots,\hat M_{Z,n})^T\).
\(V_{Z,PCD}\) is called the weighted V statistic for panel count data (WV-PCD).

We approximate \(\xi\) by \(\hat{\xi} = \hat{M}_Z^T\hat{M}_Z/n\) and use
a matrix eigen-decomposition of \((I-H)^TF(I-H)\) to obtain a finite-dimensional approximation to \(\nu_t\)'s.
Denote the eigenvalues and eigenvectors of \((I-H)^TF(I-H)\) by
\(\tilde{\nu}_{t}(t=1,\ldots,n)\) and \(\tilde{\phi}_{t}(t=1,...n)\)
respectively. Because \(\tilde{\phi}_{t}\), which is \(t\)-th
eigen-vector satisfies \(\sum_{i=1}^n \tilde{\phi}_{t,i}^2 =1\) instead
of \(n^{-1}\sum_{i=1}^n \tilde{\phi}_{t,i}^2 =1\), a finite‐dimensional
approximation for \(\nu_{t}\) is \(\hat {\nu_{t}} = \tilde{\nu_{t}}/n\). Then, the large-sample null distribution of \(nV_{Z,PCD}\)is approximately

\begin{equation}
nV_{Z,PCD}\sim \hat\xi\sum_{t=1}^{n}\hat\nu_t\chi_{1t}^2.
\label{formula3}\tag{3}
\end{equation}

Based on the approximation, we use Davies' method \cite{davies1980algorithm} to compute
the p-value \( P(nV_{Z,PCD} \geq nV_{Z,PCD,obs})\), where
\(V_{Z,PCD,obs}\) is the observed value of \(V_{Z,PCD}\).

\subsection{Weighted V tests for genetic association considering genetic heterogeneity}\label{subsec2.3}

The effects of a set of genetic markers (e.g., genes or SNPs) on
panel count outcomes may differ across subpopulations (e.g., races, sexes,
or genetic profiles). To jointly test the effect of genetic markers
across different subpopulations under heterogeneous effects, following
Wu et al. \cite{wu2021interval}, we extend the weighted V statistics \(V_{Z,PCD}\) by
replacing \(f(G_i, G_j)\) with \(w_{ij}=(1+\kappa_{ij})f(G_i,G_j)\)
where \(\kappa_{ij}\) is a measurement of subpopulation similarity between two individuals. \(w_{ij}\) is considered as a heterogeneity
weighted function to capture the latent structure. The resulting
weighted \(V\) statistic is

\[\bf{V^H_{Z,PCD}=\hat M_Z^T(I-H)^TW(I-H)\hat M_Z},\]
where \(\bf{W=(\mathbf 1+K)\odot F}\), \(\bf{1} = \{1\}_{n\times n}\),
\(\bf{K}=\{\kappa_{ij} \}_{n\times n}\) and \(\odot\) is the Hadamard product
operator. \(V_{Z,PCD}^H\) is called heterogeneity-weighted V
statistic under panel count data (HWV-PCD). It can account for both the
population structure that is fully determined by a vector of observable
variables \(\bf{X} = (X_{1},...,X_D)^T\) and the latent population
structure (e.g., subpopulations with different ancestry backgrounds)
that cannot be fully determined but can be inferred by a vector of
observable variables \textbf{X} (e.g., a large number of SNPs from the
GWAS data). In either case, we can choose \(\kappa_{ij}\) to be \(\kappa_{ij}=\exp\{-D^{-1}\sum_{d=1}^D(X_{i,d}^s-X_{j,d}^s)^2\}\),
where \(X_{id}^s\) is the standardized \(X_{id}\), i.e. ,
\(n^{-1}\sum_{i=1}^n X_{i,d}^s=0\) and
\(n^{-1}\sum_{i=1}^n (X_{i,d}^s)^2=1\). If \(X\) is a set of SNPs, we
can also use the IBS kernel for \(\kappa_{ij}\), i.e.,
\(\kappa_{ij}=\sum_{d=1}^D(2-|X_{i,d}-X_{j,d}|)/(2D)\). If \(X\) is a categorical or binary subpopulation indicator, such as sex or dichotomized environmental exposure, we use the identity kernel \(\kappa_{ij}=I(X_i=X_j)\). Under the null hypothesis that \(\bf{G}\) has no effect on the panel count outcome in any
subpopulation adjusting for \(\bf{Z}\), we use \((\ref{formula3})\) to
approximate the distributions of \(nV_{Z,PCD}^H\), replacing \(F\) by
\(W\) to compute \(\hat\nu_t\)'s. Based on this
distribution, we use Davies' method \cite{davies1980algorithm} to compute the
p-value \(P(nV^H_{Z,PCD}\geq nV^H_{Z,PCD,\mathrm{obs}})\).

\subsection{Small-sample correction}

The genetic association test works well when the sample size is relatively large. However, the asymptotic null distribution may not accurately approximate the finite-sample null distribution when the sample size
is relatively small. As shown in the simulation results,
the proposed tests tend to be conservative when the sample size is
small. To address this issue, motivated by Chen et al. \cite{chen2016smallSampleKAT}, we develop a small-sample correction for \({V_{Z,PCD}}\), \({V^H_{Z,PCD}}\) to
\[
V^c_{Z,PCD}\equiv
\frac{\hat M_Z^T(I-H)^TF(I-H)\hat M_Z}{\hat M_Z^T\hat M_Z},
\]
\[
V^{H,c}_{Z,PCD}\equiv
\frac{\hat M_Z^T(I-H)^TW(I-H)\hat M_Z}{\hat M_Z^T\hat M_Z},
\]
respectively, where the denominator \(\hat M_Z^T\hat M_Z\) standardizes the statistic by the total residual variation. The p-value,
\(P({V^c_{Z,PCD}} \geq V^c_{Z,PCD,\mathrm{obs}})\) can be written as
\[
\begin{aligned}
&P\left(V^c_{Z,\mathrm{PCD}}\ge V^c_{Z,\mathrm{PCD},\mathrm{obs}}\right)  \\
&\quad =
P\left[\hat M_Z^\top \left\{(I-H)^\top F(I-H)-V^c_{Z,\mathrm{PCD},\mathrm{obs}}I_n\right\}
\hat M_Z \ge 0 \right],
\end{aligned}
\]
where \(I_n\)
is an \(n \times n\) identity matrix. Following the same argument used for \(V^*_{Z,PCD}\), the large-sample null distribution of
\(n^{-1}\hat M_Z^T\{(I-H)^TF(I-H)-V^c_{Z,PCD,\mathrm{obs}}I_n\}\hat M_Z\) is approximately

\begin{equation}
\xi \sum_{t=1}^{n}d_{t}\chi_{1t}^2 ,
\label{4}\tag{4}
\end{equation}

where \(\xi=E(M_{Z,i}^2)\) and \(d_t=\tilde d_t/n\) , \(\tilde d_t\)'s are the eigenvalues of \((I-H)^TF(I-H)-V^c_{Z,PCD,\mathrm{obs}}I_n\). Here,
\(\chi_{1t}^2\)'s are independent chi-square random variables with one degree of freedom. Therefore, based on this distribution, we use
Davies' method \cite{davies1980algorithm} to compute the p-value. The p-value of
\(P({V^{H,c}_{Z,PCD}} \geq {V^{H,c}_{Z,PCD,obs}})\) can be obtained
similarly.

\section{Numerical experiments}\label{sec3}

We performed Monte Carlo simulations to assess the finite-sample performance of the (heterogeneity) weighted V tests for panel count data. In all simulation scenarios, subjects' baseline covariates were adjusted. Unless otherwise specified, two sample sizes, \(n=400\) and \(n=800\), and two marker-set sizes, \(p=15\) and \(p=25\), were considered. In all simulation scenarios, 1000 Monte Carlo replicates were generated, and the significance level was set to 0.05 unless otherwise specified.

For each subject, two adjustment covariates were generated: a binary covariate
\(Z_{i1}\sim \mathrm{Bernoulli}(0.5)\) and a continuous covariate
\(Z_{i2}\sim U(-2,2)\). The covariate effect was set to
\(\beta_Z=(0.01,0.1)^\top\) in the empirical size and power simulations. Each
subject had two visit times, \(T_{i1}\sim U(0.05,1)\) and
\(T_{i2}=T_{i1}+U(0.05,1)\). A subject-specific frailty was generated as
\(u_i\sim \mathrm{Gamma}(\mathrm{shape}=2,\mathrm{scale}=0.5)\), so that
\(E(u_i)=1\).

The counting process was generated from a mixed Poisson process with mean function
\[
E\{N_i(t)\mid Z_i,G_i,u_i\}=2t\exp(Z_i^T\beta+G_i^T\gamma)u_i,
\]
where \(u_i\sim\Gamma(2,0.5)\), with shape 2 and scale 0.5. We set \(\beta=(0.01,0.1)^T\) in the empirical size and power simulations. In the actual data generation, we generated the event counts at two observation intervals as
\[
N_i(T_{i1})\sim \mathrm{Poisson}\{2T_{i1}u_i\exp(Z_i^T\beta+G_i^T\gamma)\},
\]
\[
N_i(T_{i2})-N_i(T_{i1})\sim \mathrm{Poisson}\{2(T_{i2}-T_{i1})u_i\exp(Z_i^T\beta+G_i^T\gamma)\}.
\]

Thus, the cumulative count at the last observation time was
\[
N_i(T_{i2})=N_i(T_{i1})+\{N_i(T_{i2})-N_i(T_{i1})\}.
\]
The null proportional mean model was fitted using the \texttt{panelReg} function
with the AEEX method in the \texttt{spef} package. The terminal residual used in
the weighted V statistic was computed as
\[
\hat M_i=N_i(T_{i2})-\hat\Lambda_0(T_{i2})\exp(Z_i^\top\hat\beta).
\]

\subsection{Testing genetic association in the absence of genetic
heterogeneity}\label{subsec3.1}

In this series of simulations, we evaluated the performance of WV-PCD in testing the association between a SNP set and a panel count outcome, without involving genetic heterogeneity. The SNP data were sampled from the genotype data of the 1000 Genomes Project Phase 3 \cite{genomesProject2015}. For each subject \(i=1,\ldots,n\), the counting process was generated from a mixed Poisson process with conditional cumulative mean function

\[
\begin{aligned}
\Lambda(t\mid Z_i,G_i,u_i)
&=E\{N_i(t)\mid Z_i,G_i,u_i\} \\
&=2t\exp\left\{
\sum_{k=1}^2\beta_kZ_{ik}
+\sum_{j=1}^p\gamma_jG_{ij}
\right\}u_i ,
\end{aligned}
\]

where \(t_i\) is the visit time. \(\gamma_{j} =0 ,(j=1,...,p)\) for empirical size and
\(\gamma_{j} = 0.05\) for power assessment. IBS kernel was used to
measure the genetic similarity in WV-PCD. Table \ref{tab:wvpcd} shows that the empirical sizes of WV-PCD were close to the nominal level across different values of \(n\) and \(p\). Additionally, the power of WV-PCD increased with the sample size, and WV-PCD consistently showed greater power than WV-IC.

% =========================
% Table 1
% =========================
\begin{table}[!htbp]
\centering
\caption{Empirical size and power of WV-PCD in testing genetic effects in the absence of genetic heterogeneity.}
\label{tab:wvpcd}
\small
\begin{tabular}{ccccc}
\toprule
\(n\) & \(p\) & Empirical size & \multicolumn{2}{c}{Power} \\
\cmidrule(lr){4-5}
 & & & WV-PCD & WV-IC \\
\midrule
400 & 15 & 0.060 & 0.886 & 0.650 \\
400 & 25 & 0.052 & 0.966 & 0.769 \\
800 & 15 & 0.047 & 0.958 & 0.816 \\
800 & 25 & 0.041 & 0.993 & 0.897 \\
\bottomrule
\end{tabular}
\end{table}

\FloatBarrier

\subsection{Testing genetic association in the presence of genetic heterogeneity}\label{subsec3.2}

In this series of simulations, we investigated the performance of HWV-PCD in testing the association of a SNP set with a panel count outcome across three types of heterogeneity settings: (1) two observable subpopulations, (2) two latent subpopulations, and (3) individual genome profiles. These scenarios were designed to assess whether incorporating subpopulation similarity improves power when genetic effects are heterogeneous. The performance of \(V_{Z,PCD}\) and \(V_{Z,IC}\) was also evaluated under the same settings for comparison.

\subsubsection{Empirical size and power of HWV-PCD in the presence of genetic heterogeneity across two observable subpopulations}\label{subsubsec1}

In this simulation, we investigated the empirical size and power of \(V^H_{Z,PCD}\) under genetic heterogeneity across two observable subpopulations. The SNP data \(G\) were sampled from the 1000 Genomes Project Phase 3 genotype data. We used males and females as two observable subpopulations. Sex information was obtained from the 1000 Genomes data set and denoted by \(Z_{i1}\).

The counting process was generated according to the general simulation framework described above, with the genetic effect allowed to vary by \(Z_{i1}\):
\[
E\{N_i(t)\mid Z_i,G_i,u_i\}
=2t\exp\left\{\sum_{k=1}^2\beta_kZ_{ik}
+\sum_{j=1}^p(\gamma_{1j}+\gamma_{2j}Z_{i1})G_{ij}\right\}u_i.
\]

We set \(\beta=(0.01,0.1)^T\). For empirical size assessment, we set \(\gamma_{1j}=\gamma_{2j}=0\), \(j=1,\ldots,p\). For power assessment, we set \(\gamma_{1j}=0.01\) and \(\gamma_{2j}=0.02\), \(j=1,\ldots,p\). The IBS kernel was used to measure genetic similarity, and the identity kernel was used to measure subpopulation similarity in HWV-PCD.

Table \ref{tab:obs_heter} shows that the empirical sizes of HWV-PCD and WV-PCD were both close to the nominal level. HWV-PCD exhibited higher power by accounting for genetic heterogeneity across the two observable subpopulations. Additionally, HWV-PCD outperformed both HWV-IC and WV-IC.

% =========================
% Table 2
% =========================
\begin{table*}[!t]
\centering
\caption{Empirical size and power of HWV-PCD across two observable subpopulations under different sample sizes and marker-set sizes.}
\label{tab:obs_heter}
\small
\setlength{\tabcolsep}{4pt}
\renewcommand{\arraystretch}{1.1}
\begin{threeparttable}
\begin{tabular*}{\textwidth}{@{\extracolsep{\fill}}cccccccc}
\toprule
\(n\) & \(p\) & \multicolumn{2}{c}{Empirical size} & \multicolumn{4}{c}{Power} \\
\cmidrule(lr){3-4}\cmidrule(lr){5-8}
 & & HWV-PCD & WV-PCD & HWV-PCD & WV-PCD & HWV-IC & WV-IC \\
\midrule
400 & 15 & 0.059 & 0.061 & 0.500 & 0.479 & 0.179 & 0.166 \\
400 & 25 & 0.054 & 0.058 & 0.706 & 0.677 & 0.312 & 0.291 \\
800 & 15 & 0.052 & 0.059 & 0.670 & 0.648 & 0.323 & 0.310 \\
800 & 25 & 0.045 & 0.056 & 0.834 & 0.828 & 0.497 & 0.484 \\
\bottomrule
\end{tabular*}
\end{threeparttable}
\end{table*}

\subsubsection{Empirical size and power of HWV‐PCD in the presence of genetic heterogeneity across two latent subpopulations}\label{subsubsec2}

In this simulation, we investigated the empirical size and power of HWV-PCD under genetic heterogeneity across two latent subpopulations. The SNP data \(G\) were sampled from the 1000 Genomes Project Phase 3 genotype data.

The counting process was generated according to the general simulation framework described above, with genetic effects allowed to differ across the two latent subpopulations. For subject \(i\) in latent subpopulation \(j\), \(j=1,2\), the mean function was
\[
\begin{aligned}
E\{N_i(t)\mid Z_i,G_i,u_i,a_i=j\}
&=2t\exp\left\{
\sum_{k=1}^2\beta_kZ_{ik}
+\sum_{m=1}^p\gamma_{jm}G_{im}
\right\}u_i .
\end{aligned}
\]

We set \(\beta=(0.01,0.1)^T\). The genetic effects \(\gamma_{jm}\), \(j=1,2\), \(m=1,\ldots,p\), represent subpopulation-specific genetic effects. For empirical size assessment, we set \(\gamma_{jm}=0\) for all \(j\) and \(m\). For power assessment, \(\gamma_{jm}\)'s were assigned different values to represent different heterogeneity scenarios.
To infer the latent subpopulation structure, we generated variable \(X\in\mathbb{R}^{n\times 1}\). Specifically,
\[
X_i=\mathrm{gender}_i+1+e_i,
\]
where \(e_i\sim N(0,0.5)\), and \(\mathrm{gender}_i\in\{0,1\}\) was obtained from the 1000 Genomes data set, such that \(gender_{i} = 0\)
indicates that the \(i\)th subject is from the first subpopulation.
\(\gamma_{j,m}=0 , (m=1,...,p: j = 1,2)\) for empirical size. For power evaluation, \(\gamma_{j,m}\) were assigned to different values for different heterogeneity scenarios. The IBS kernel was used to measure genetic similarity, and the Gaussian kernel for subpopulation similarity. 

Table \ref{tab:latent_size} shows that the empirical size of HWV-PCD is close to the nominal level. Table \ref{2latent_power} demonstrates that the power of HWV-PCD increases as the sample size grows. Moreover, HWV-PCD generally outperforms WV-PCD under heterogeneous-effect scenarios, whereas WV-PCD performs better in T5, where the two subpopulations have equal effects in the same direction.

% =========================
% Table 3
% =========================
\begin{table}[!htbp]
\centering
\caption{Empirical size of HWV-PCD in testing genetic effects under genetic heterogeneity across two latent subpopulations.}
\label{tab:latent_size}
\small
\begin{tabular}{ccc}
\toprule
\(n\) & \(p\) & Empirical size \\
\midrule
400 & 15 & 0.045 \\
400 & 25 & 0.049 \\
800 & 15 & 0.043 \\
800 & 25 & 0.045 \\
\bottomrule
\end{tabular}
\end{table}

% =========================
% Table 4
% =========================

\begin{table*}[!t]
\centering
\caption{Empirical power under genetic heterogeneity across two latent
subpopulations.}
\label{2latent_power}
\setlength{\tabcolsep}{4pt}
\small

\begin{threeparttable}
\begin{tabular}{llcccc llcccc}
\toprule

\multicolumn{6}{c}{\textbf{\(n=400,\ p=15\)}} &
\multicolumn{6}{c}{\textbf{\(n=400,\ p=25\)}} \\
\cmidrule(lr){1-6} \cmidrule(lr){7-12}

Scen. & \(\gamma_1,\gamma_2\)
& HWV-PCD & WV-PCD & HWV-IC & WV-IC
&
Scen. & \(\gamma_1,\gamma_2\)
& HWV-PCD & WV-PCD & HWV-IC & WV-IC \\
\midrule

T1 & $-0.01, 0.01$ & 0.183 & 0.050 & 0.110 & 0.057 &
T1 & $-0.01, 0.01$ & 0.431 & 0.052 & 0.198 & 0.032 \\

T2 & $-0.01, 0.02$ & 0.412 & 0.105 & 0.172 & 0.054 &
T2 & $-0.01, 0.02$ & 0.800 & 0.201 & 0.421 & 0.042 \\

T3 & $0, 0.02$ & 0.312 & 0.167 & 0.121 & 0.067 &
T3 & $0, 0.02$ & 0.638 & 0.315 & 0.242 & 0.103 \\

T4 & $0.01, 0.03$ & 0.550 & 0.431 & 0.186 & 0.168 &
T4 & $0.01, 0.03$ & 0.848 & 0.666 & 0.388 & 0.298 \\

T5 & $0.02, 0.02$ & 0.361 & 0.411 & 0.157 & 0.184 &
T5 & $0.02, 0.02$ & 0.545 & 0.616 & 0.246 & 0.314 \\

\midrule

\multicolumn{6}{c}{\textbf{\(n=800,\ p=15\)}} &
\multicolumn{6}{c}{\textbf{\(n=800,\ p=25\)}} \\
\cmidrule(lr){1-6} \cmidrule(lr){7-12}

Scen. & \(\gamma_1,\gamma_2\)
& HWV-PCD & WV-PCD & HWV-IC & WV-IC
&
Scen. & \(\gamma_1,\gamma_2\)
& HWV-PCD & WV-PCD & HWV-IC & WV-IC \\
\midrule

T1 & $-0.01, 0.01$ & 0.312 & 0.055 & 0.159 & 0.046 &
T1 & $-0.01, 0.01$ & 0.716 & 0.076 & 0.385 & 0.062 \\

T2 & $-0.01, 0.02$ & 0.667 & 0.184 & 0.322 & 0.057 &
T2 & $-0.01, 0.02$ & 0.966 & 0.345 & 0.689 & 0.083 \\

T3 & $0, 0.02$ & 0.550 & 0.305 & 0.240 & 0.131 &
T3 & $0, 0.02$ & 0.888 & 0.501 & 0.470 & 0.190 \\

T4 & $0.01, 0.03$ & 0.784 & 0.672 & 0.404 & 0.353 &
T4 & $0.01, 0.03$ & 0.956 & 0.826 & 0.690 & 0.494 \\

T5 & $0.02, 0.02$ & 0.584 & 0.640 & 0.314 & 0.359 &
T5 & $0.02, 0.02$ & 0.742 & 0.798 & 0.467 & 0.548 \\

\bottomrule
\end{tabular}

\begin{tablenotes}
\small
\item Note: Various heterogeneity scenarios were considered, determined by the values of \(\gamma_1\) and \(\gamma_2\), including the same effect size but opposite effect direction (T1), different sizes and opposite directions (T2), no effect in one subpopulation while positive effect in the other (T3), different sizes but same directions (T4), and same sizes and the same direction (T5).
\end{tablenotes}
\end{threeparttable}
\end{table*}

\subsubsection{Empirical size and power of HWV‐PCD in the presence of genetic heterogeneity across individual genome profiles}\label{subsubsec3.2.3}

In this simulation, we investigated the performance of HWV-PCD when the genetic effect varied across individual genome profiles. The SNP set \(G\) under testing was generated to mimic linkage disequilibrium (LD) between SNPs within a marker set. Specifically, genotypes were generated by a two-step procedure. First, we sampled \(n\) independent vectors from a multivariate normal distribution with zero mean and covariance matrix \(\Sigma_G=\{0.3^{|k-l|}\}_{p\times p}\), where \(k,l=1,\ldots,p\) are SNP indices. Second, each component of the multivariate normal vectors was categorized into three levels, labeled 0, 1, and 2, using cut-off values selected to achieve Hardy-Weinberg equilibrium (HWE), with the minor allele frequency (MAF) for each SNP randomly sampled from \(U(0.1,0.4)\).

The counting process followed the mixed Poisson process described in the general simulation framework. Specifically, for each subject \(i=1,\ldots,n\), the mean function was
\[
E\{N_i(t)\mid Z_i,G_i,u_i\}
=2t\exp\left(\sum_{k=1}^2\beta_kZ_{ik}+\sum_{m=1}^p\gamma_{im}G_{im}\right)u_i,
\]
where \(u_i\sim\Gamma(2,0.5)\), with shape 2 and scale 0.5, and \(\beta=(0.01,0.1)^T\). We set \(\gamma_{i1}=\cdots=\gamma_{ip}=\gamma_i\). For empirical size assessment, we set \(\gamma_i=0\) for all subjects. For power assessment, \(\gamma_i\)'s were randomly sampled from a uniform distribution with mean \(\mu_\gamma\) and variance \(\sigma_\gamma^2\). The values of \(\mu_\gamma\) and \(\sigma_\gamma^2\) varied across simulation scenarios to represent different levels of mean genetic effect and genetic heterogeneity.

To represent individual genome profiles, we simulated a set of 1000 SNPs for each subject, denoted by \(\{X_{id}\}_{d=1}^{1000}\), \(i=1,\ldots,n\). The genome profile was generated in two steps. First, for each \(d=1,\ldots,1000\), we sampled
\[
\tilde X_d=(\tilde X_{1d},\ldots,\tilde X_{nd})^T
\]
from a multivariate normal distribution \(N(0,\Sigma_X)\), where \(\Sigma_X\) is an \(n\times n\) covariance matrix. Under the null hypothesis of no genetic association, the \((i,j)\)-th element was \(\Sigma_{X,ij}=I(i=j)\). Under the alternative hypothesis, the \((i,j)\)-th element was
\( \Sigma_{X,ij}=\exp\{-|\gamma_i-\gamma_j|/\sigma_\gamma\}. \)
Second, \(X_{id}\) was obtained by categorizing \(\tilde X_{id}\) into three levels, 0, 1, and 2, using rank-based cut-off values. The cut-off values were selected to achieve HWE, with the pre-specified MAF randomly sampled from \(U(0.05,0.2)\).

The IBS kernel was used to measure both genetic similarity based on \(G\) and subpopulation similarity based on the genome profile \(X\) in HWV-PCD. Table \ref{genome-profile-p} shows that the empirical size of HWV-PCD was close to the nominal level. Table \ref{tab:genome-profile-power} shows that the power of HWV-PCD increased with the mean effect size \(\mu_\gamma\). In addition, for a fixed mean effect size, the power of HWV-PCD improved as the genetic heterogeneity \(\sigma_\gamma\) increased.

\begin{table}[htbp]
\caption{Empirical size of HWV‐PCD in testing genetic effects under genetic heterogeneity across individual genome profiles}
\label{genome-profile-p}
\centering
\begin{tabular}{llc}
\toprule
Sample Size  & SNP size & Empirical Size (HWV) \\
\midrule
n = 400 & p = 15 & 0.045 \\
n = 400 & p = 25 & 0.042 \\
n = 800 & p = 15 & 0.050 \\
n = 800 & p = 25 & 0.046 \\
\bottomrule
\end{tabular}
\end{table}
\FloatBarrier

\begin{table*}[!t]
\centering
\caption{Power comparison under genetic heterogeneity across individual genome profiles.}
\label{tab:genome-profile-power}
\setlength{\tabcolsep}{3pt}
\small
\begin{threeparttable}
\begin{tabular}{lcccc@{\hspace{10pt}}lcccc}
\toprule
\multicolumn{10}{l}{\textbf{Number of subjects : } \(n=400\)} \\
\midrule
\multicolumn{5}{c}{\(p = 15\)} & \multicolumn{5}{c}{\(p = 25\)} \\
\cmidrule(lr){1-5} \cmidrule(lr){6-10}
\(\mu_\gamma,\sigma_\gamma\) & HWV-PCD & WV-PCD & HWV-IC & WV-IC
& \(\mu_\gamma,\sigma_\gamma\) & HWV-PCD & WV-PCD & HWV-IC & WV-IC \\
\midrule
0, 0.01    & 0.083 & 0.062 & 0.054 & 0.039 & 0, 0.01    & 0.166 & 0.046 & 0.084 & 0.046 \\
0, 0.02    & 0.236 & 0.070 & 0.110 & 0.049 & 0, 0.02    & 0.772 & 0.054 & 0.315 & 0.048 \\
0, 0.04    & 0.888 & 0.083 & 0.403 & 0.051 & 0, 0.04    & 1.000 & 0.151 & 0.978 & 0.059 \\
0.02, 0.01 & 0.155 & 0.114 & 0.063 & 0.055 & 0.02, 0.01 & 0.316 & 0.127 & 0.090 & 0.052 \\
0.02, 0.02 & 0.280 & 0.124 & 0.093 & 0.056 & 0.02, 0.02 & 0.873 & 0.207 & 0.315 & 0.078 \\
0.02, 0.04 & 0.910 & 0.227 & 0.328 & 0.059 & 0.02, 0.04 & 1.000 & 0.353 & 0.934 & 0.065 \\
0.04, 0.01 & 0.444 & 0.390 & 0.118 & 0.106 & 0.04, 0.01 & 0.725 & 0.542 & 0.182 & 0.124 \\
0.04, 0.02 & 0.589 & 0.395 & 0.149 & 0.106 & 0.04, 0.02 & 0.987 & 0.605 & 0.406 & 0.146 \\
0.04, 0.04 & 0.969 & 0.512 & 0.353 & 0.100 & 0.04, 0.04 & 1.000 & 0.729 & 0.954 & 0.077 \\
0.08, 0.01 & 0.982 & 0.973 & 0.399 & 0.397 & 0.08, 0.01 & 0.999 & 0.997 & 0.436 & 0.419 \\
0.08, 0.02 & 0.993 & 0.977 & 0.458 & 0.402 & 0.08, 0.02 & 1.000 & 0.998 & 0.574 & 0.401 \\
0.08, 0.04 & 1.000 & 0.983 & 0.657 & 0.321 & 0.08, 0.04 & 1.000 & 0.997 & 0.925 & 0.301 \\
\midrule
\multicolumn{10}{l}{\textbf{Number of subjects: } \(n=800\)} \\
\midrule
\multicolumn{5}{c}{\(p = 15\)} & \multicolumn{5}{c}{\(p = 25\)} \\
\cmidrule(lr){1-5} \cmidrule(lr){6-10}
\(\mu_\gamma,\sigma_\gamma\) & HWV-PCD & WV-PCD & HWV-IC & WV-IC
& \(\mu_\gamma,\sigma_\gamma\) & HWV-PCD & WV-PCD & HWV-IC & WV-IC \\
\midrule
0, 0.01    & 0.098 & 0.045 & 0.056 & 0.041 & 0, 0.01    & 0.421 & 0.054 & 0.152 & 0.041 \\
0, 0.02    & 0.418 & 0.063 & 0.161 & 0.048 & 0, 0.02    & 0.975 & 0.044 & 0.642 & 0.040 \\
0, 0.04    & 0.996 & 0.088 & 0.700 & 0.044 & 0, 0.04    & 1.000 & 0.171 & 0.999 & 0.055 \\
0.02, 0.01 & 0.343 & 0.204 & 0.126 & 0.085 & 0.02, 0.01 & 0.698 & 0.284 & 0.226 & 0.096 \\
0.02, 0.02 & 0.847 & 0.265 & 0.286 & 0.090 & 0.02, 0.02 & 0.999 & 0.369 & 0.651 & 0.098 \\
0.02, 0.04 & 0.999 & 0.396 & 0.649 & 0.073 & 0.02, 0.04 & 1.000 & 0.637 & 1.000 & 0.063 \\
0.04, 0.01 & 0.796 & 0.721 & 0.247 & 0.221 & 0.04, 0.01 & 0.976 & 0.889 & 0.407 & 0.266 \\
0.04, 0.02 & 0.962 & 0.781 & 0.411 & 0.242 & 0.04, 0.02 & 1.000 & 0.907 & 0.727 & 0.243 \\
0.04, 0.04 & 1.000 & 0.877 & 0.855 & 0.194 & 0.04, 0.04 & 1.000 & 0.980 & 1.000 & 0.142 \\
0.08, 0.01 & 1.000 & 1.000 & 0.790 & 0.776 & 0.08, 0.01 & 1.000 & 1.000 & 0.856 & 0.813 \\
0.08, 0.02 & 1.000 & 1.000 & 0.830 & 0.735 & 0.08, 0.02 & 1.000 & 1.000 & 0.961 & 0.794 \\
0.08, 0.04 & 1.000 & 1.000 & 0.938 & 0.633 & 0.08, 0.04 & 1.000 & 1.000 & 1.000 & 0.620 \\
\bottomrule
\end{tabular}
\begin{tablenotes}
\footnotesize
\item \(\mu_\gamma\) and \(\sigma_\gamma\) denote the mean and standard deviation of the subject-specific genetic effects, respectively.
\end{tablenotes}
\end{threeparttable}
\end{table*}

\subsection{Small-sample correction}\label{subsec3.3}

In this series of simulations, we evaluated the performance of the small-sample corrected tests proposed in Section 2.4. The purpose of these simulations was to examine whether the correction can improve finite-sample performance when the sample size is relatively small. We considered two settings. First, in the absence of genetic heterogeneity, we compared WV-PCD, \(V_{Z,PCD}\), with its small-sample corrected version, \(V^c_{Z,PCD}\). Second, under genetic heterogeneity across two observable subpopulations, we compared WV-PCD and HWV-PCD with their small-sample corrected versions, \(V^c_{Z,PCD}\) and \(V^{H,c}_{Z,PCD}\), respectively.

The simulation settings followed those in Sections 3.1 and 3.2.1, except that the sample size was reduced to \(n=150\) and the marker-set size was fixed at \(p=25\). Tables~\ref{tab:small_noheter} and~\ref{tab:small_heter} report the empirical sizes and powers of the original and small-sample corrected tests. In the absence of genetic heterogeneity, both WV-PCD and WV-PCD-SC maintained empirical sizes close to the nominal level, while WV-PCD-SC achieved slightly higher power. Under genetic heterogeneity, all four tests had empirical sizes close to 0.05. HWV-PCD-SC achieved the highest power, indicating that the small-sample correction can slightly improve power while preserving size control in small samples.

\begin{figure*}[!t]
\centering
\footnotesize
\setlength{\tabcolsep}{4pt}

\begin{tabular}{cc}
\textbf{(A) WV-PCD} & \textbf{(B) WV-PCD-SC} \\[-1mm]
\includegraphics[width=0.36\textwidth]{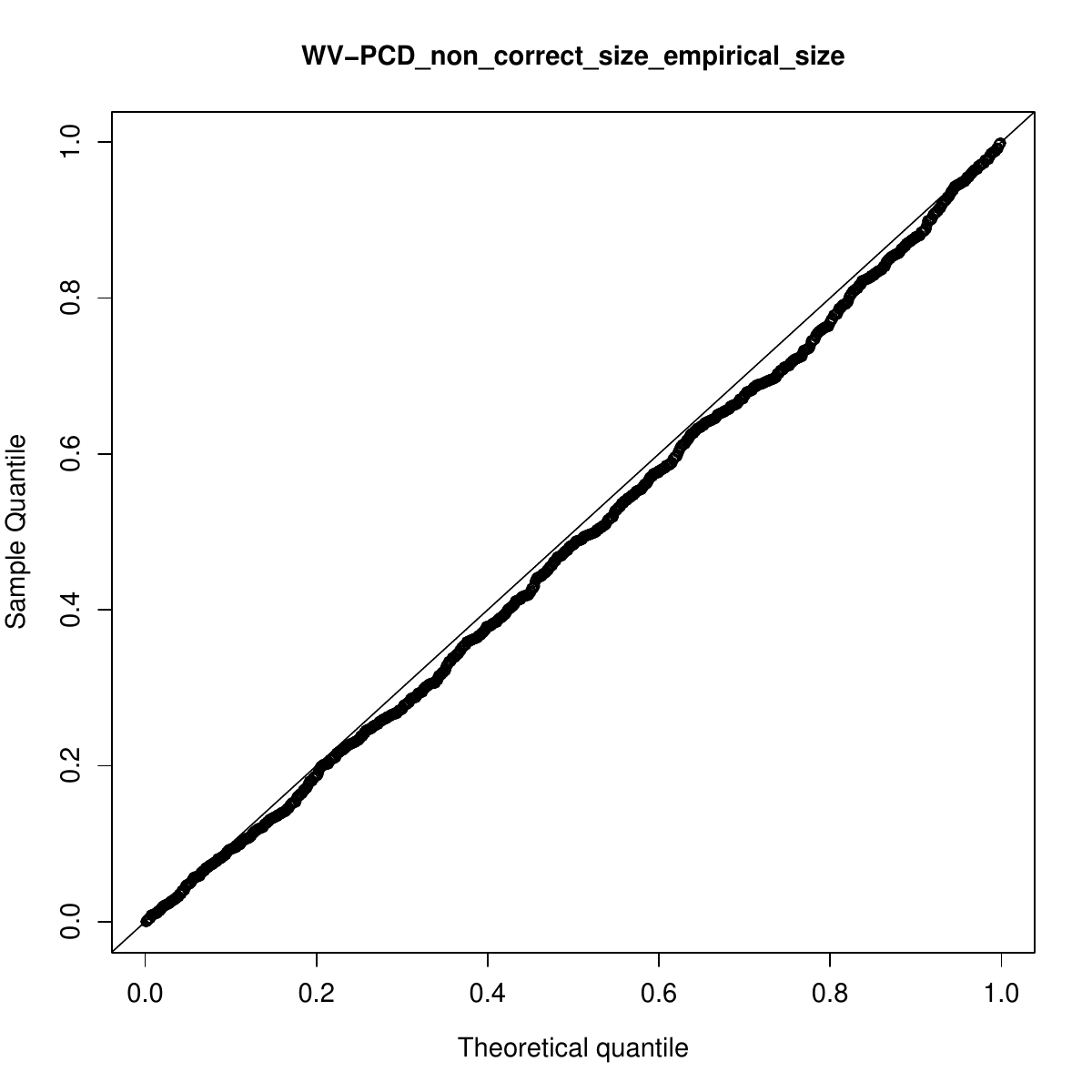} &
\includegraphics[width=0.36\textwidth]{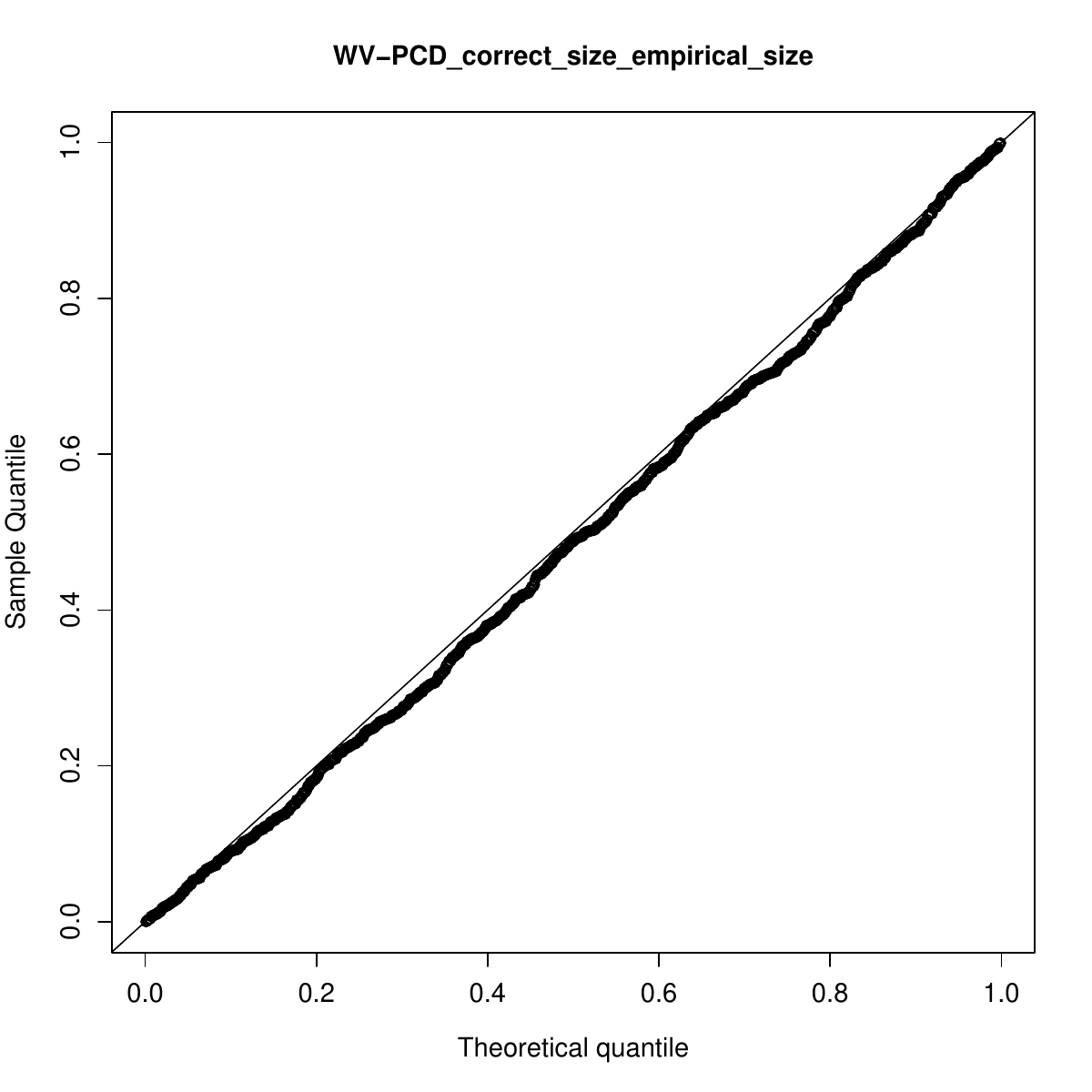}
\end{tabular}

\vspace{-2mm}
\caption{Q--Q plots under the null setting without genetic heterogeneity for the small-sample scenario with $n=150$ and $p=25$.}
\label{fig:qq-sc-noheter1}
\end{figure*}

\begin{figure*}[!t]
\centering
\footnotesize
\setlength{\tabcolsep}{3pt}

\begin{tabular}{cc}
\textbf{(A) HWV-PCD} & \textbf{(B) HWV-PCD-SC} \\[-1mm]
\includegraphics[width=0.32\textwidth]{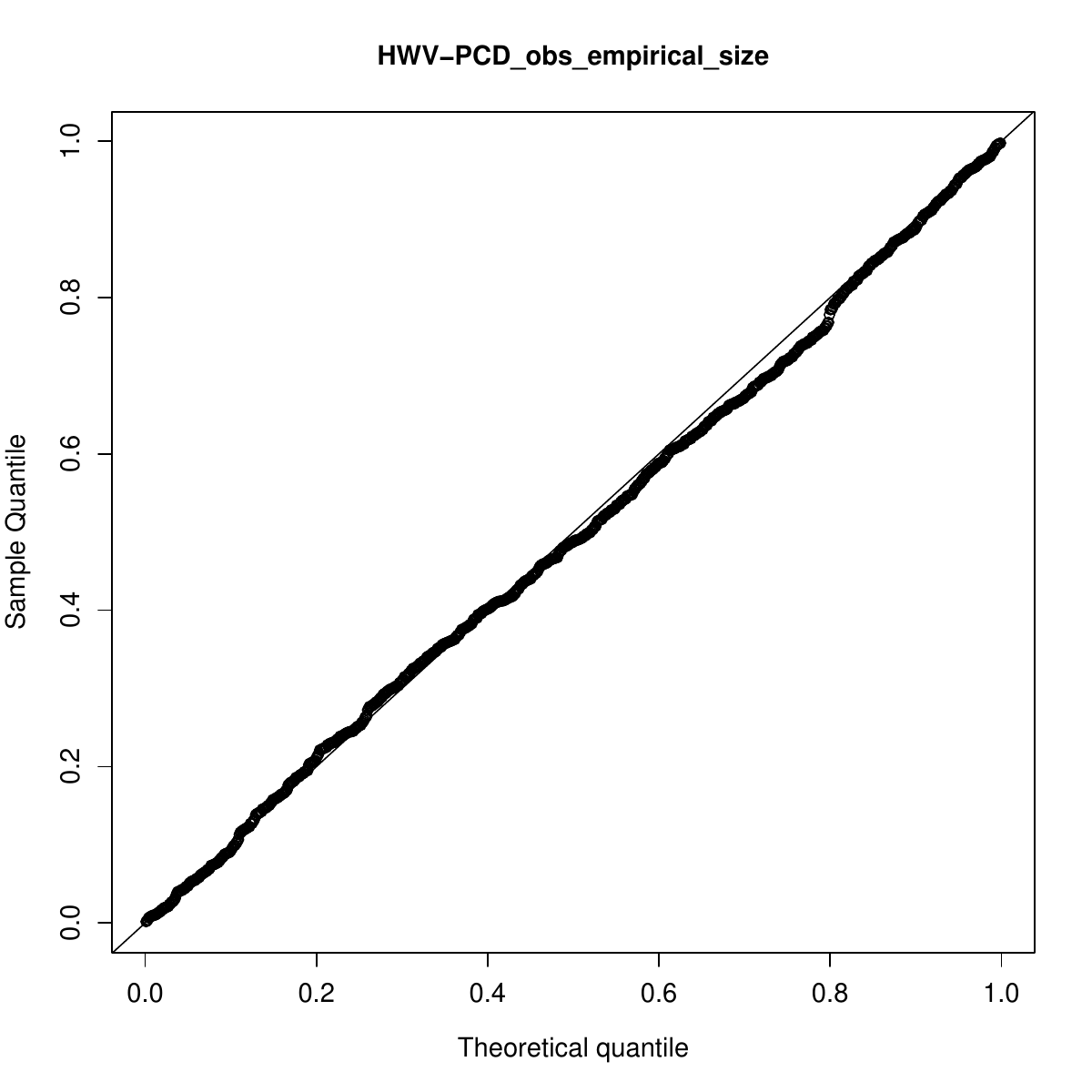} &
\includegraphics[width=0.32\textwidth]{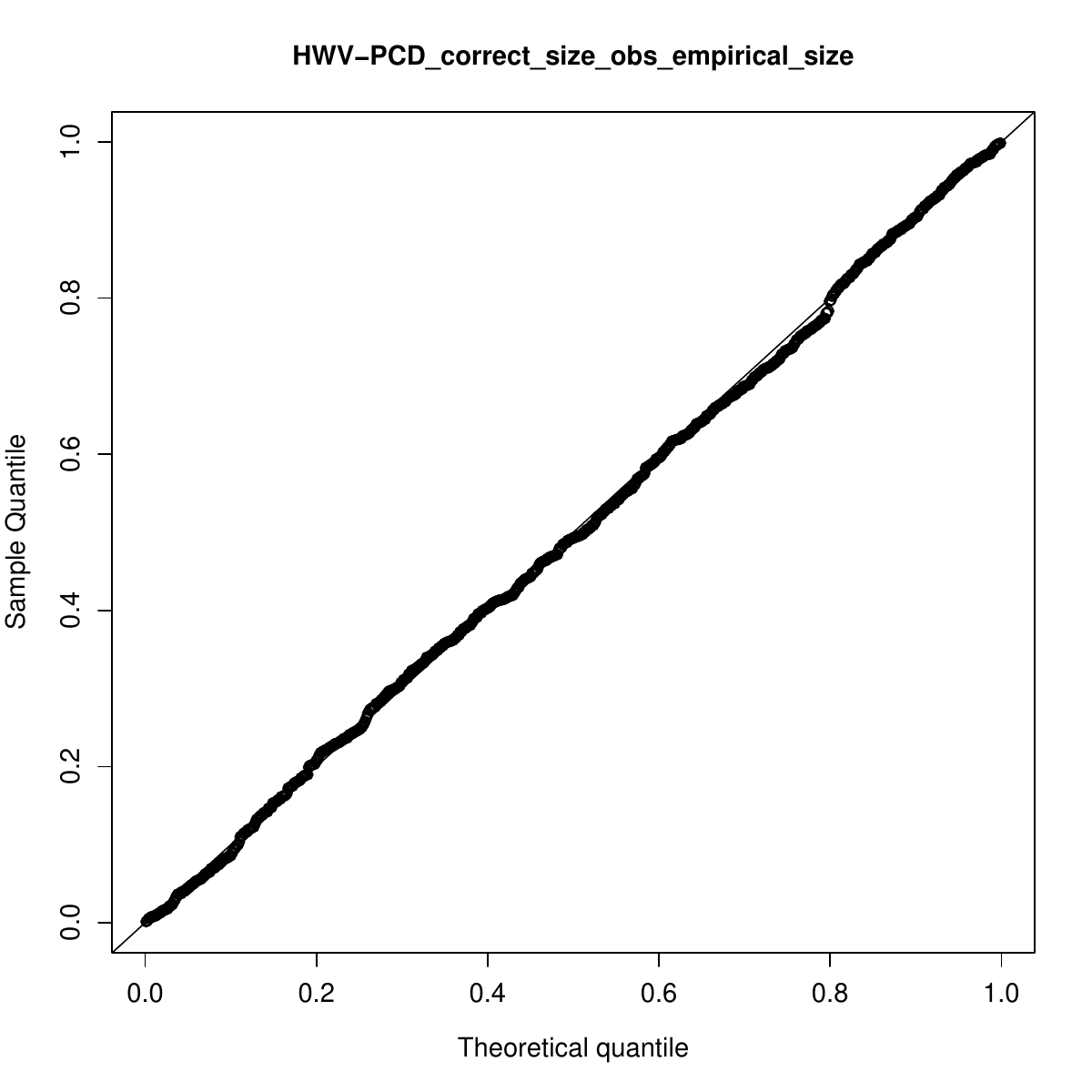} \\[-1mm]

\textbf{(C) WV-PCD} & \textbf{(D) WV-PCD-SC} \\[-1mm]
\includegraphics[width=0.32\textwidth]{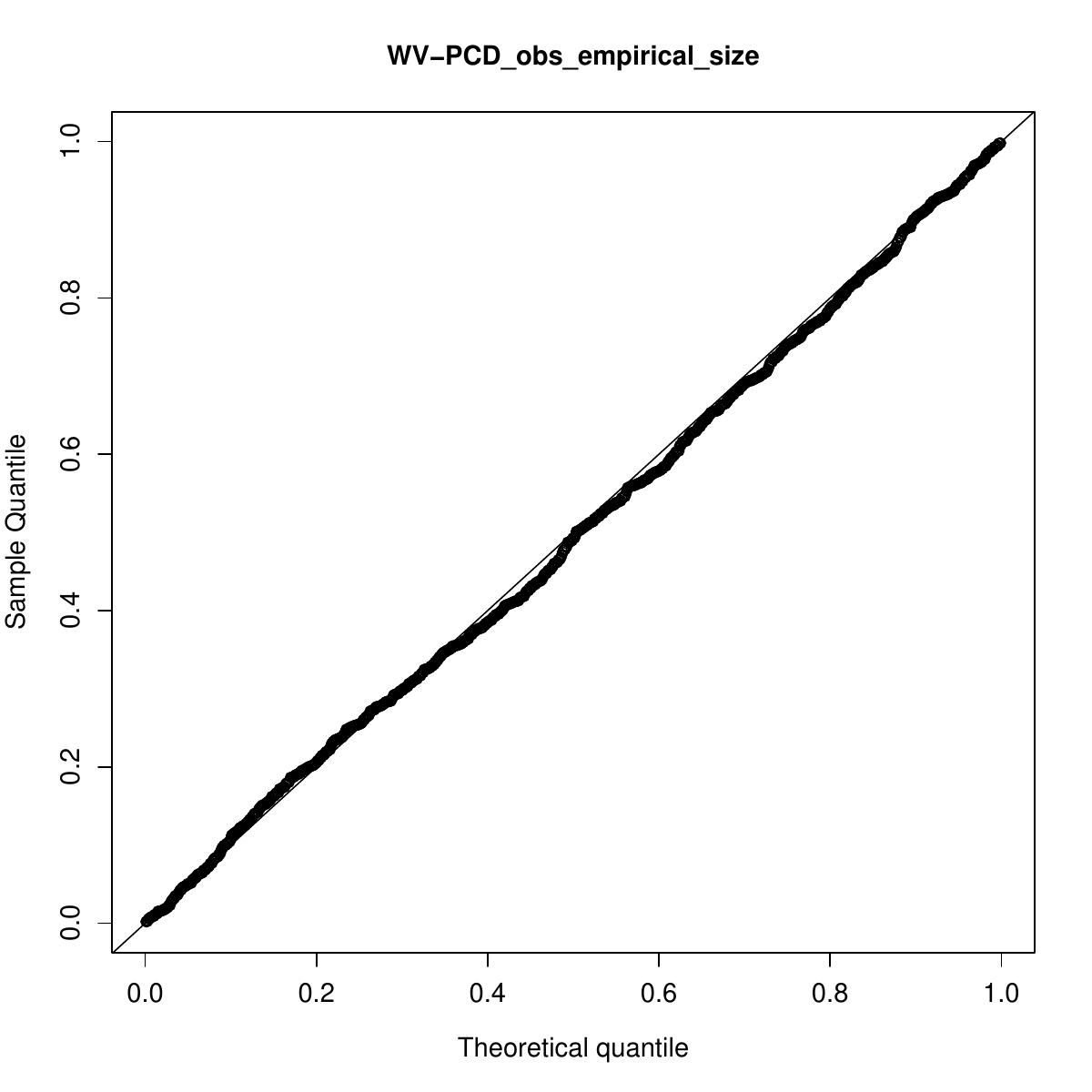} &
\includegraphics[width=0.32\textwidth]{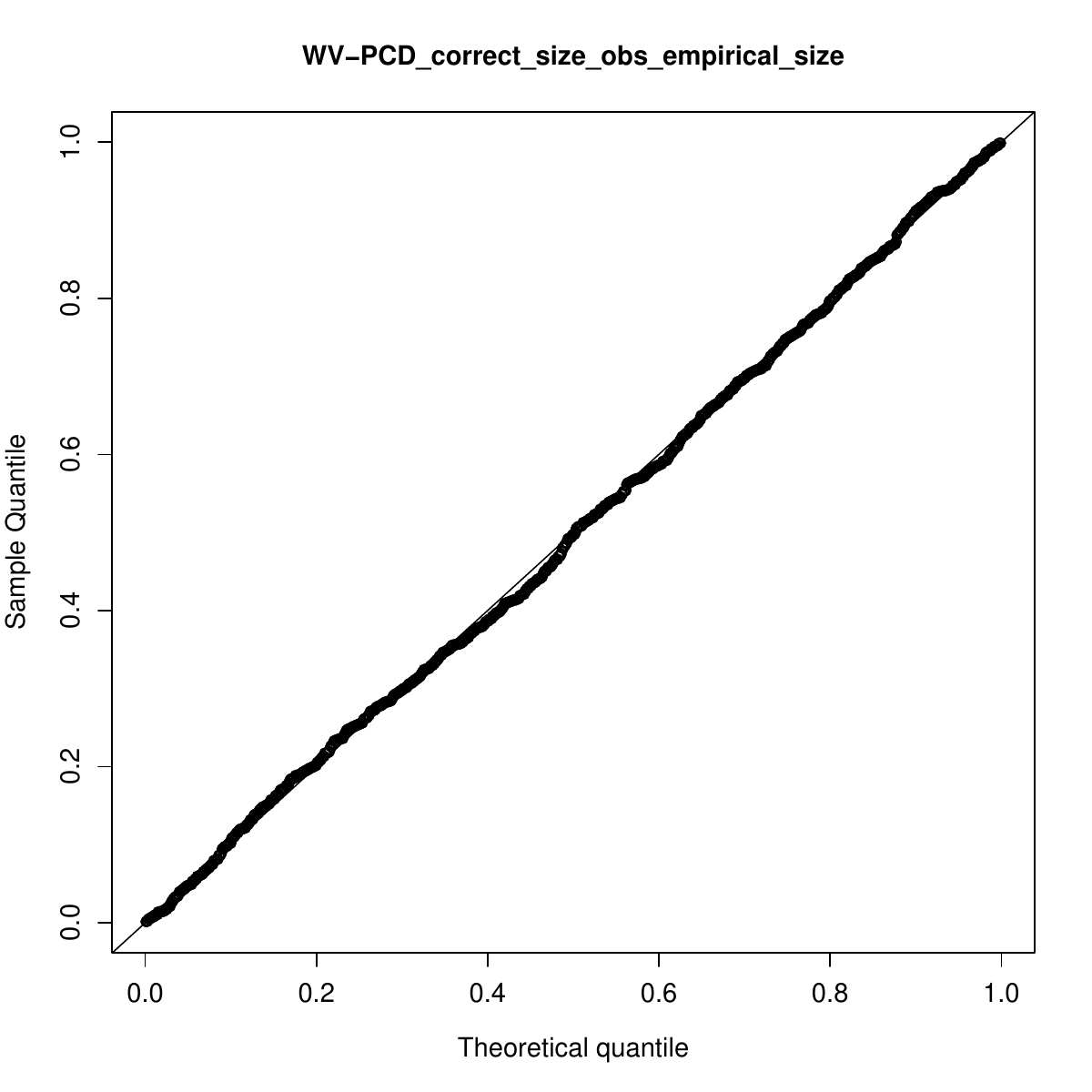}
\end{tabular}

\vspace{-2mm}
\caption{Q--Q plots for the two-observed-subpopulation setting under the small-sample scenario with $n=150$ and $p=25$.}
\label{fig:qq-sc-2obs2}
\end{figure*}

Figures~\ref{fig:qq-sc-noheter1} and \ref{fig:qq-sc-2obs2} further compare the empirical distributions of the p-values before and after small-sample correction using Q--Q plots. After correction, the empirical p-value quantiles were closer to the theoretical uniform quantiles, especially in the lower tail, indicating improved finite-sample calibration of the proposed tests.

% =========================
% Table 7: Small-sample correction without heterogeneity
% =========================
\begin{table}[!t]
\centering
\caption{Empirical size and power of WV-PCD and its small-sample corrected version in the absence of genetic heterogeneity, with \(n=150\) and \(p=25\).}
\label{tab:small_noheter}
\small
\begin{tabular}{lcc}
\toprule
Measure & WV-PCD & WV-PCD-SC \\
\midrule
Empirical size & 0.052 & 0.054 \\
Power          & 0.854 & 0.860 \\
\bottomrule
\end{tabular}
\end{table}
\FloatBarrier

% =========================
% Table 8: Small-sample correction under heterogeneity
% =========================
\begin{table}[!t]
\centering
\caption{Empirical size and power of HWV-PCD and WV-PCD, compared with their small-sample corrected versions, under genetic heterogeneity across two observable subpopulations, with \(n=150\) and \(p=25\).}
\label{tab:small_heter}
\small
\begin{tabular}{lcccc}
\toprule
Measure & HWV-PCD & WV-PCD & HWV-PCD-SC & WV-PCD-SC \\
\midrule
Empirical size & 0.051 & 0.049 & 0.056 & 0.054 \\
Power          & 0.412 & 0.397 & 0.425 & 0.402 \\
\bottomrule
\end{tabular}
\end{table}
\FloatBarrier

% Q-Q plot: no heterogeneity, n=150, p=25

% Q-Q plot: two observed subpopulations, n=150, p=25

\subsection{Empirical sizes under stringent p-value thresholds}\label{subsec3.4}

Genome-wide association studies with genotyping or sequencing data usually involve testing the associations between hundreds of thousands of genetic variants and a phenotype, which may cause a severe multiple-testing problem. Common approaches to address multiple testing, such as the Bonferroni correction and the Benjamini-Hochberg procedure \cite{benjamini1995fdr}, lead to stringent p-value thresholds when applied to large-scale genetic association analyses. In this simulation, we examined the empirical sizes of WV-PCD and HWV-PCD, with and without small-sample correction, under stringent p-value thresholds much smaller than 0.05.

The simulation setting for WV-PCD was the same as that used to assess its empirical size in the absence of genetic heterogeneity in Section 3.1. The simulation setting for HWV-PCD was the same as that used to assess its empirical size in the presence of genetic heterogeneity across two observable subpopulations in Section 3.2.1. In both scenarios, 150,000 Monte Carlo replicates with \(n=400\) and \(p=25\) were generated to compute the empirical sizes.

Tables \ref{tab_stringent_p1} and \ref{tab_stringent_p2} show that the empirical sizes of the tests were close to the corresponding stringent p-value thresholds. These results suggest that WV-PCD and HWV-PCD, with or without small-sample correction, maintain appropriate size control under stringent thresholds and are suitable for large-scale genetic association analyses.

\begin{table}[!t]
\centering
\caption{Empirical sizes of WV and WV-SC under stringent p-value thresholds in the absence of genetic heterogeneity, with $n = 400$ and $p = 25$.}
\label{tab_stringent_p1}
\begin{tabular}{lll}
\toprule
& \multicolumn{2}{c}{Empirical size} \\
\cmidrule(lr){2-3}
Threshold & WV & WV-SC \\
\midrule
0.005 & 0.004727 & 0.004980 \\
0.0005 & 0.0003867 & 0.0004400 \\
0.00005 & 0.00004667 & 0.00006000 \\
\bottomrule
\end{tabular}
\end{table}

\begin{table}[!t]
\centering
\caption{Empirical sizes of HWV and WV, HWV-SC, WV-SC under stringent p-value thresholds, in the presence of genetic heterogeneity across two observed subpopulations, with $n = 400$ and $p = 25$.}
\label{tab_stringent_p2}
\begin{tabular}{lllll}
\toprule
& \multicolumn{4}{c}{Empirical size} \\
\cmidrule(lr){2-5}
Threshold & HWV & WV & HWV-SC & WV-SC \\
\midrule
0.005 & 0.005033 & 0.004967 & 0.005467 & 0.005287 \\
0.0005 & 0.0004867 & 0.0004667 & 0.0005667 & 0.0005733 \\
0.00005 & 0.00003333 & 0.00002667 & 0.00004667 & 0.00004667 \\
\bottomrule
\end{tabular}
\end{table}

\section{Real Data Analysis}\label{sec4}

Dental caries is a complex disease with genetic, microbial, behavioral, and environmental components. Previous studies have shown evidence of genetic contributions to caries and have suggested that genetic effects may vary across dentitions, tooth surfaces, and population groups \cite{bretz2005twins,wang2010dentitions,shaffer2012surfaces,
vieira2014review,orlova2019heterogeneity}. In this study, we applied the proposed WV-PCD and HWV-PCD tests to the ZOE 2.0 study, a community-based genetic epidemiologic study of early childhood oral health \cite{divaris2020zoe2}.The goal of this real-data analysis was to identify genes associated with the cumulative number of decayed, missing, or filled primary tooth surfaces.

The original phenotype file contained 6,161 subjects. The phenotype of interest was ECC experience measured at the ICDAS $\geq 3$ threshold. Let \(N_i(t)\) denote the cumulative number of decayed, missing, or filled tooth surfaces for subject \(i\) by age \(t\). In the ZOE2 data, each child had a clinical examination at age \(V_i\), recorded in months as \texttt{AGE\_MONTHS}. The observed count outcome was \texttt{t6c}, defined as the cumulative number of ICDAS $\geq 3$ carious tooth surfaces by the examination age. Therefore, the observed data were represented as panel count data \((V_i, N_i(V_i))\), where \(V_i\) is \texttt{AGE\_MONTHS} and \(N_i(V_i)\) is \texttt{t6c}.

We used the covariate-adjusted model in the real-data analysis. The covariates included sex, race indicators, optimal home water fluoride exposure, sugar-sweetened beverage score (SSBS), and the first eight ancestry principal components. The first eight ancestry principal components were included to adjust for population structure, consistent with the covariate adjustment strategy used in Shrestha et al. \cite{shrestha2025eccgwas}.

To increase genomic coverage, genotype data were derived from two sources: directly genotyped SNP array data and dbGaP-provided imputed untyped genotype data. The directly genotyped array data were first quality-controlled using PLINK 2.0 \cite{chang2015plink}. Specifically, following standard GWAS quality-control procedures \cite{reed2015gwasqc}, SNPs with genotype missingness greater than 2\% or minor allele frequency less than 1\% were removed; subjects with genotype missingness greater than 5\% were excluded; non-standard variants were removed by retaining only A/C/G/T alleles; related individuals were filtered using KING with a kinship cutoff of 0.1; duplicate variants were removed by SNP ID and chromosome-position pairs; and Hardy--Weinberg equilibrium filtering was applied using a threshold of \(10^{-6}\). After this initial array QC, 888,057 directly genotyped SNPs and 5,805 subjects remained. LD pruning was performed only for principal component analysis, using a 50-variant window, a step size of 5 variants, and an \(r^2\) threshold of 0.2.

The QCed array genotype data were then imputed using IMPUTE2 to recover additional variants that were not directly typed on the array. IMPUTE2 was run chromosome by chromosome using approximately 5-Mb genomic windows, with adjacent sparse windows merged when fewer than 50 SNPs were present. Segment-level imputed files were merged using gtool with a genotype-call probability threshold of 0.9 and converted back to PLINK binary format. Because the array data were originally aligned to hg19/b37, whereas the externally imputed untyped data were aligned to hg38, the array-imputed data were lifted over to hg38 using UCSC liftOver \cite{kent2002ucsc}. Remaining missing genotypes after liftOver were imputed chromosome by chromosome using \texttt{bigsnpr} \cite{prive2018bigsnpr}.

The dbGaP-provided imputed untyped genotype data were separately subjected to simple quality control using minor allele frequency and imputation-quality \(R^2\) filters. Non-SNP and non-standard variants were removed by retaining only A/C/G/T SNPs. After QC, the externally imputed untyped data included 13,052,839 SNPs and 6,103 subjects. The processed array-imputed data and the QCed externally imputed untyped data were then merged. The final merged genotype dataset included 13,098,866 SNPs and 5,731 subjects before phenotype matching.

After matching the final merged genotype dataset with phenotype data, ancestry principal components, and covariates, and removing subjects with missing values, 5,587 subjects remained for the real-data analysis.

Gene-level SNP sets were constructed using UCSC hg38 gene annotations. Variants located within a gene region or within 5 kb upstream or downstream of a gene were assigned to that gene according to their hg38 genomic positions. After phenotype matching and within-gene SNP filtering, 23,210 genes with 7,001,566 gene-level SNP entries were included in the final analysis. Within each gene, SNPs with missing genotypes among the analyzed subjects or without genotype variation were removed before constructing the gene-level kernel.

Using the covariate-adjusted panel count model, we first applied the standard WV-PCD test and its small-sample corrected version without explicitly modeling genetic heterogeneity. We then applied HWV-PCD to account for potential heterogeneous genetic effects across subpopulations. The heterogeneity sources considered included sex, home water fluoride exposure, SSBS, and genome-wide genetic background. The IBS kernel was used to measure gene-level genetic similarity among subjects. For observed categorical heterogeneity variables, an identity kernel was used to measure subpopulation similarity. For genetic-background heterogeneity, 200,000 SNPs were randomly selected from the whole genome to construct an IBS similarity kernel among subjects.

For each gene, p-values were computed using Davies' method \cite{davies1980algorithm} based on the
asymptotic null distribution of the corresponding weighted V statistic.
Multiple testing was controlled separately for each gene-based association
test using the Benjamini--Yekutieli false discovery rate (FDR) procedure \cite{benjamini2001control}
under arbitrary dependence. For $m=23{,}210$ tested genes, the FDR-based
p-value threshold for the $i$th ordered p-value was defined as
$\alpha i /\{m\sum_{k=1}^{m}(1/k)\}$, $i=1,\ldots,m$, where $\alpha=0.10$
was the target FDR level. No individual gene reached FDR significance in
any of the gene-based association tests. Therefore, the top-ranked genes
were interpreted as suggestive signals rather than statistically significant discoveries. Raw gene-level p-values for the top-ranked genes are reported in Table~\ref{tab:top5_genes_full}.

Because no individual gene reached multiple-testing corrected significance, we further performed gene set enrichment analysis using ranked gene-level p-values from the WV-PCD and HWV-PCD tests. Gene-set enrichment analysis was performed using a ranked-list GSEA framework. Genes were ranked by \(-\log_{10}(p)\), where \(p\) was the gene-level association p-value from each test. Because genes were ranked by the unsigned statistic \(-\log_{10}(p)\), the ranking reflects association strength rather than effect direction. Thus, NES values should not be interpreted as up- or down-regulation, or as risk-increasing or protective effects. The enrichment results are interpreted as pathway-level association signals. Gene Ontology enrichment analyses were conducted separately for Biological Process, Molecular Function, and Cellular Component terms using the \texttt{gseGO} function in the \texttt{clusterProfiler} R package \cite{subramanian2005gsea,ashburner2000go,yu2012clusterprofiler,wu2021clusterprofiler}. Human gene annotations were obtained from \texttt{org.Hs.eg.db}. Multiple testing was controlled using the Benjamini--Hochberg procedure \cite{benjamini1995fdr}. 

In the model, the most interpretable enrichment signals were observed for HWV-PCD using sugar-sweetened beverage score (SSBS) as the heterogeneity source. Significant GO biological process terms included negative regulation of cellular glucuronidation, natural killer cell activation involved in immune response, humoral immune response, antimicrobial humoral response, and regulation of cellular glucuronidation. Significant GO molecular function terms included type I interferon receptor binding, bitter taste receptor activity, taste
receptor activity, GTPase regulator activity, nucleoside-triphosphatase
regulator activity, N-acyltransferase activity, and vinculin binding. These signals were largely consistent after applying the small-sample correction. Detailed significant GO enrichment results are provided in Appendix Table~\ref{tab:appendix-gsea-go}.

The enrichment of taste receptor activity and bitter taste receptor activity is particularly noteworthy because Shrestha et al \cite{shrestha2025eccgwas} also reported taste receptor activity as a significant gene-set finding. This overlap supports the biological relevance of taste-related pathways in early childhood caries and suggests that modeling SSBS-related heterogeneity may help identify pathway-level genetic signals related to sugar-associated ECC risk. The immune-related enrichment terms further suggest that host immune and antimicrobial defense pathways may contribute to ECC susceptibility. Enrichment results from the other WV-PCD and HWV-PCD tests were less consistent and are provided in the supplementary materials.

\begin{table*}[!t]
\centering
\caption{Top five ranked genes from each gene-based association test for ZOE2.0 data. Raw gene-level p-values are shown below each gene symbol. None of the listed genes remained significant after FDR correction.}
\label{tab:top5_genes_full}
\small
\renewcommand{\arraystretch}{1.4}

\begin{tabular*}{\textwidth}{l@{\extracolsep{\fill}}ccccc}
\toprule
Method & Rank 1 & Rank 2 & Rank 3 & Rank 4 & Rank 5 \\
\midrule

% === Wald Variance ===
WV 
& \makecell{VIP \\ \scriptsize $6.85 \times 10^{-5}$} 
& \makecell{HR \\ \scriptsize $8.71 \times 10^{-5}$} 
& \makecell{ZNF142 \\ \scriptsize $8.91 \times 10^{-5}$} 
& \makecell{FAM234A \\ \scriptsize $1.62 \times 10^{-4}$} 
& \makecell{PLCD4 \\ \scriptsize $2.05 \times 10^{-4}$} \\

WV-SC 
& \makecell{VIP \\ \scriptsize $6.44 \times 10^{-5}$} 
& \makecell{HR \\ \scriptsize $7.99 \times 10^{-5}$} 
& \makecell{ZNF142 \\ \scriptsize $8.71 \times 10^{-5}$} 
& \makecell{FAM234A \\ \scriptsize $1.51 \times 10^{-4}$} 
& \makecell{PLCD4 \\ \scriptsize $2.05 \times 10^{-4}$} \\
\cmidrule(lr){1-6}

% === HWV-FLU ===
HWV-FLU 
& \makecell{VIP \\ \scriptsize $3.22 \times 10^{-5}$} 
& \makecell{ZNF142 \\ \scriptsize $5.31 \times 10^{-5}$} 
& \makecell{FAAP100 \\ \scriptsize $6.40 \times 10^{-5}$} 
& \makecell{PLCD4 \\ \scriptsize $7.85 \times 10^{-5}$} 
& \makecell{VIL1 \\ \scriptsize $8.60 \times 10^{-5}$} \\

HWV-FLU-SC 
& \makecell{VIP \\ \scriptsize $2.81 \times 10^{-5}$} 
& \makecell{ZNF142 \\ \scriptsize $3.63 \times 10^{-5}$} 
& \makecell{FAAP100 \\ \scriptsize $5.14 \times 10^{-5}$} 
& \makecell{PLCD4 \\ \scriptsize $6.42 \times 10^{-5}$} 
& \makecell{VIL1 \\ \scriptsize $8.82 \times 10^{-5}$} \\
\cmidrule(lr){1-6}

% === HWV-SSBS ===
HWV-SSBS 
& \makecell{EGLN1 \\ \scriptsize $8.48 \times 10^{-4}$} 
& \makecell{NUDT18 \\ \scriptsize $9.09 \times 10^{-4}$} 
& \makecell{C1orf68 \\ \scriptsize $5.04 \times 10^{-3}$} 
& \makecell{SPRTN \\ \scriptsize $8.24 \times 10^{-3}$} 
& \makecell{CTAGE15 \\ \scriptsize $1.03 \times 10^{-2}$} \\

HWV-SSBS-SC 
& \makecell{EGLN1 \\ \scriptsize $8.46 \times 10^{-4}$} 
& \makecell{NUDT18 \\ \scriptsize $8.87 \times 10^{-4}$} 
& \makecell{C1orf68 \\ \scriptsize $5.01 \times 10^{-3}$} 
& \makecell{SPRTN \\ \scriptsize $8.21 \times 10^{-3}$} 
& \makecell{CTAGE15 \\ \scriptsize $1.02 \times 10^{-2}$} \\
\cmidrule(lr){1-6}

% === HWV-SEX ===
HWV-SEX 
& \makecell{VIP \\ \scriptsize $9.67 \times 10^{-5}$} 
& \makecell{HR \\ \scriptsize $1.09 \times 10^{-4}$} 
& \makecell{ZNF142 \\ \scriptsize $1.36 \times 10^{-4}$} 
& \makecell{SSMEM1 \\ \scriptsize $2.02 \times 10^{-4}$} 
& \makecell{CCL25 \\ \scriptsize $2.29 \times 10^{-4}$} \\

HWV-SEX-SC 
& \makecell{VIP \\ \scriptsize $9.02 \times 10^{-5}$} 
& \makecell{HR \\ \scriptsize $1.08 \times 10^{-4}$} 
& \makecell{ZNF142 \\ \scriptsize $1.21 \times 10^{-4}$} 
& \makecell{SSMEM1 \\ \scriptsize $1.87 \times 10^{-4}$} 
& \makecell{CCL25 \\ \scriptsize $2.25 \times 10^{-4}$} \\
\cmidrule(lr){1-6}

% === HWV-GENOME ===
HWV-GENOME 
& \makecell{VIP \\ \scriptsize $6.73 \times 10^{-5}$} 
& \makecell{HR \\ \scriptsize $8.28 \times 10^{-5}$} 
& \makecell{ZNF142 \\ \scriptsize $8.57 \times 10^{-5}$} 
& \makecell{FAM234A \\ \scriptsize $1.46 \times 10^{-4}$} 
& \makecell{PLCD4 \\ \scriptsize $1.99 \times 10^{-4}$} \\

HWV-GENOME-SC 
& \makecell{VIP \\ \scriptsize $6.58 \times 10^{-5}$} 
& \makecell{HR \\ \scriptsize $7.96 \times 10^{-5}$} 
& \makecell{ZNF142 \\ \scriptsize $8.56 \times 10^{-5}$} 
& \makecell{FAM234A \\ \scriptsize $1.49 \times 10^{-4}$} 
& \makecell{PLCD4 \\ \scriptsize $2.00 \times 10^{-4}$} \\
\bottomrule
\end{tabular*}
\end{table*}

\section{Discussion}\label{subsec5}

In this paper, we have developed a suite of set-based genetic association tests for panel count data under a weighted V-statistic framework. The proposed WV-PCD test evaluates the joint effect of a marker set on a recurrent-event process while adjusting for covariates. The HWV-PCD test further accounts for genetic heterogeneity and can increase power when genetic effects vary across subpopulations. We developed small-sample corrected versions of the proposed tests to improve finite-sample accuracy. These methods extend existing multi-marker testing frameworks from interval-censored outcomes to panel count outcomes, which arise naturally in many biomedical studies.

The simulation studies confirmed the practical advantages of using a panel count framework for recurrent-event genetic association testing. Compared with interval-censored analyses, WV-PCD achieved higher power by retaining information on the cumulative number of events, and HWV-PCD further improved power when genetic effects varied across subpopulations or individual genome profiles. In the ZOE 2.0 early childhood caries application, no individual gene remained significant after multiple-testing correction; however, enrichment analysis suggested several biologically plausible patterns. The SSBS-based analysis may therefore capture variation in caries risk that is missed by standard gene-level tests.

Several limitations should be noted. First, the results of the proposed tests may depend on the selected adjustment covariates and heterogeneity sources. In practice, the optimal choice of these components is often unknown and may vary across applications. Second, the gene-level analysis did not identify any gene that remained significant after multiple-testing correction. The enrichment findings should therefore be interpreted as exploratory and require validation in independent ECC cohorts. Third, the gene set enrichment analysis was based on genes ranked by unsigned gene-level association \(p\)-values. As a result, the enrichment results indicate whether genes within a GO category tend to show stronger association evidence as a group, but they do not provide directional biological interpretation. In particular, the NES sign should not be interpreted
as indicating up- or down-regulation, or risk-increasing versus protective genetic effects. Future work incorporating signed gene-level effect estimates or variant-level directionality may allow directional pathway interpretation. Fourth, the performance of HWV-PCD depends on the choice of the heterogeneity source and similarity kernel. If the chosen heterogeneity variable does not capture the true variation in genetic effects, the power gain may be limited.

Future work can extend the proposed framework in several directions. One important direction is to develop gene-environment interaction tests for panel count outcomes, which would allow direct testing of whether genetic effects vary with different environmental risk factors.

\section{Key Points}

\begin{itemize}
\item We developed kernel-based multi-marker association tests for panel count outcomes under proportional means models.

\item The proposed tests can account for genetic effect heterogeneity through heterogeneity similarity kernels to improve power for association detection.

\item We developed small-sample corrections for the proposed weighted V-statistic tests to improve small-sample accuracy.

\item The proposed tests have explicit asymptotic null distributions and were evaluated through simulations and an application to the ZOE 2.0 early childhood caries study.
\end{itemize}

\section{Conflicts of interest}
None declared.

\section{Funding}
This work was supported in part by the National Institutes of Health (R03DE032357 to K.X., J.Z., Q.L. and C.L.).

\section{Acknowledgments}
The data for the application presented in this work were from the ZOE 2.0 study, "Genome-wide association study of early childhood caries", which was supported by a grant
from the National Institute of Dental \& Craniofacial Research: U01-DE025046. Genotyping was performed by
CIDR, supported by a resource-allocation grant X01-HG010871, funded by NIDCR. We are grateful to the Principal Investigator, Kimon Divaris, for facilitating the data transfer and for helpful discussions on data aspects.

\section{Data availability}
The data that support the findings of this study are available in dbGaP
(Accession No.: phs002232.v1.p1, “TOPDECC-Trans-omics for Precision 
Dentistry and Early Childhood Caries: Genome-Wide Genotyping 
(CIDR) and Microbiome in the ZOE 2.0 Study”).

%\begin{thebibliography}{10}
%\bibliographystyle{oup-plain}
%\bibliography{reference}
%\end{thebibliography}

%%%%%%%%%%%%%%

\begin{appendices}
\clearpage
\onecolumn

\section{Supplementary tables and figures}\label{app:supplement}

\subsection{Simulation results without genetic heterogeneity}\label{app:noheter}

% qq plot of Without Genetic Heterogeneity
\begin{figure}[H]
\centering

\begin{minipage}{0.48\textwidth}
\centering
\textbf{(A) } $n=400,\ p=15$\\
\includegraphics[width=\linewidth]{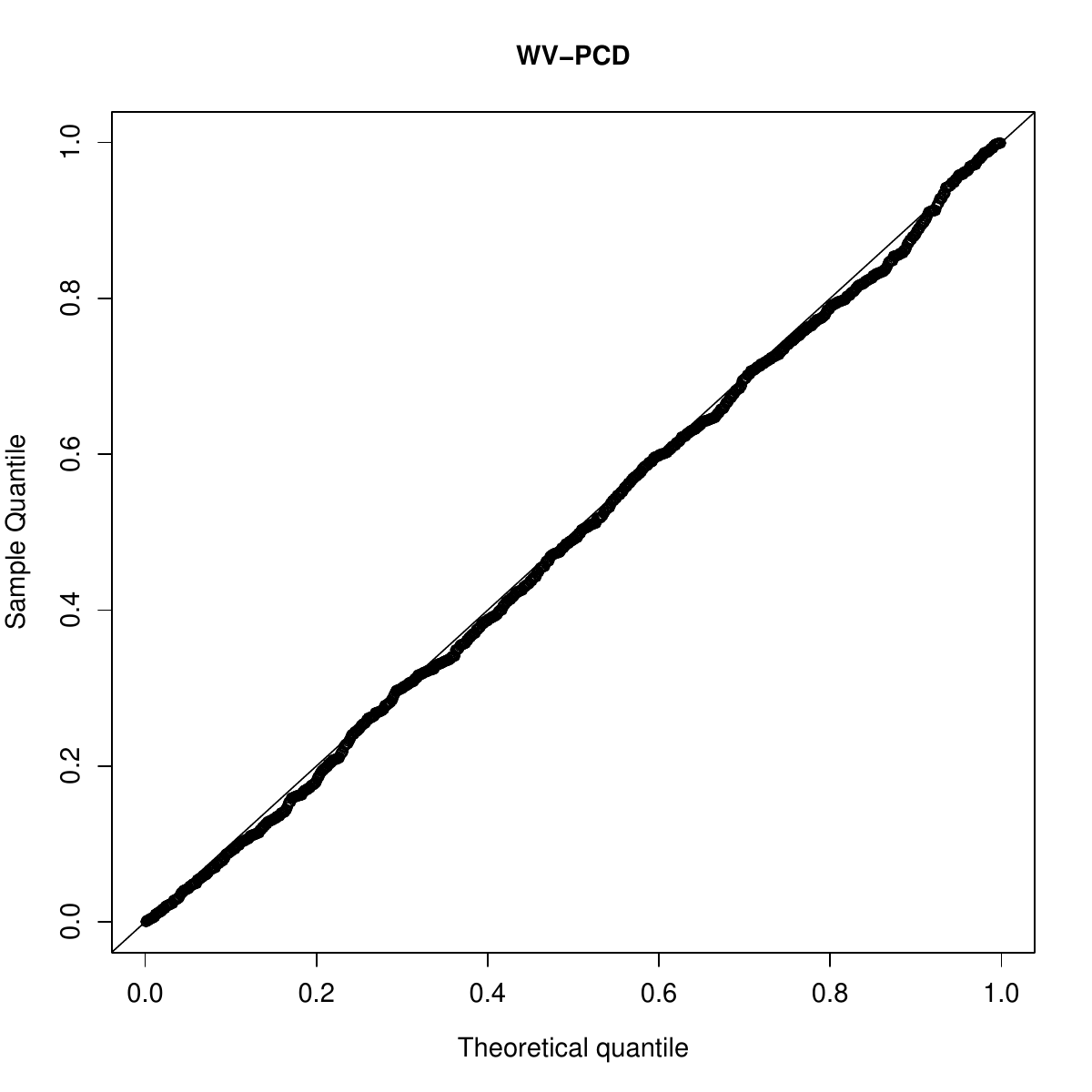}
\end{minipage}
\hfill
\begin{minipage}{0.48\textwidth}
\centering
\textbf{(B) } $n=400,\ p=25$\\
\includegraphics[width=\linewidth]{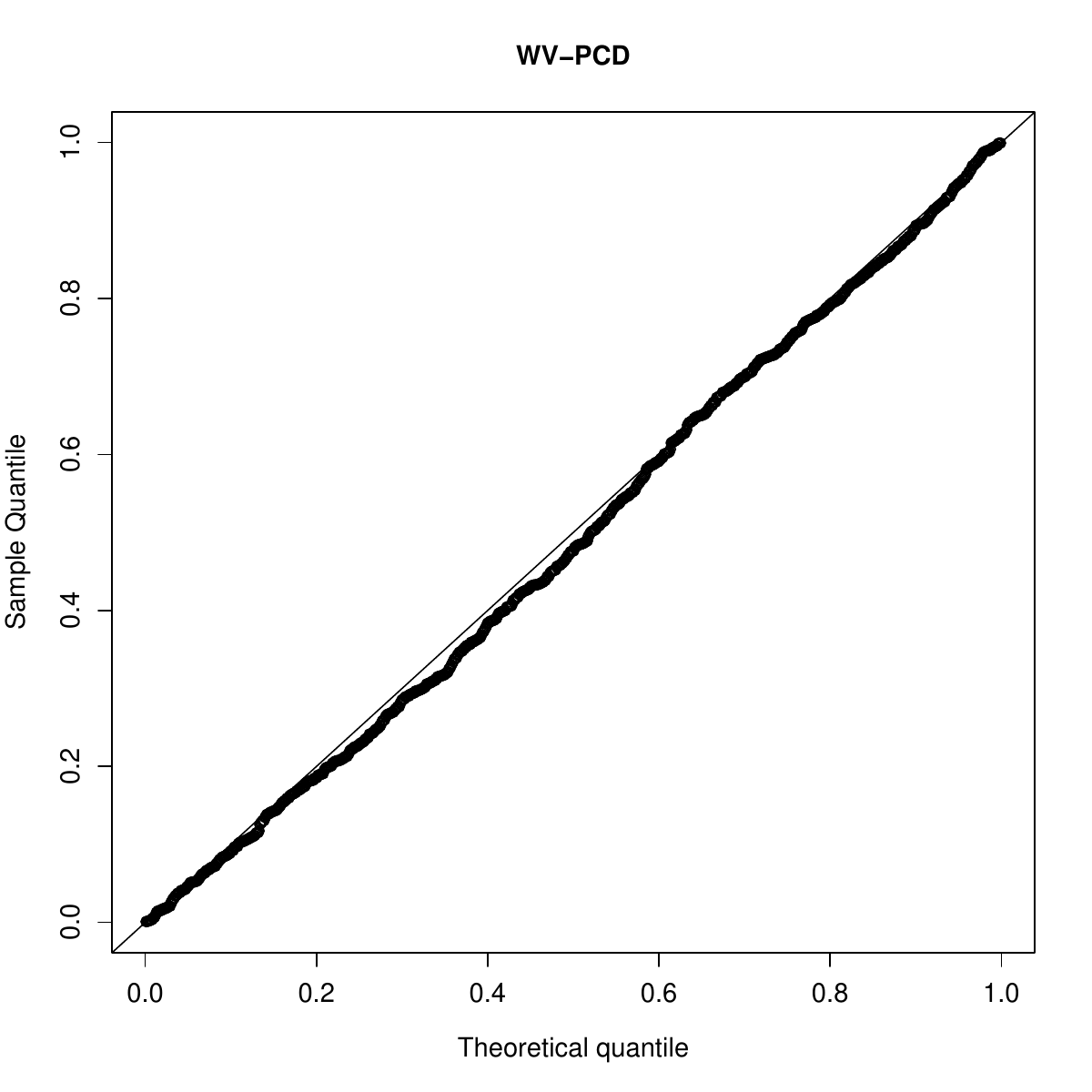}
\end{minipage}

\vspace{0.3cm}

\begin{minipage}{0.48\textwidth}
\centering
\textbf{(C) } $n=800,\ p=15$\\
\includegraphics[width=\linewidth]{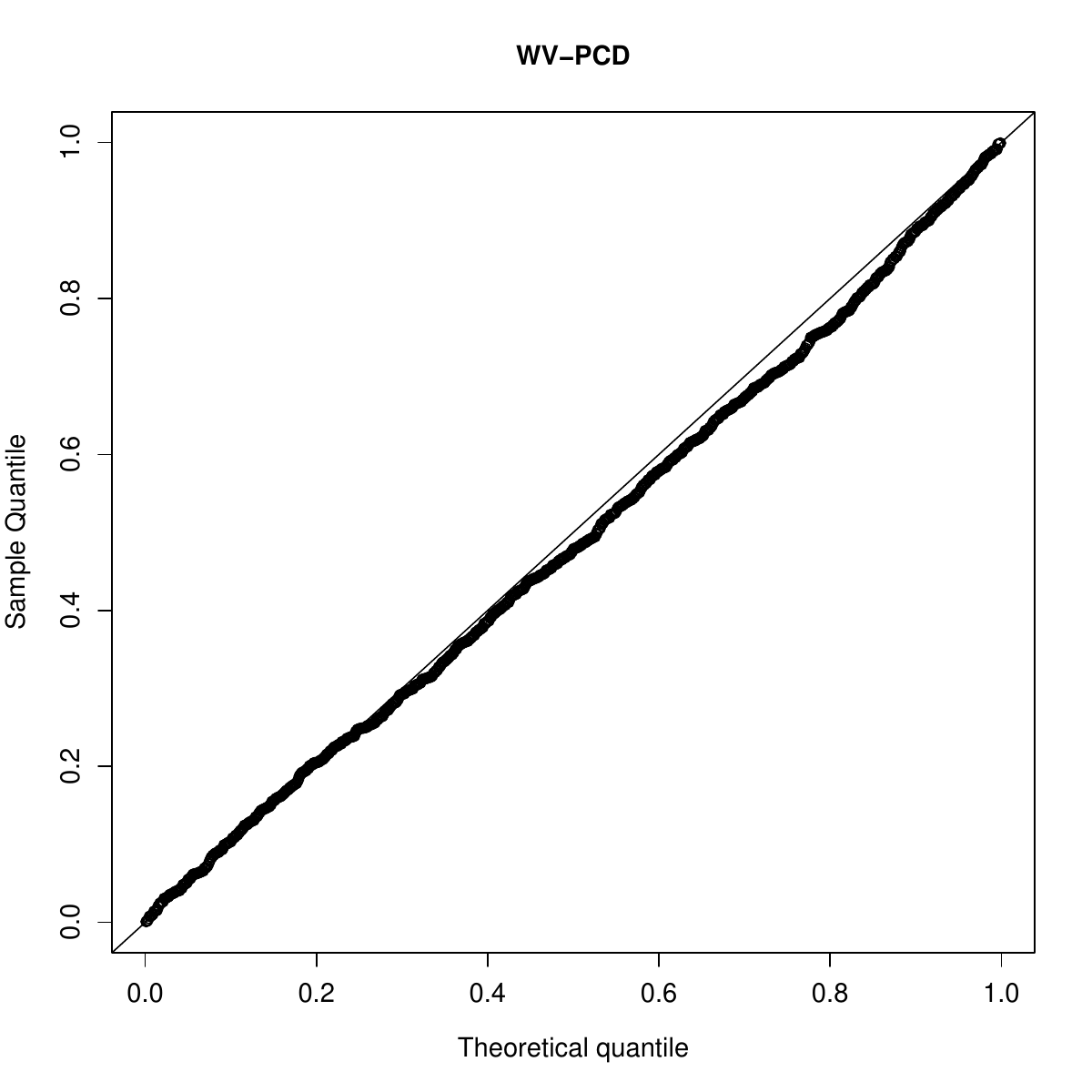}
\end{minipage}
\hfill
\begin{minipage}{0.48\textwidth}
\centering
\textbf{(D) } $n=800,\ p=25$\\
\includegraphics[width=\linewidth]{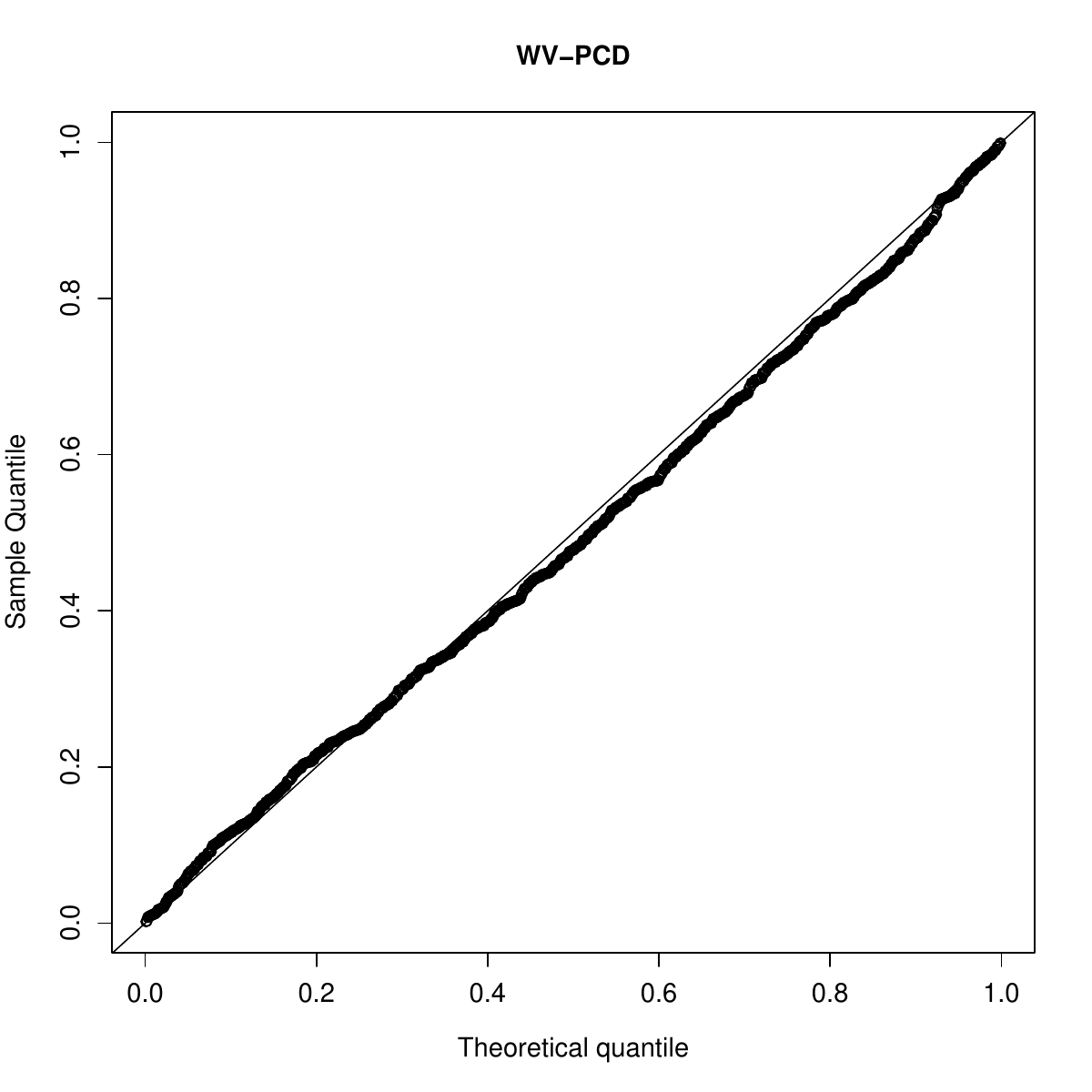}
\end{minipage}

\caption{Q--Q plots of WV-PCD p-values under the null setting without genetic heterogeneity. Panels correspond to (A) $n=400,\ p=15$, (B) $n=400,\ p=25$, (C) $n=800,\ p=15$, and (D) $n=800,\ p=25$. The points closely follow the 45-degree reference line, suggesting that the WV-PCD test is well calibrated under the null model.}
\label{fig:appendix-noheter-wv-qq}
\end{figure}

\clearpage
\subsection{Simulation Results With Two Observed Subpopulations}
% qq plot of HWV-PCD of 2 observed
\begin{figure}[H]
\centering

\begin{minipage}{0.47\textwidth}
\centering
\textbf{(A) } HWV-PCD, $n=400,\ p=15$\\
\includegraphics[width=0.92\linewidth]{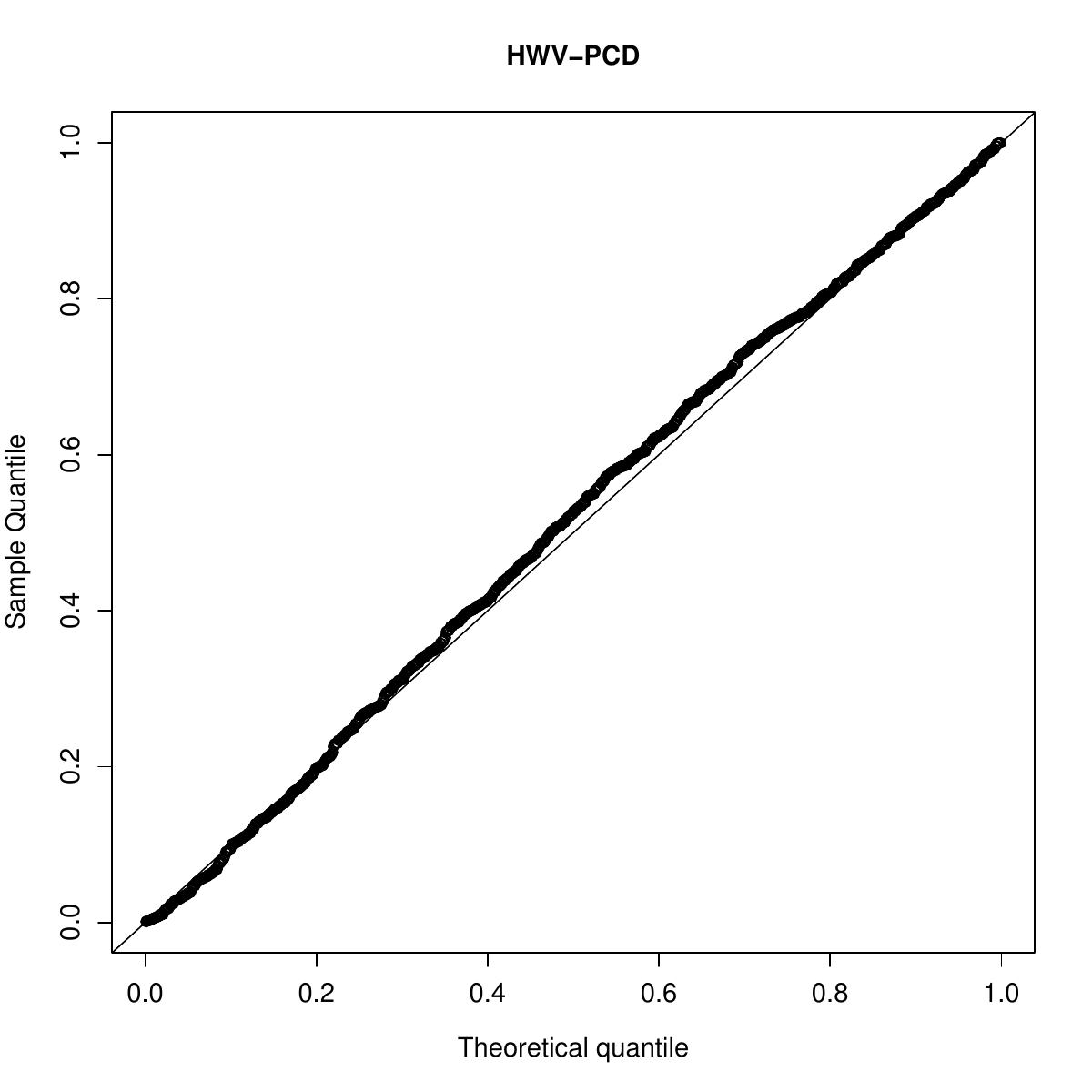}
\end{minipage}
\hfill
\begin{minipage}{0.47\textwidth}
\centering
\textbf{(B) } HWV-PCD, $n=400,\ p=25$\\
\includegraphics[width=0.92\linewidth]{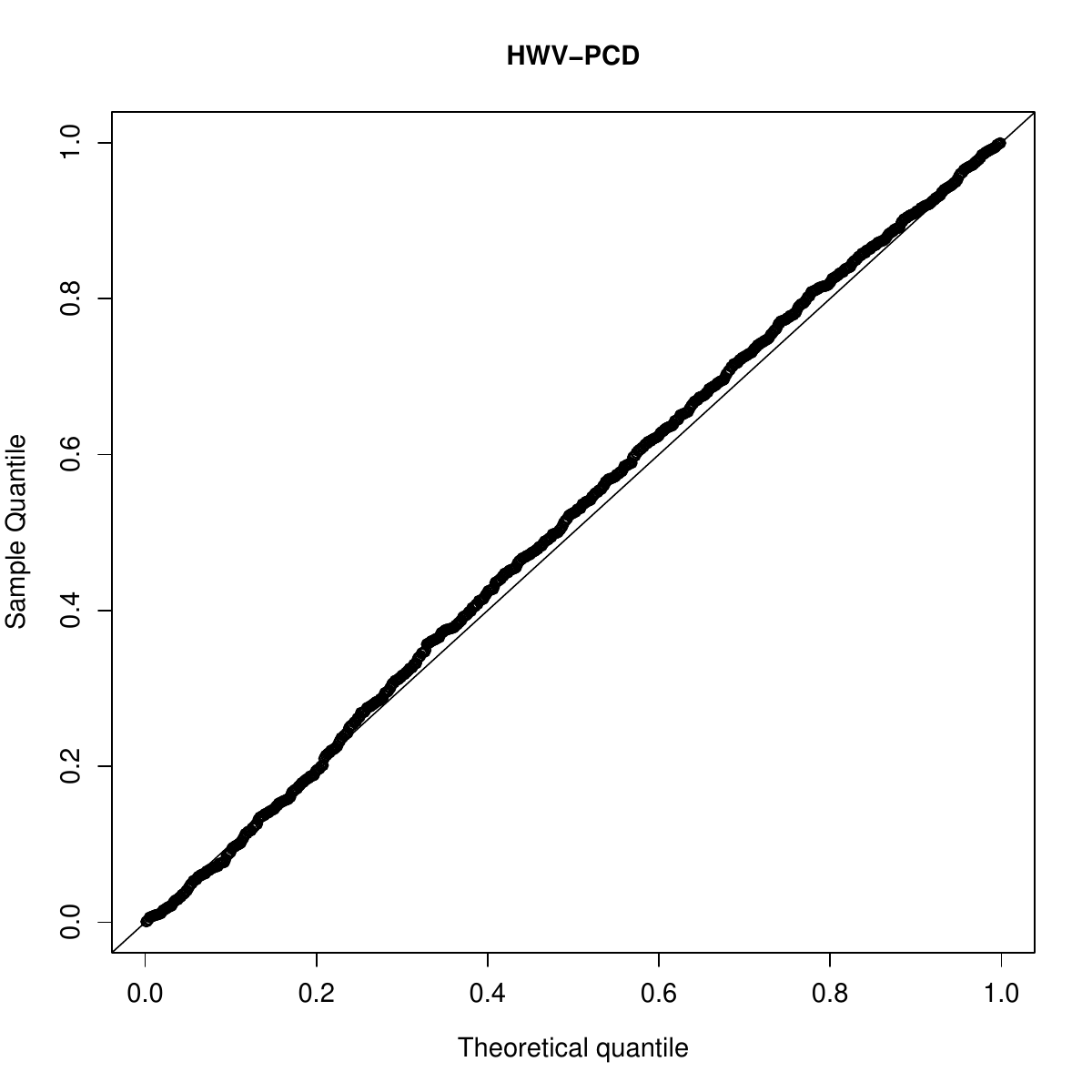}
\end{minipage}

\vspace{0.25cm}

\begin{minipage}{0.47\textwidth}
\centering
\textbf{(C) } HWV-PCD, $n=800,\ p=15$\\
\includegraphics[width=0.92\linewidth]{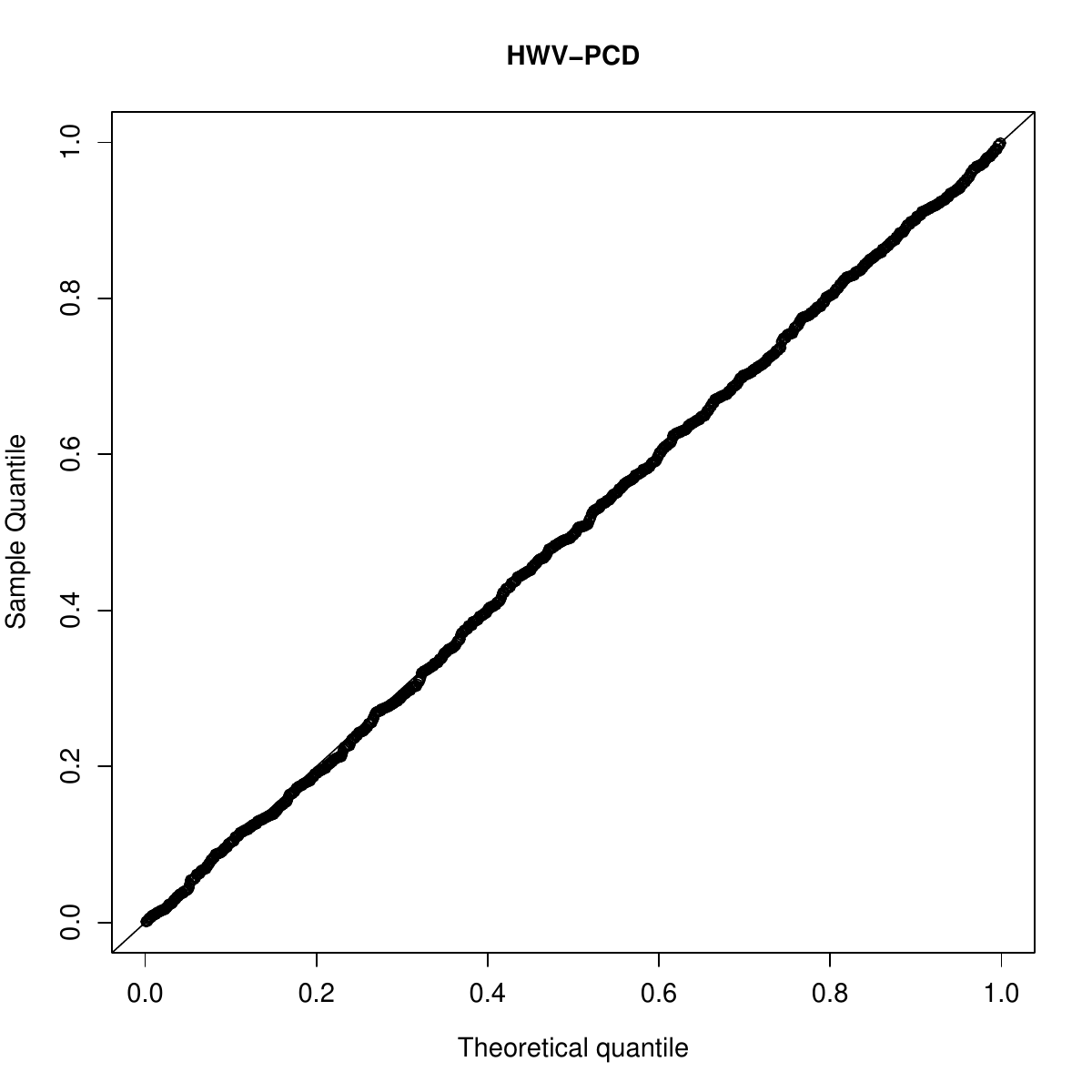}
\end{minipage}
\hfill
\begin{minipage}{0.47\textwidth}
\centering
\textbf{(D) } HWV-PCD, $n=800,\ p=25$\\
\includegraphics[width=0.92\linewidth]{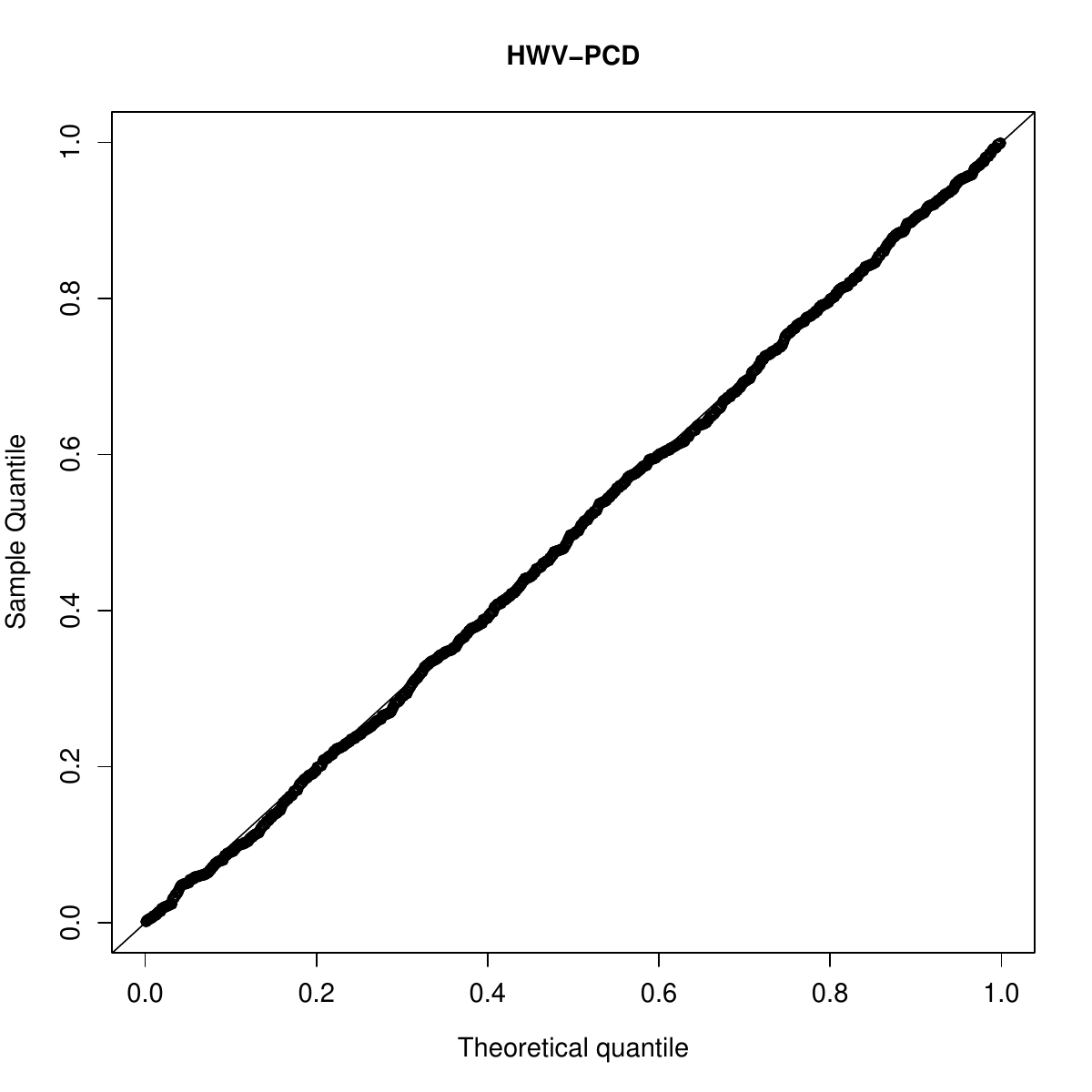}
\end{minipage}

\caption{Q--Q plots of HWV-PCD p-values under the setting with two observed subpopulations. Panels correspond to (A) $n=400,\ p=15$, (B) $n=400,\ p=25$, (C) $n=800,\ p=15$, and (D) $n=800,\ p=25$.}
\label{fig:qq-hwv-2obs}
\end{figure}

% qq plot of WV-PCD of 2 observed
\begin{figure}[H]
\centering

\begin{minipage}{0.47\textwidth}
\centering
\textbf{(A) } WV-PCD, $n=400,\ p=15$\\
\includegraphics[width=0.92\linewidth]{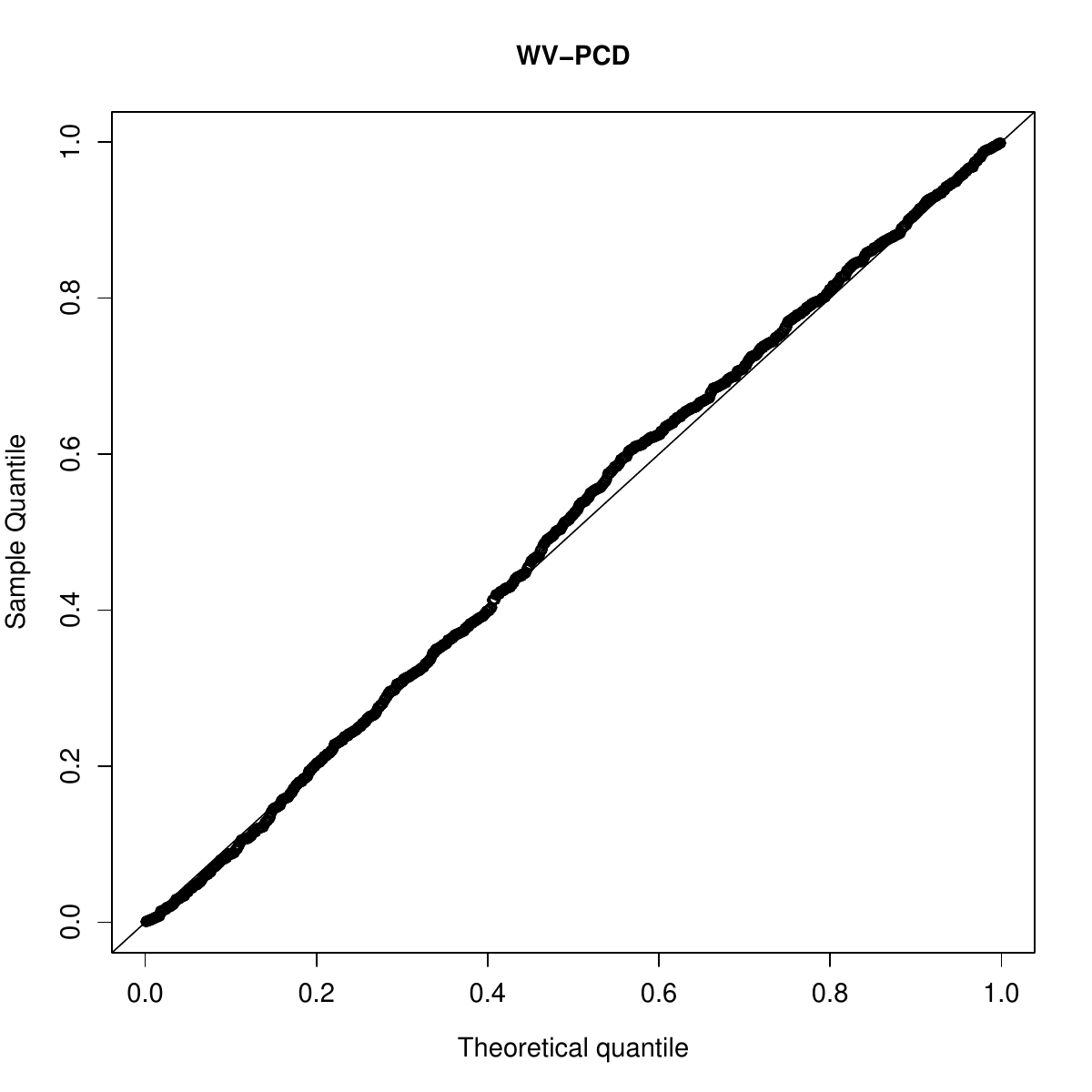}
\end{minipage}
\hfill
\begin{minipage}{0.47\textwidth}
\centering
\textbf{(B) } WV-PCD, $n=400,\ p=25$\\
\includegraphics[width=0.92\linewidth]{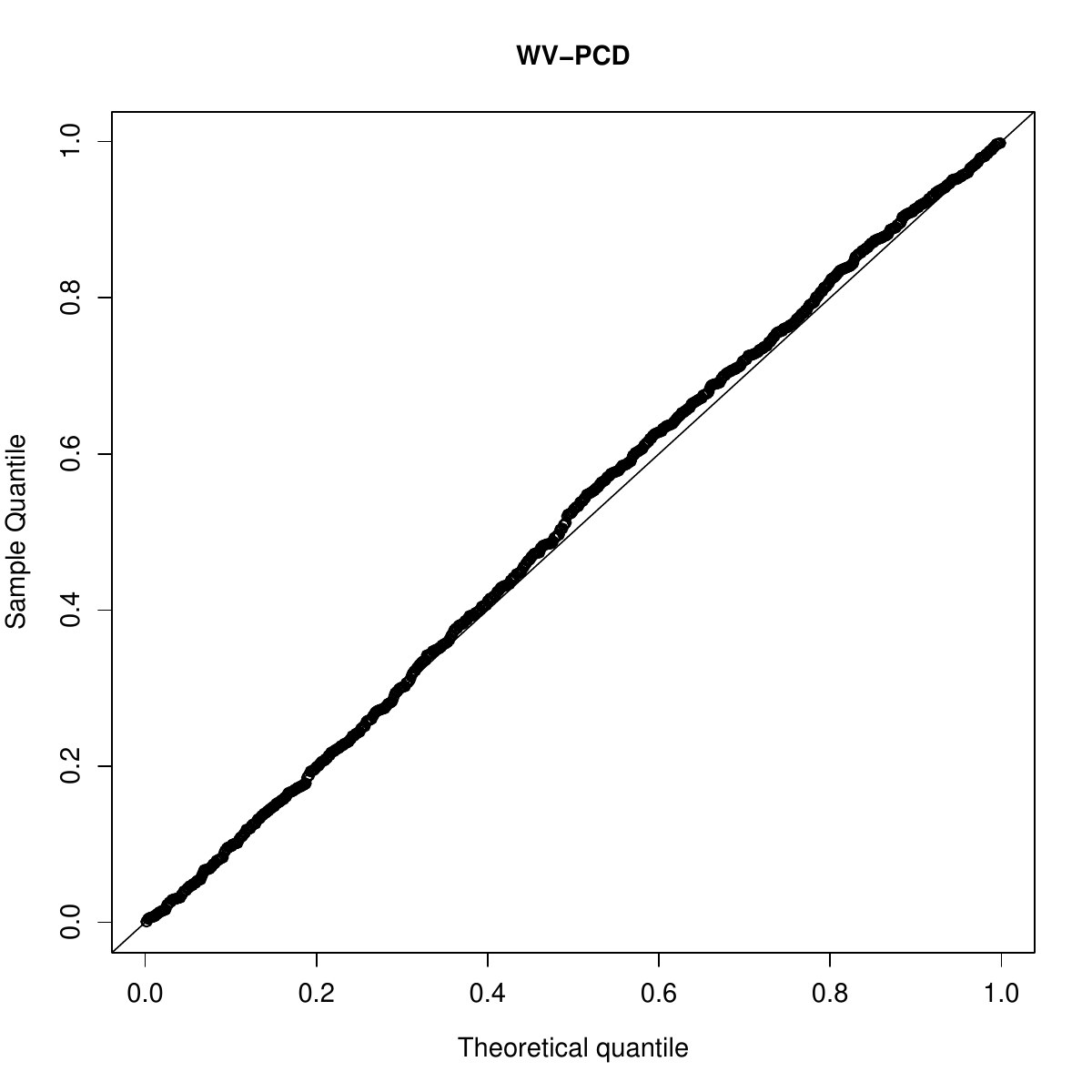}
\end{minipage}

\vspace{0.25cm}

\begin{minipage}{0.47\textwidth}
\centering
\textbf{(C) } WV-PCD, $n=800,\ p=15$\\
\includegraphics[width=0.92\linewidth]{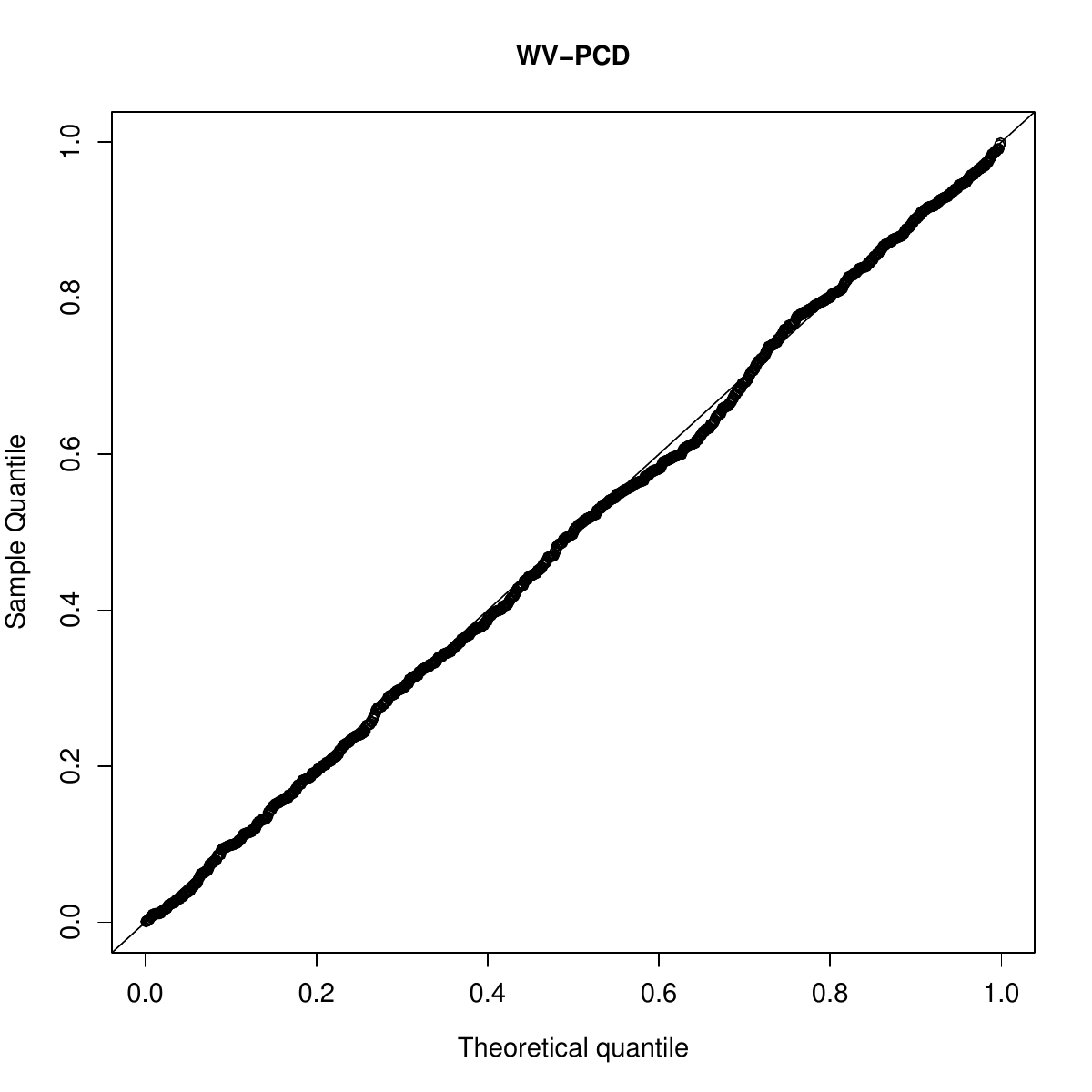}
\end{minipage}
\hfill
\begin{minipage}{0.47\textwidth}
\centering
\textbf{(D) } WV-PCD, $n=800,\ p=25$\\
\includegraphics[width=0.92\linewidth]{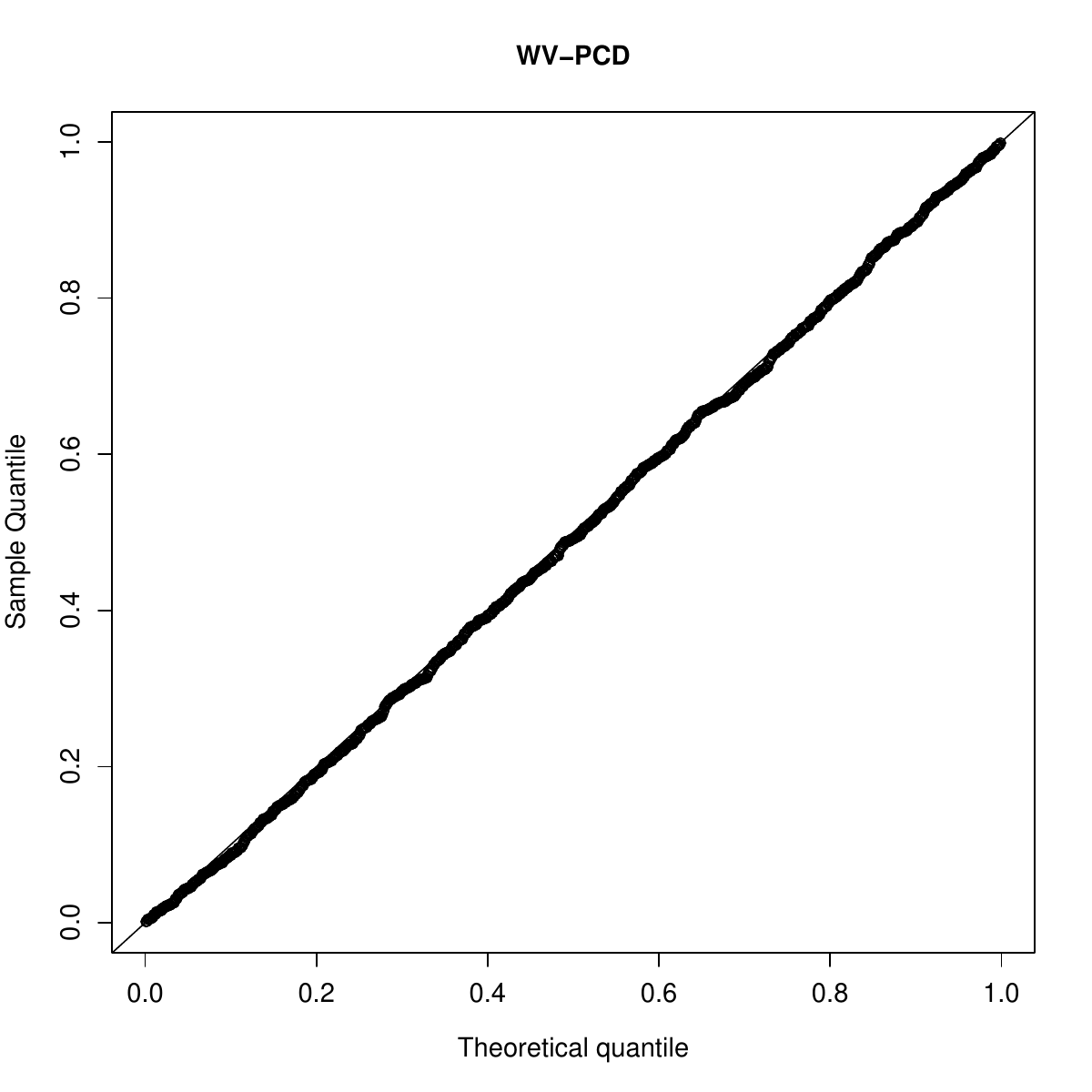}
\end{minipage}

\caption{Q--Q plots of WV-PCD p-values under the setting with two observed subpopulations. Panels correspond to (A) $n=400,\ p=15$, (B) $n=400,\ p=25$, (C) $n=800,\ p=15$, and (D) $n=800,\ p=25$.}
\label{fig:qq-wv-2obs}
\end{figure}

\clearpage
\subsection{Simulation Results With Two Latent Subpopulations}
%Figure 1: HWV-PCD
\begin{figure}[H]
\centering

\begin{minipage}{0.47\textwidth}
\centering
\textbf{(A) } HWV-PCD, $n=400,\ p=15$\\
\includegraphics[width=0.92\linewidth]{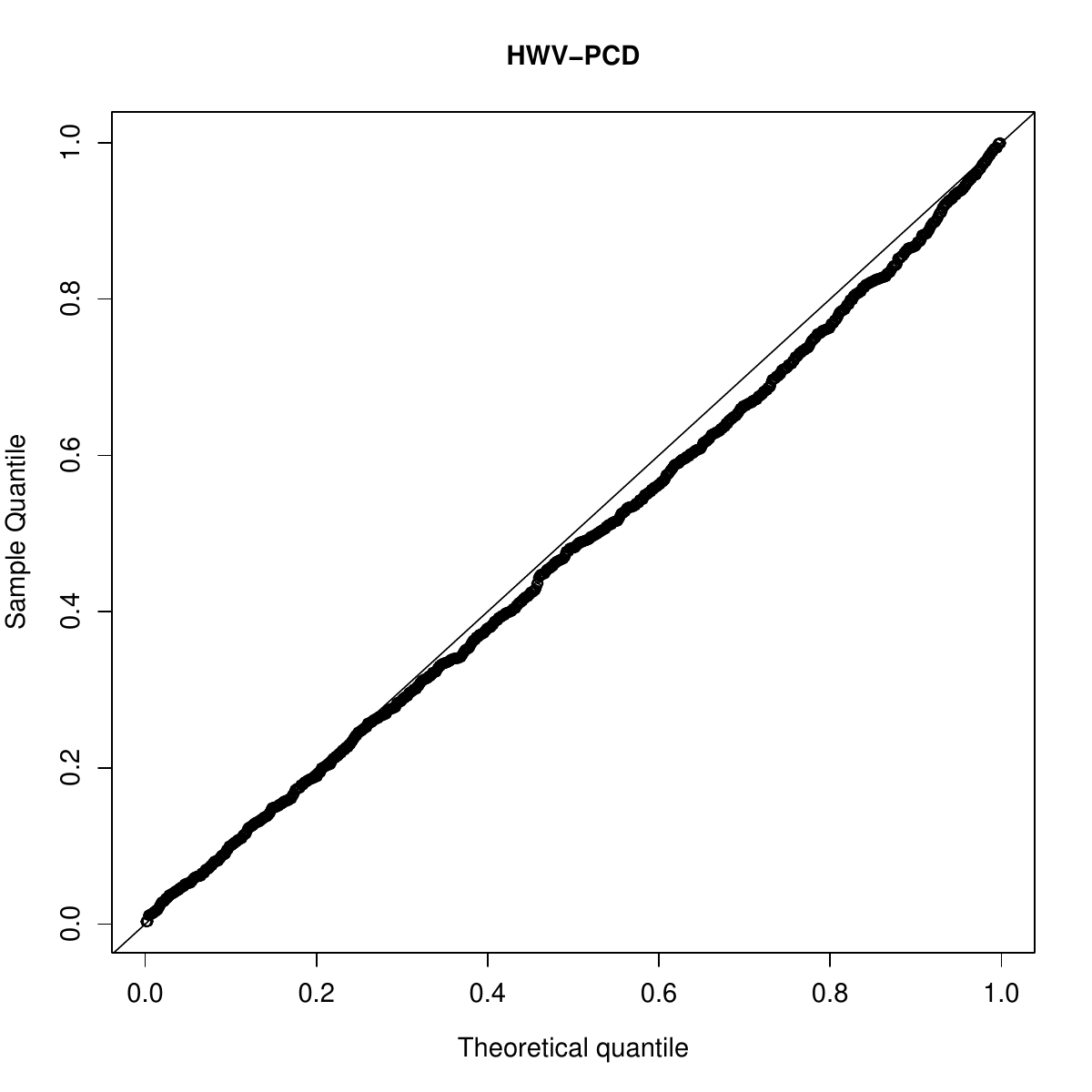}
\end{minipage}
\hfill
\begin{minipage}{0.47\textwidth}
\centering
\textbf{(B) } HWV-PCD, $n=400,\ p=25$\\
\includegraphics[width=0.92\linewidth]{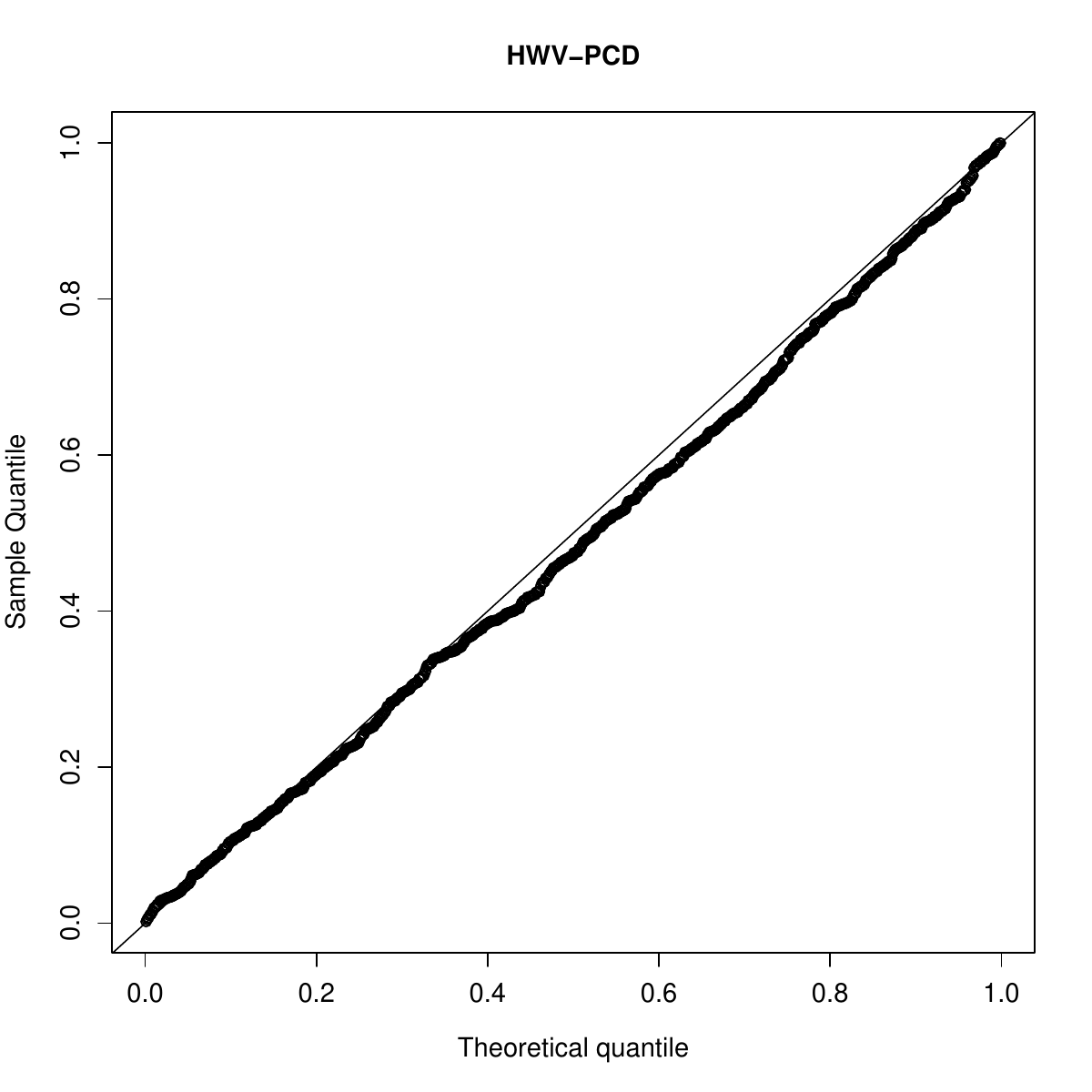}
\end{minipage}

\vspace{0.25cm}

\begin{minipage}{0.47\textwidth}
\centering
\textbf{(C) } HWV-PCD, $n=800,\ p=15$\\
\includegraphics[width=0.92\linewidth]{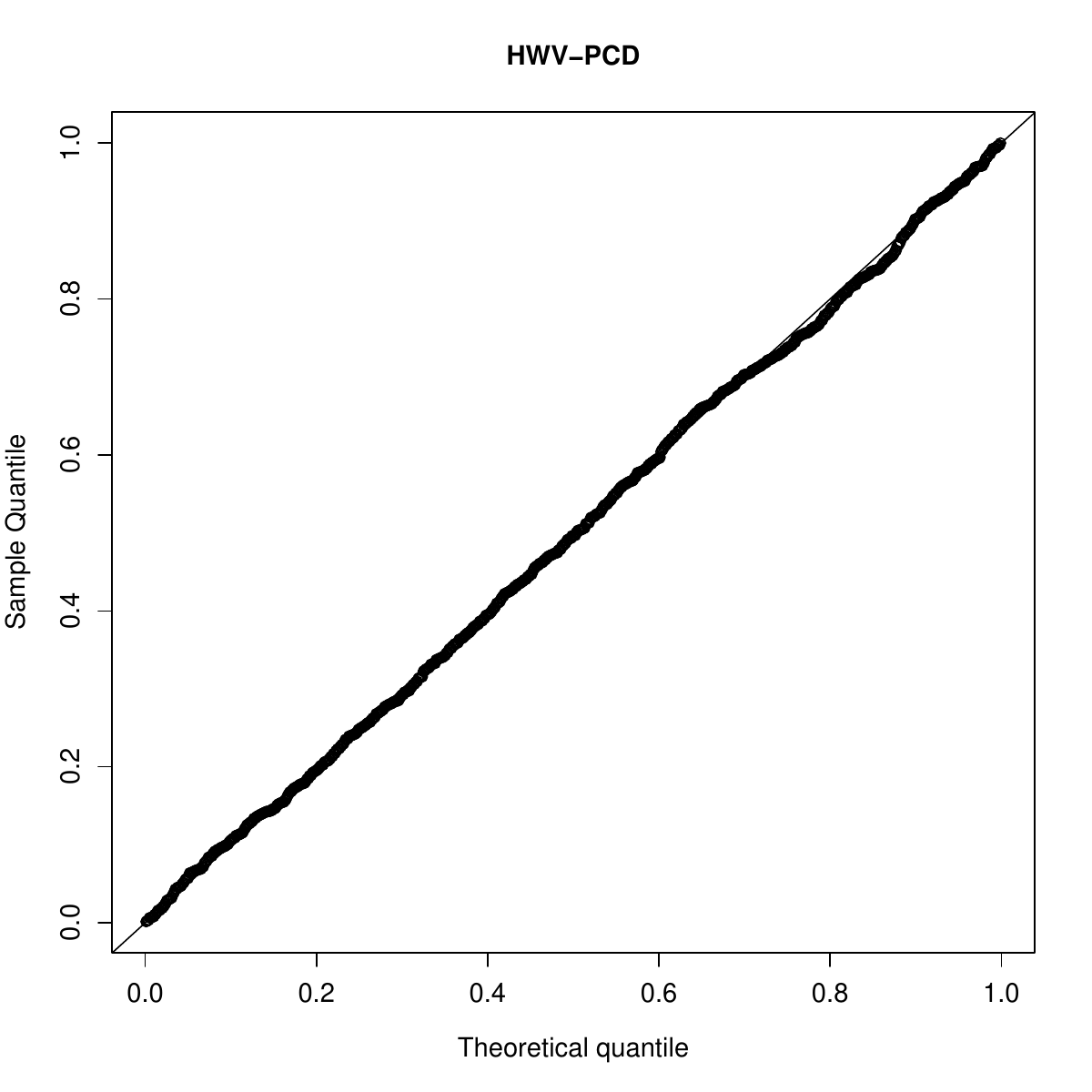}
\end{minipage}
\hfill
\begin{minipage}{0.47\textwidth}
\centering
\textbf{(D) } HWV-PCD, $n=800,\ p=25$\\
\includegraphics[width=0.92\linewidth]{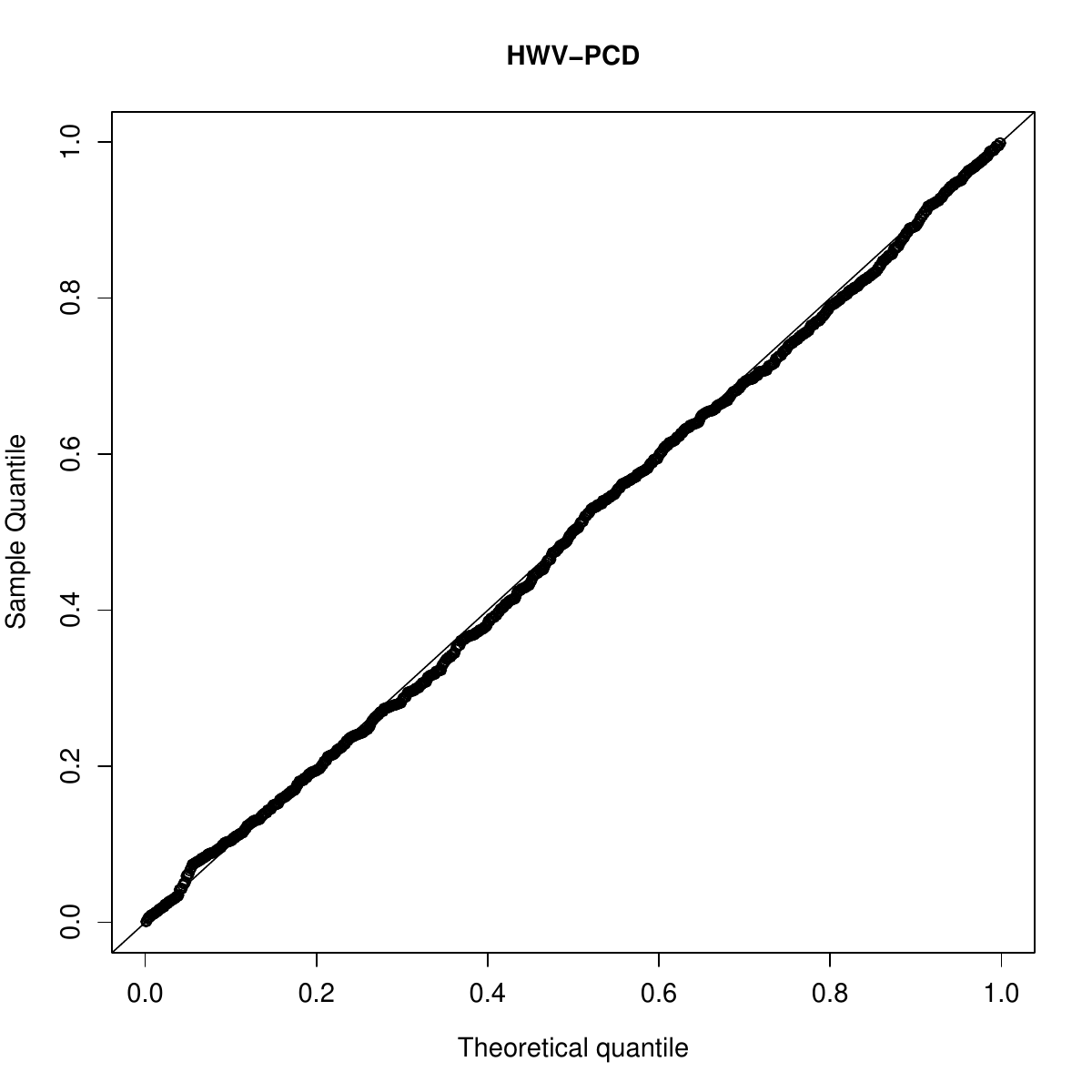}
\end{minipage}

\caption{Q--Q plots of HWV-PCD p-values under the simulation setting with two latent subpopulations. Panels correspond to (A) $n=400,\ p=15$, (B) $n=400,\ p=25$, (C) $n=800,\ p=15$, and (D) $n=800,\ p=25$.}
\label{fig:qq-hwv-2latent}
\end{figure}

%Figure 2: WV-PCD
\begin{figure}[htbp]
\centering

\begin{minipage}{0.47\textwidth}
\centering
\textbf{(A) } WV-PCD, $n=400,\ p=15$\\
\includegraphics[width=0.92\linewidth]{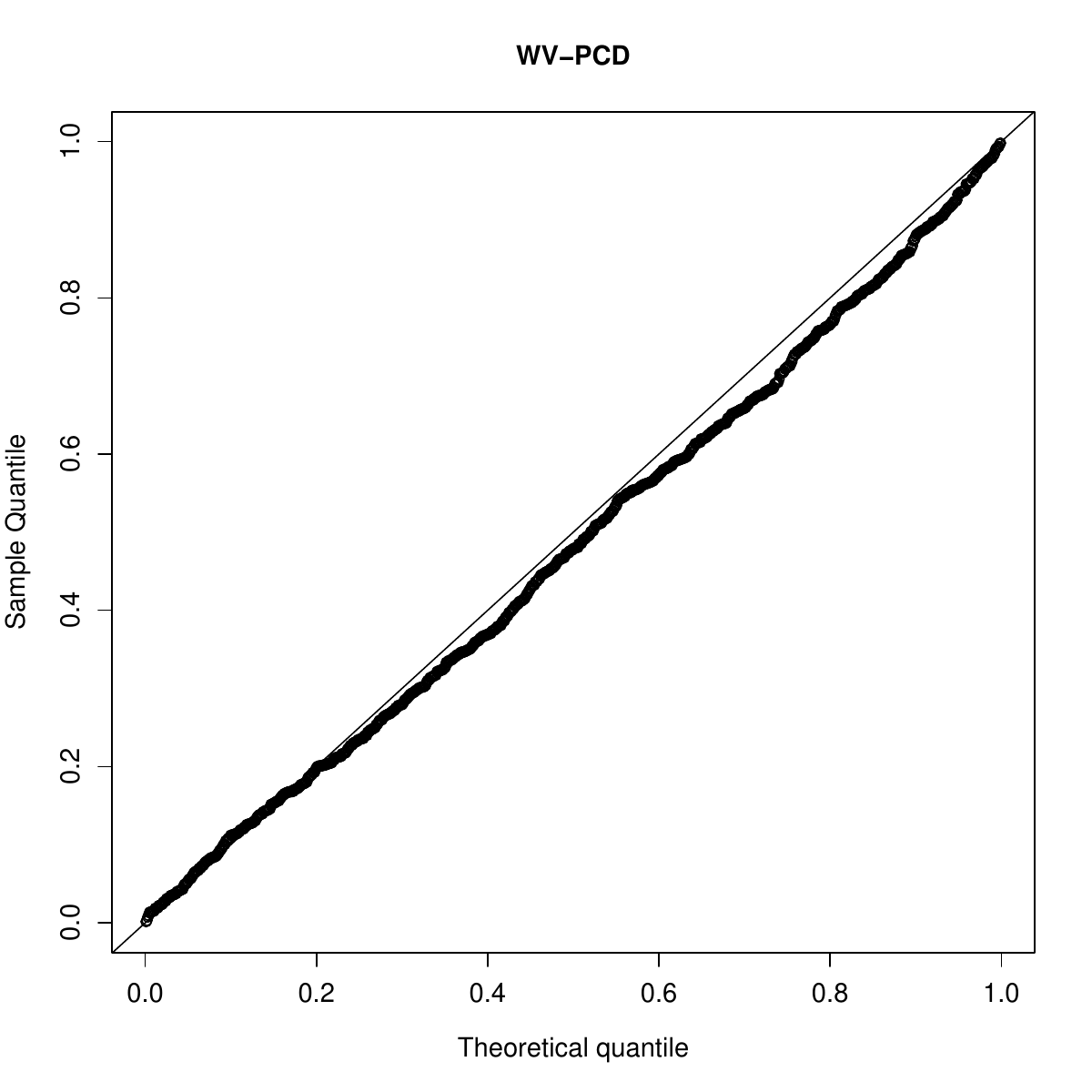}
\end{minipage}
\hfill
\begin{minipage}{0.47\textwidth}
\centering
\textbf{(B) } WV-PCD, $n=400,\ p=25$\\
\includegraphics[width=0.92\linewidth]{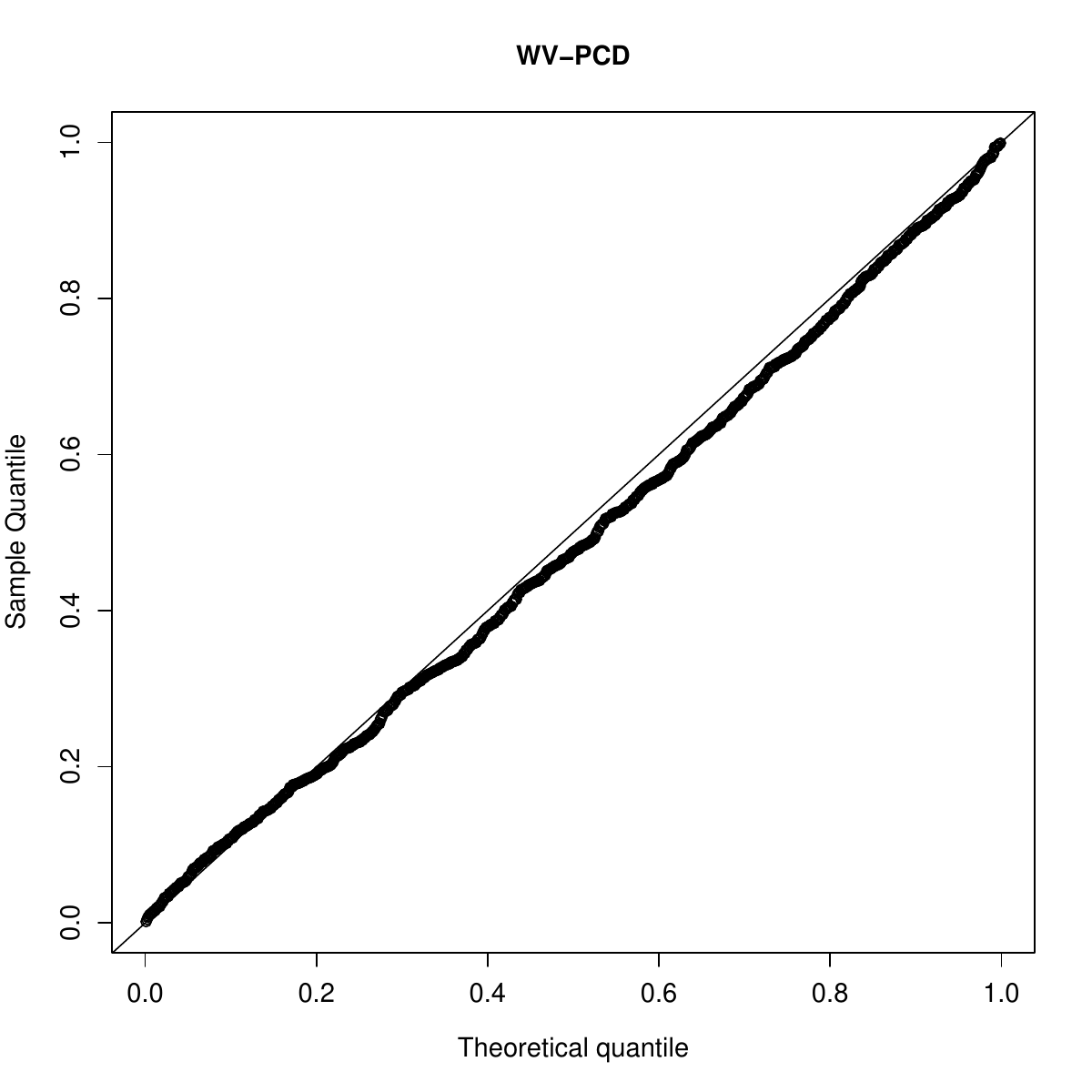}
\end{minipage}

\vspace{0.25cm}

\begin{minipage}{0.47\textwidth}
\centering
\textbf{(C) } WV-PCD, $n=800,\ p=15$\\
\includegraphics[width=0.92\linewidth]{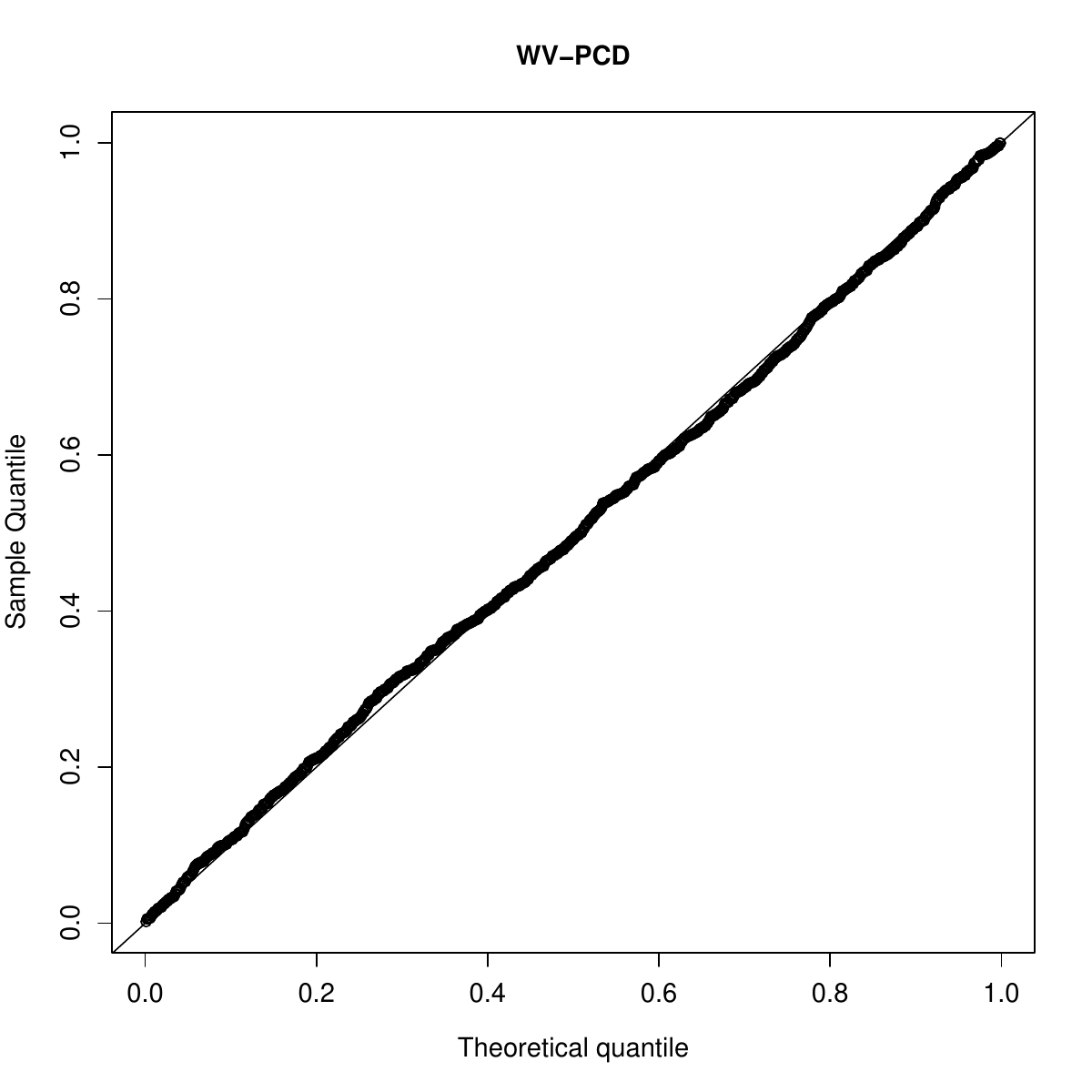}
\end{minipage}
\hfill
\begin{minipage}{0.47\textwidth}
\centering
\textbf{(D) } WV-PCD, $n=800,\ p=25$\\
\includegraphics[width=0.92\linewidth]{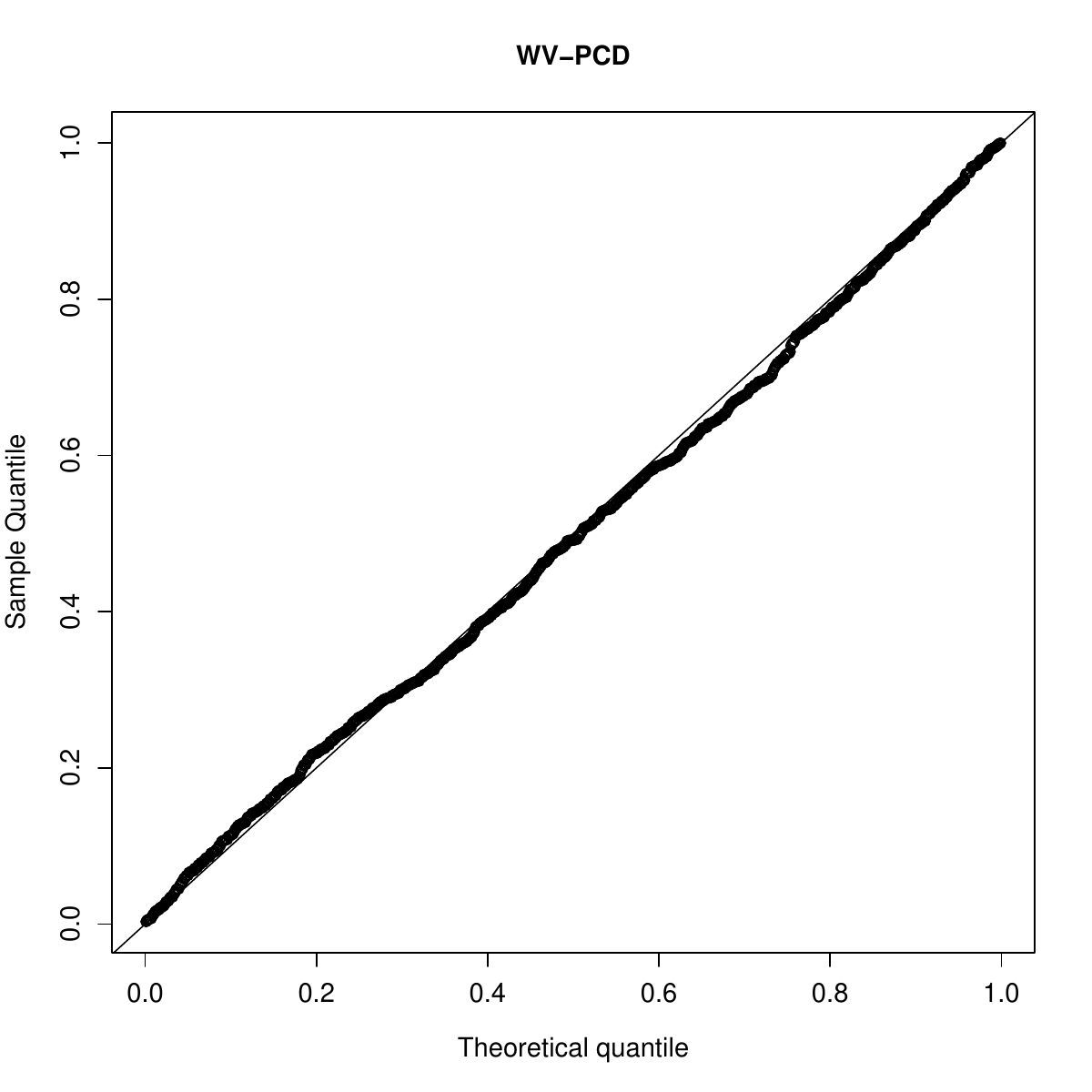}
\end{minipage}

\caption{Q--Q plots of WV-PCD p-values under the setting with two latent subpopulations. Panels correspond to (A) $n=400,\ p=15$, (B) $n=400,\ p=25$, (C) $n=800,\ p=15$, and (D) $n=800,\ p=25$.}
\label{fig:qq-wv-2latent}
\end{figure}

\clearpage

\subsection{Simulation Results With Genetic Heterogeneity Across Individual Genome Profiles}
%HWV-PCD
\begin{figure}[H]
\centering

\begin{minipage}{0.47\textwidth}
\centering
\textbf{(A) } HWV-PCD, $n=400,\ p=15$\\
\includegraphics[width=0.92\linewidth]{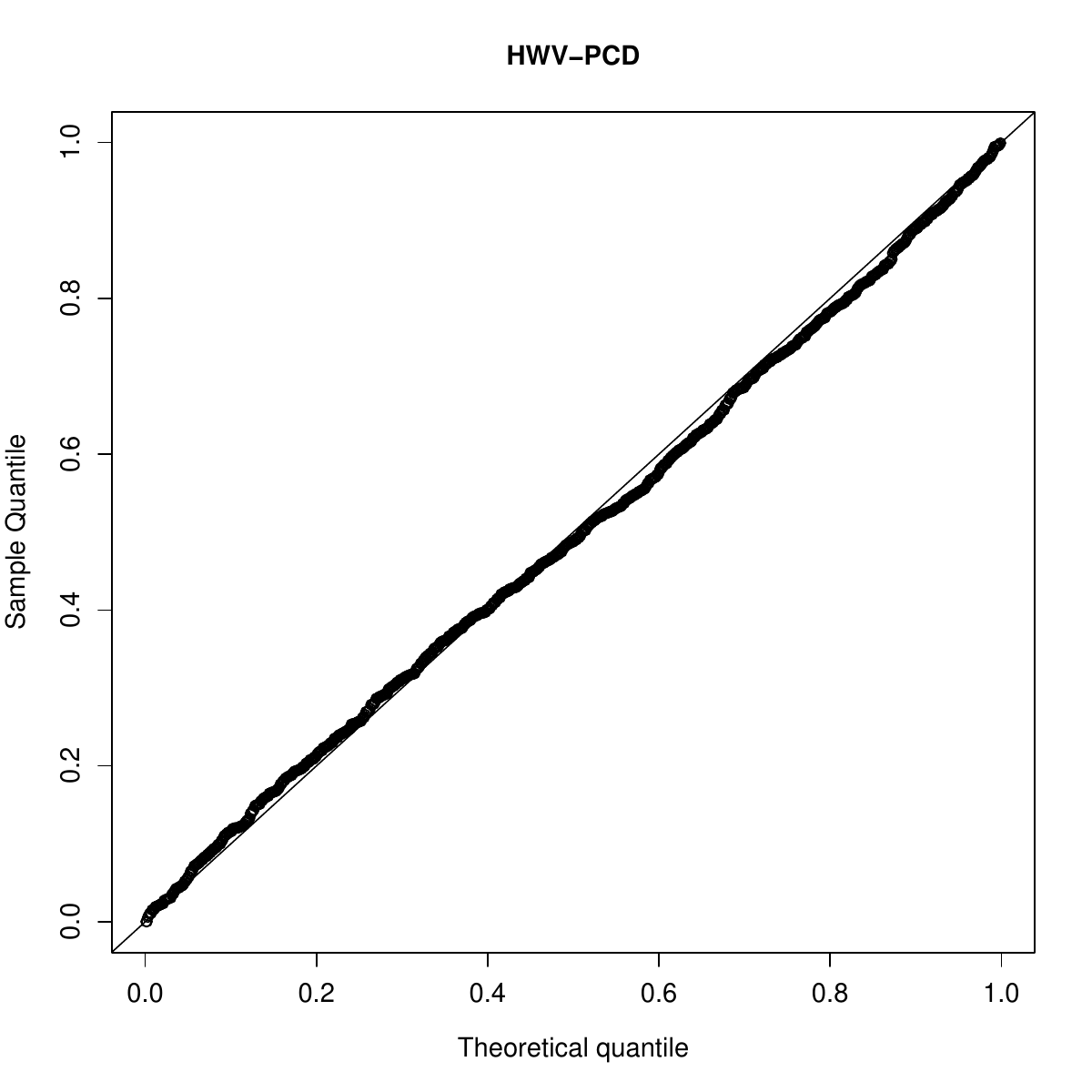}
\end{minipage}
\hfill
\begin{minipage}{0.47\textwidth}
\centering
\textbf{(B) } HWV-PCD, $n=400,\ p=25$\\
\includegraphics[width=0.92\linewidth]{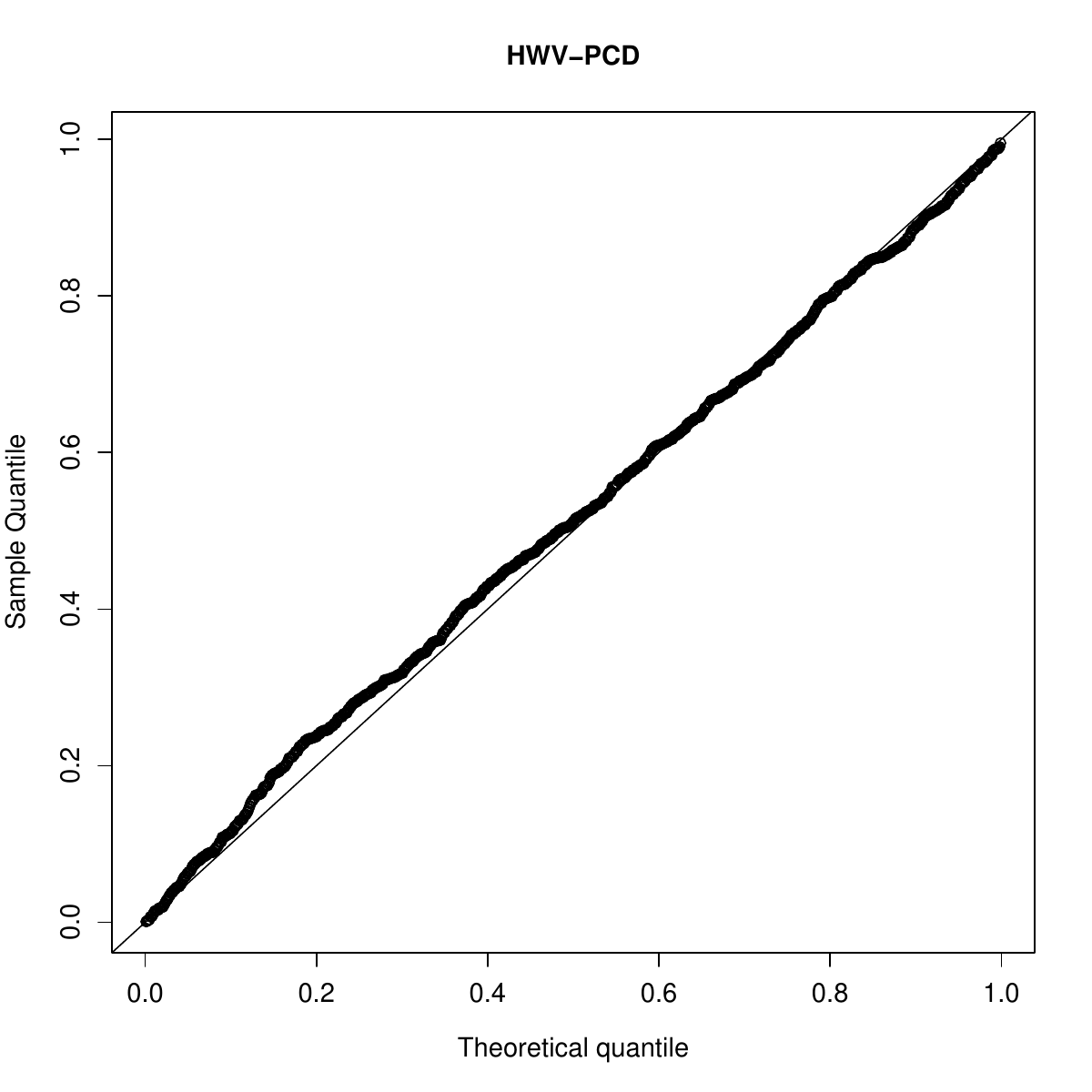}
\end{minipage}

\vspace{0.25cm}

\begin{minipage}{0.47\textwidth}
\centering
\textbf{(C) } HWV-PCD, $n=800,\ p=15$\\
\includegraphics[width=0.92\linewidth]{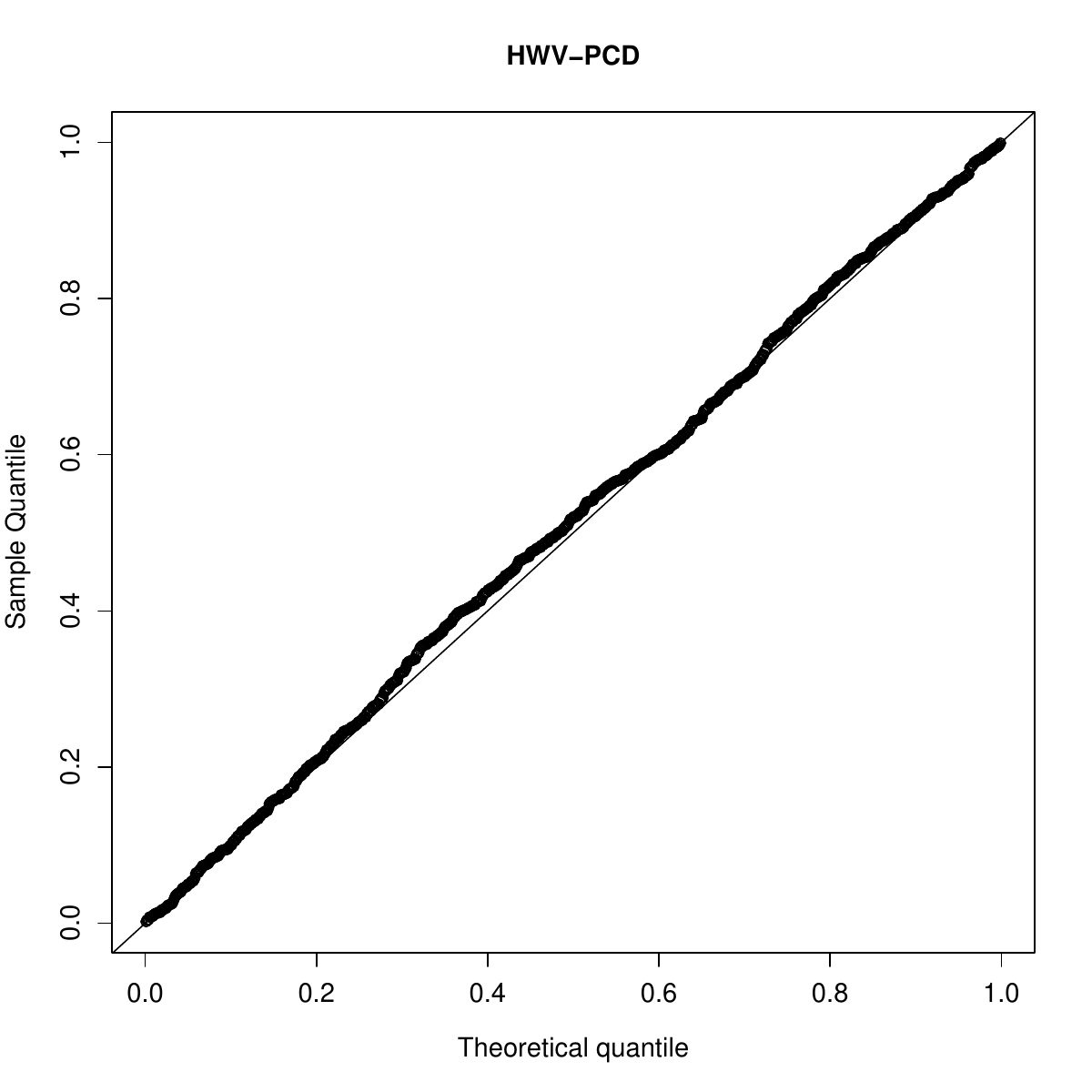}
\end{minipage}
\hfill
\begin{minipage}{0.47\textwidth}
\centering
\textbf{(D) } HWV-PCD, $n=800,\ p=25$\\
\includegraphics[width=0.92\linewidth]{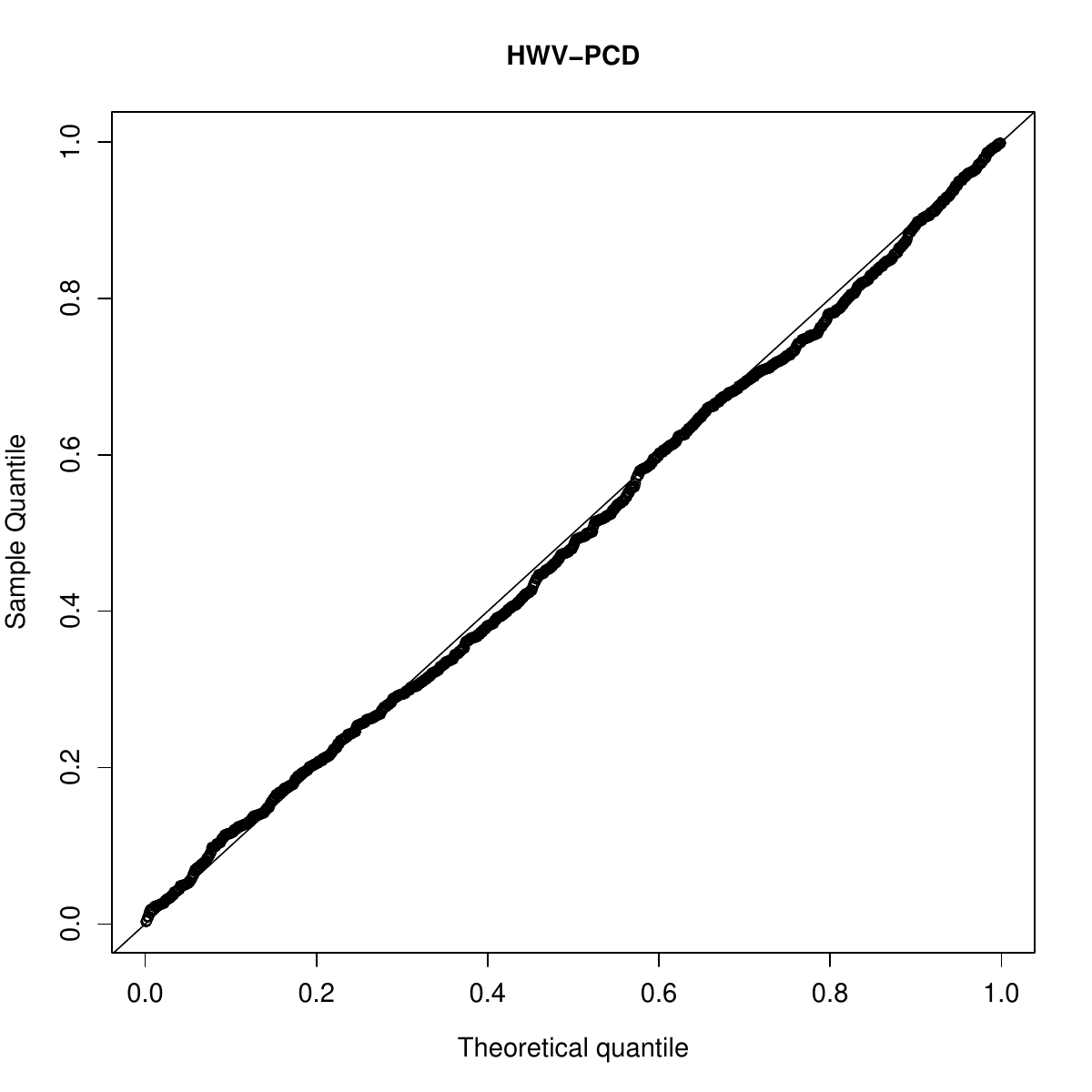}
\end{minipage}

\caption{Q--Q plots of HWV-PCD p-values under genetic heterogeneity across individual genome profiles. Panels correspond to (A) $n=400,\ p=15$, (B) $n=400,\ p=25$, (C) $n=800,\ p=15$, and (D) $n=800,\ p=25$.}
\label{fig:qq-hwv-across-individual}
\end{figure}

% WV-PCD
\begin{figure}[H]
\centering

\begin{minipage}{0.47\textwidth}
\centering
\textbf{(A) } WV-PCD, $n=400,\ p=15$\\
\includegraphics[width=0.92\linewidth]{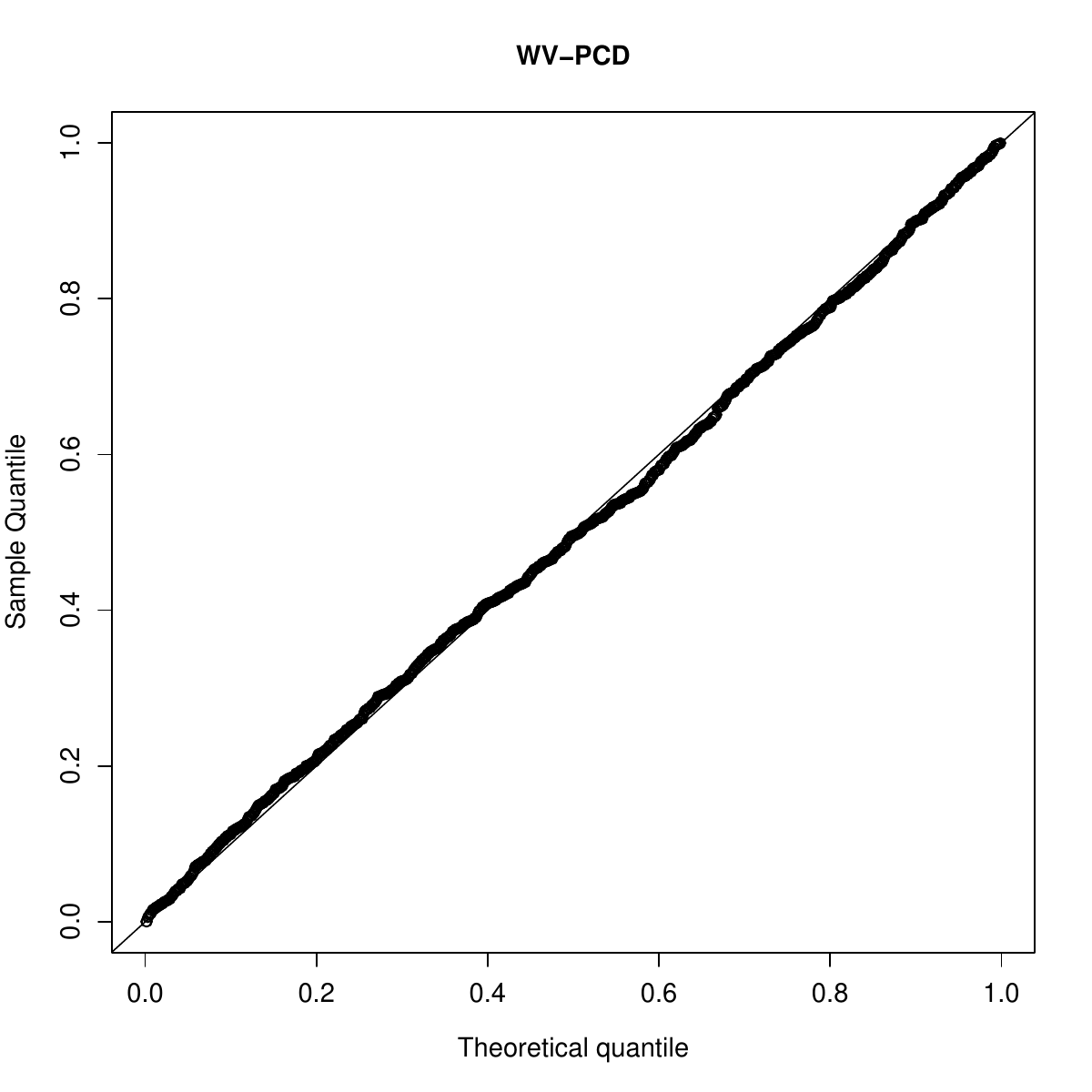}
\end{minipage}
\hfill
\begin{minipage}{0.47\textwidth}
\centering
\textbf{(B) } WV-PCD, $n=400,\ p=25$\\
\includegraphics[width=0.92\linewidth]{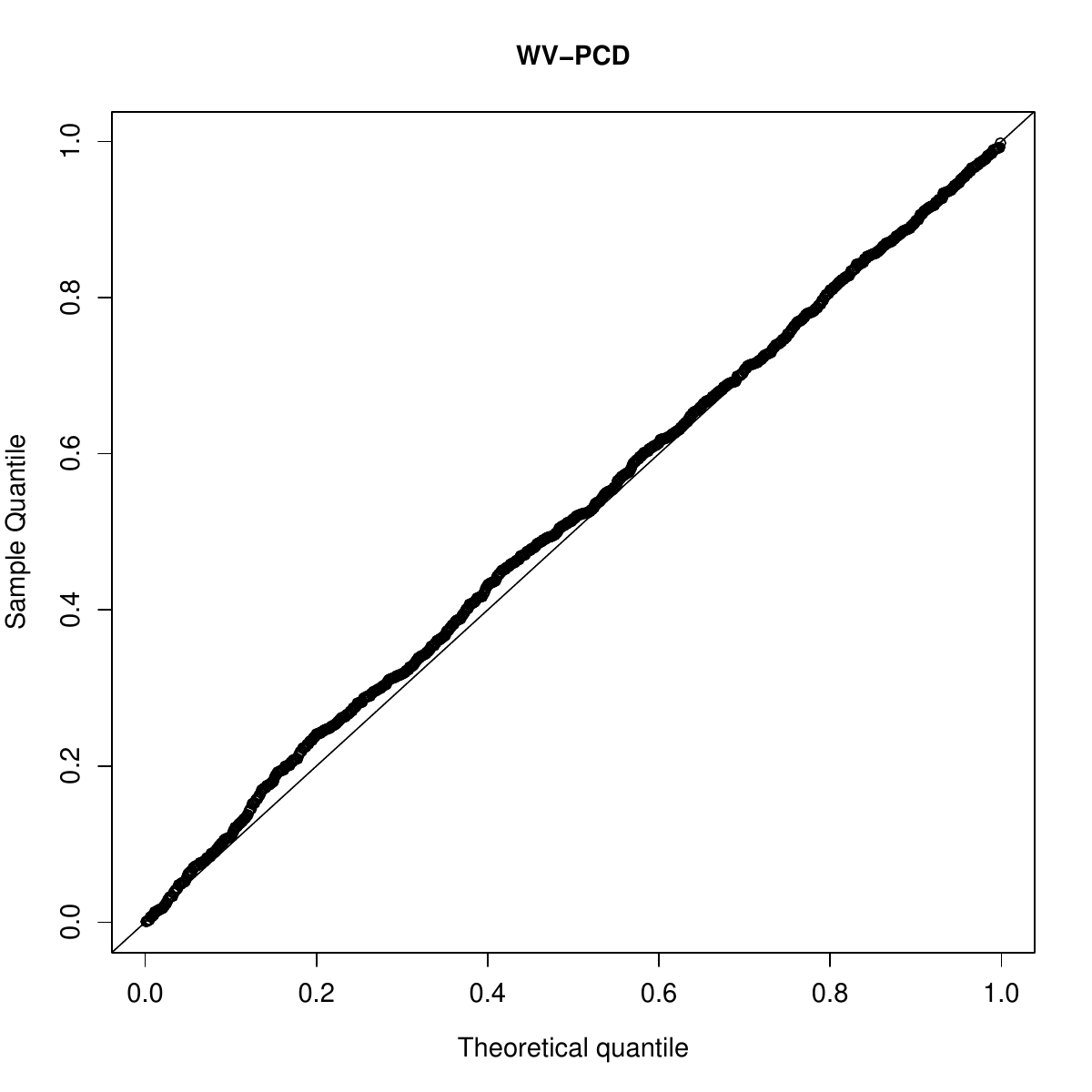}
\end{minipage}

\vspace{0.25cm}

\begin{minipage}{0.47\textwidth}
\centering
\textbf{(C) } WV-PCD, $n=800,\ p=15$\\
\includegraphics[width=0.92\linewidth]{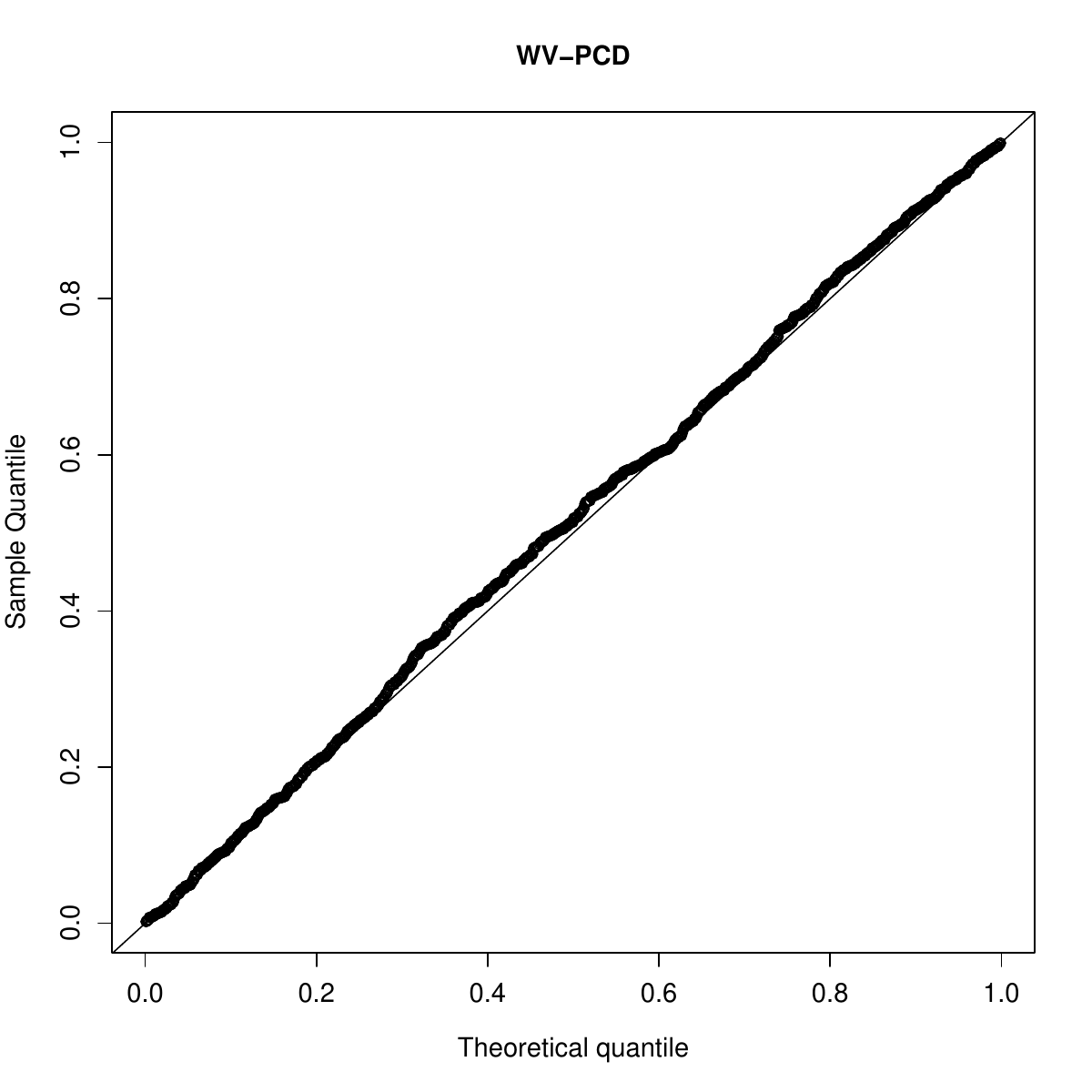}
\end{minipage}
\hfill
\begin{minipage}{0.47\textwidth}
\centering
\textbf{(D) } WV-PCD, $n=800,\ p=25$\\
\includegraphics[width=0.92\linewidth]{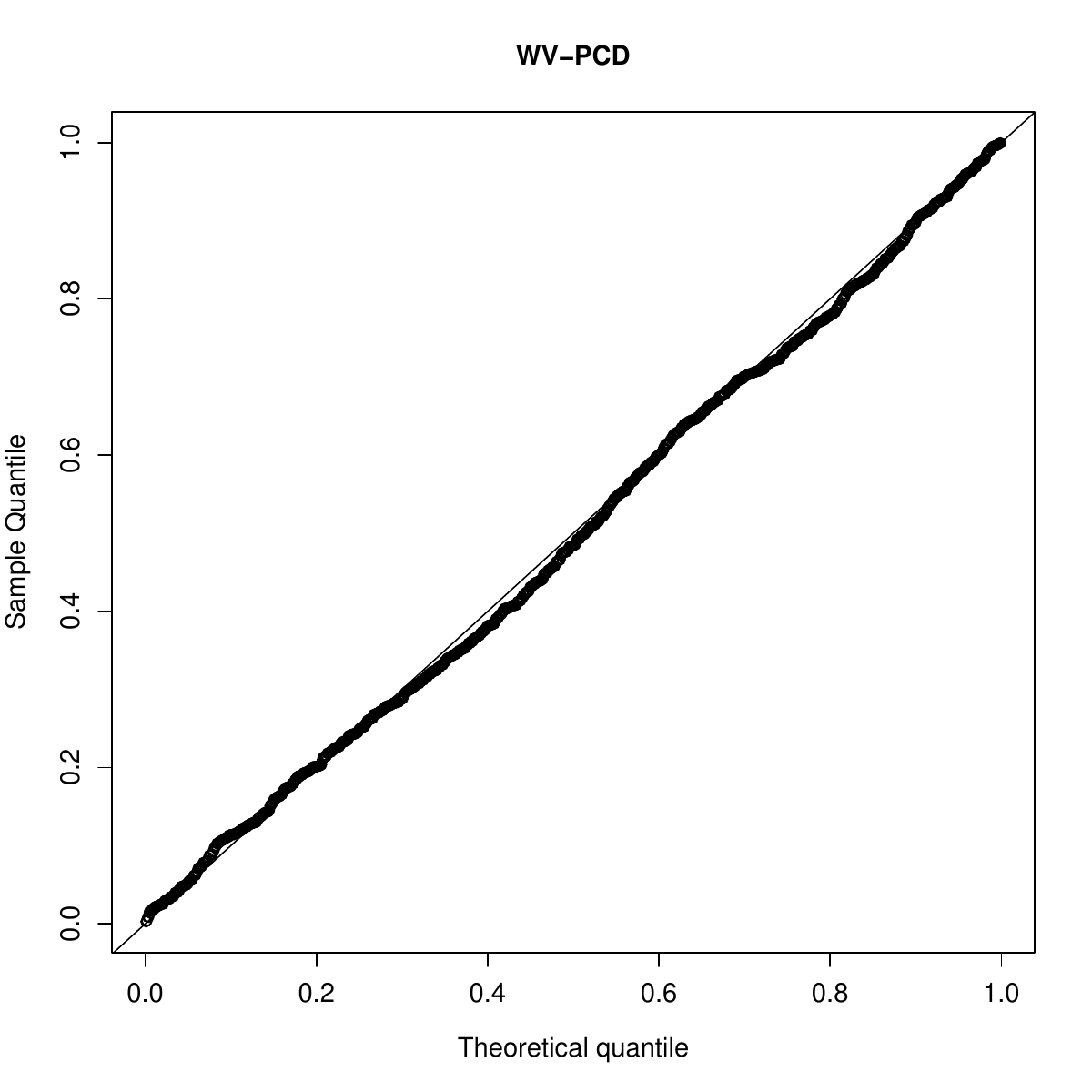}
\end{minipage}

\caption{Q--Q plots of WV-PCD p-values under genetic heterogeneity across individual genome profiles. Panels correspond to (A) $n=400,\ p=15$, (B) $n=400,\ p=25$, (C) $n=800,\ p=15$, and (D) $n=800,\ p=25$.}
\label{fig:qq-wv-across-individual}
\end{figure}

\clearpage
\subsection{Small-Sample Correction Results}

%no heterogeneity, n=150, p=25
\begin{figure}[H]
\centering

\begin{minipage}{0.47\textwidth}
\centering
\textbf{(A) } WV-PCD, $n=150,\ p=25$\\
\includegraphics[width=0.92\linewidth]{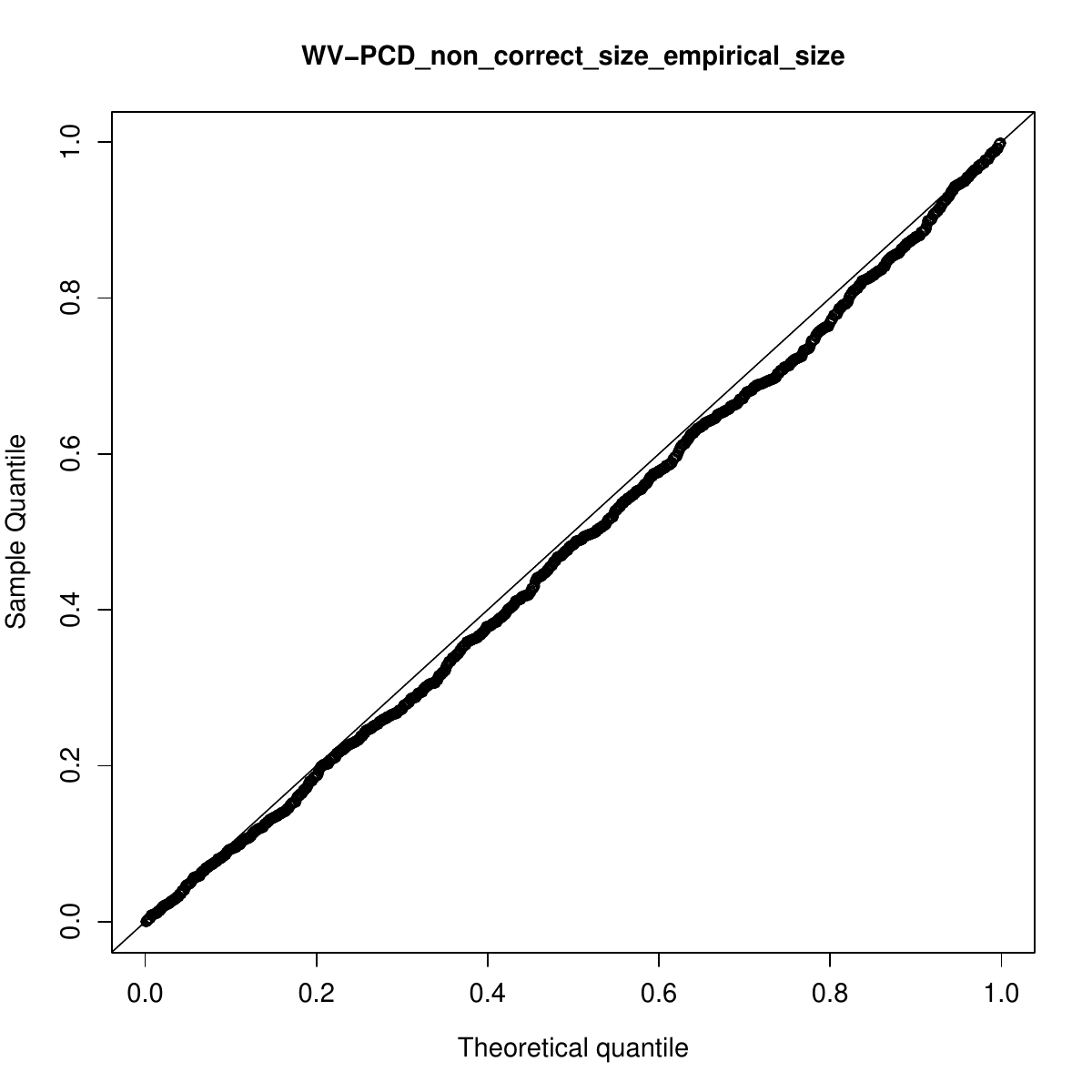}
\end{minipage}
\hfill
\begin{minipage}{0.47\textwidth}
\centering
\textbf{(B) } WV-PCD-SC, $n=150,\ p=25$\\
\includegraphics[width=0.92\linewidth]{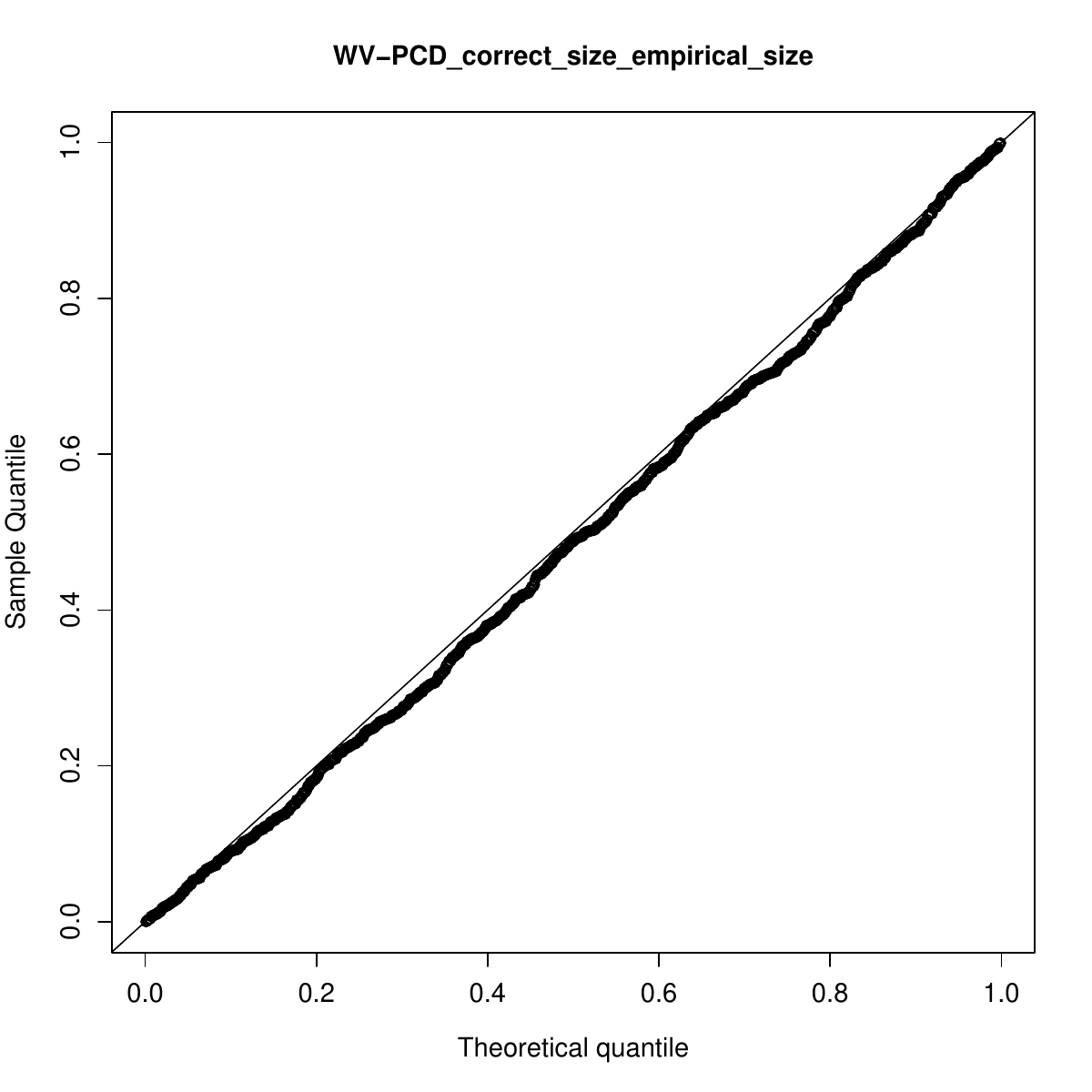}
\end{minipage}

\caption{Q--Q plots under the null setting without genetic heterogeneity for the small-sample scenario with $n=150$ and $p=25$. Panel (A) shows the WV-PCD test and panel (B) shows the small-sample corrected WV-PCD-SC test.}
\label{fig:qq-sc-noheter}
\end{figure}

%two observed subpopulations, n=150, p=25

\begin{figure}[H]
\centering

\begin{minipage}{0.47\textwidth}
\centering
\textbf{(A) } HWV-PCD, $n=150,\ p=25$\\
\includegraphics[width=0.92\linewidth]{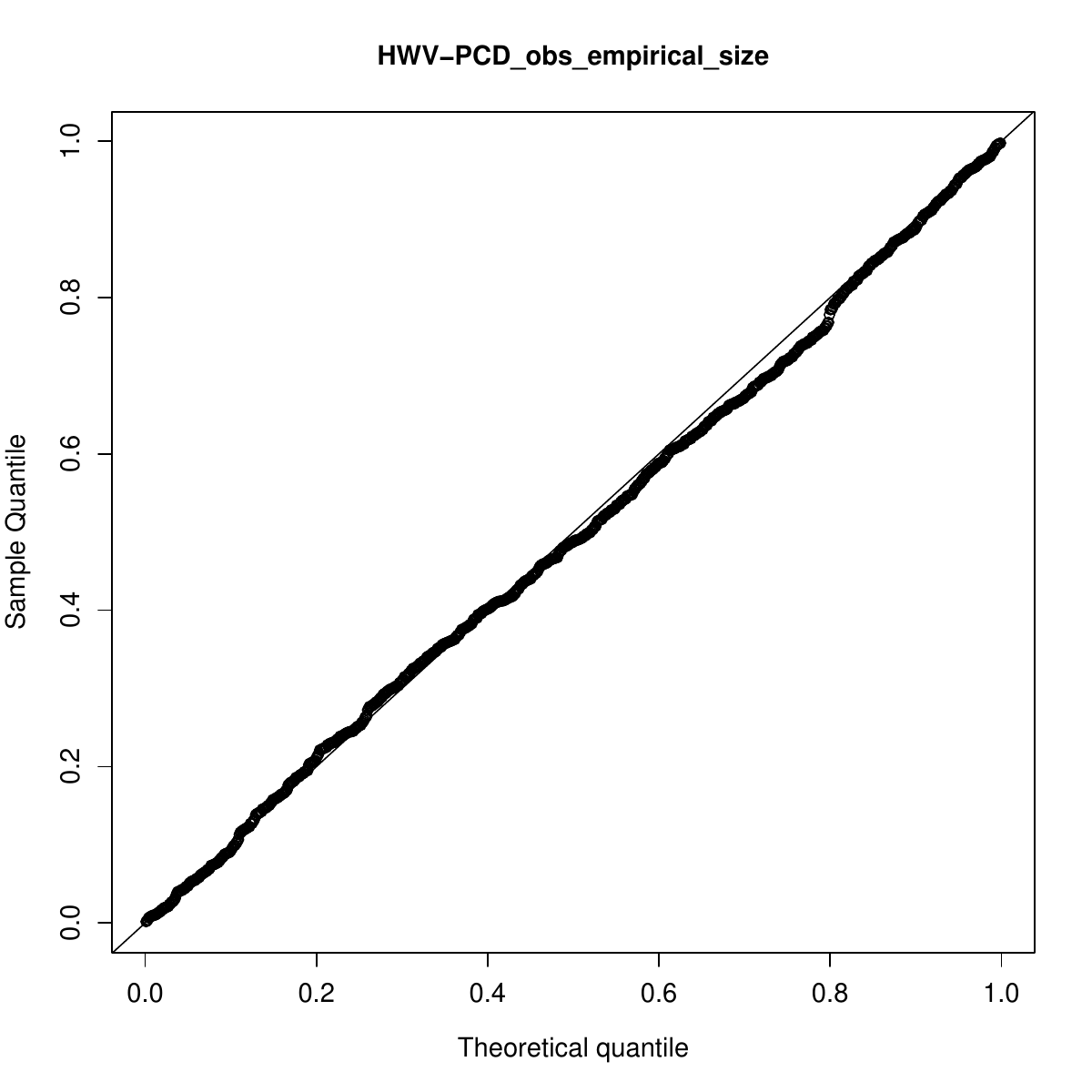}
\end{minipage}
\hfill
\begin{minipage}{0.47\textwidth}
\centering
\textbf{(B) } HWV-PCD-SC, $n=150,\ p=25$\\
\includegraphics[width=0.92\linewidth]{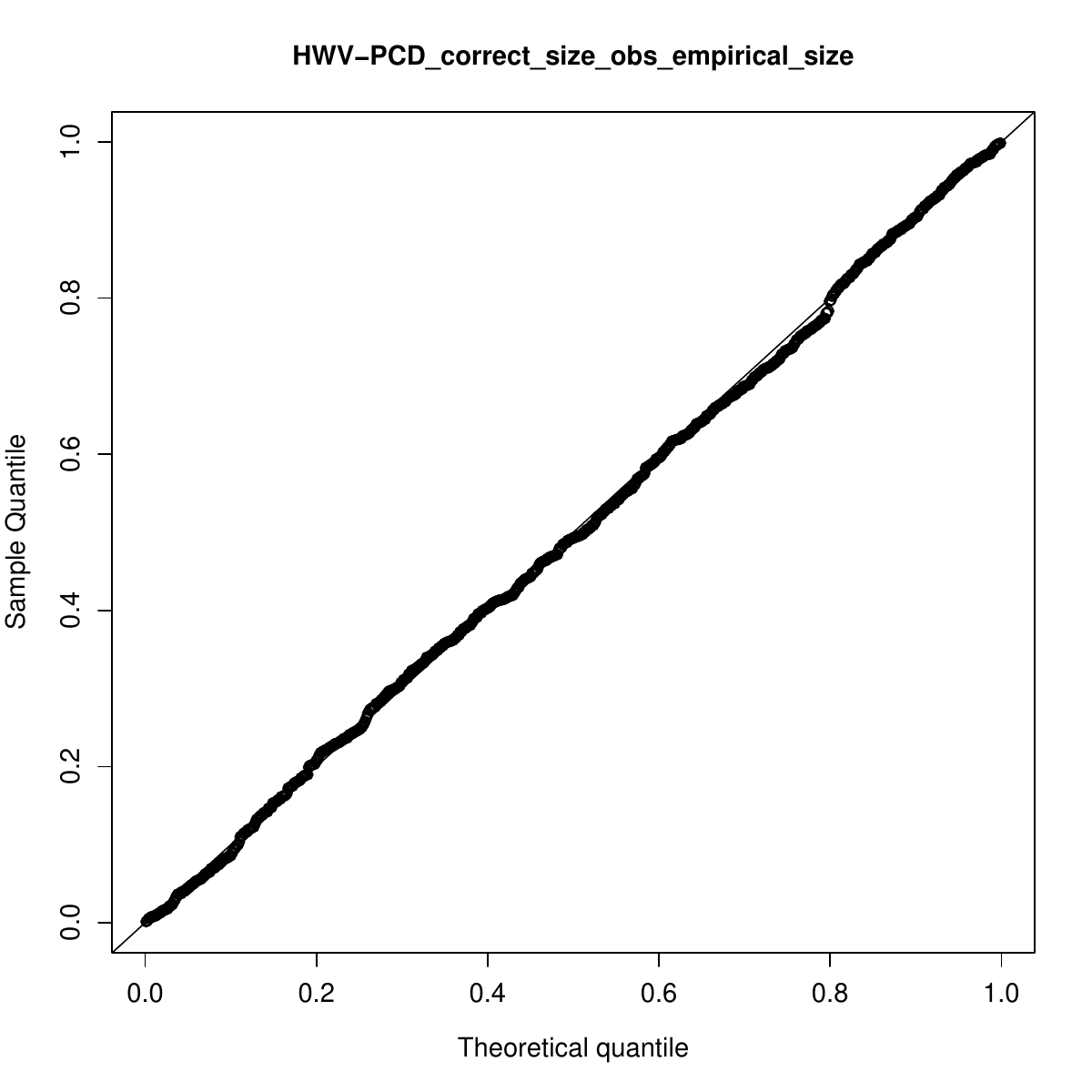}
\end{minipage}

\vspace{0.25cm}

\begin{minipage}{0.47\textwidth}
\centering
\textbf{(C) } WV-PCD, $n=150,\ p=25$\\
\includegraphics[width=0.92\linewidth]{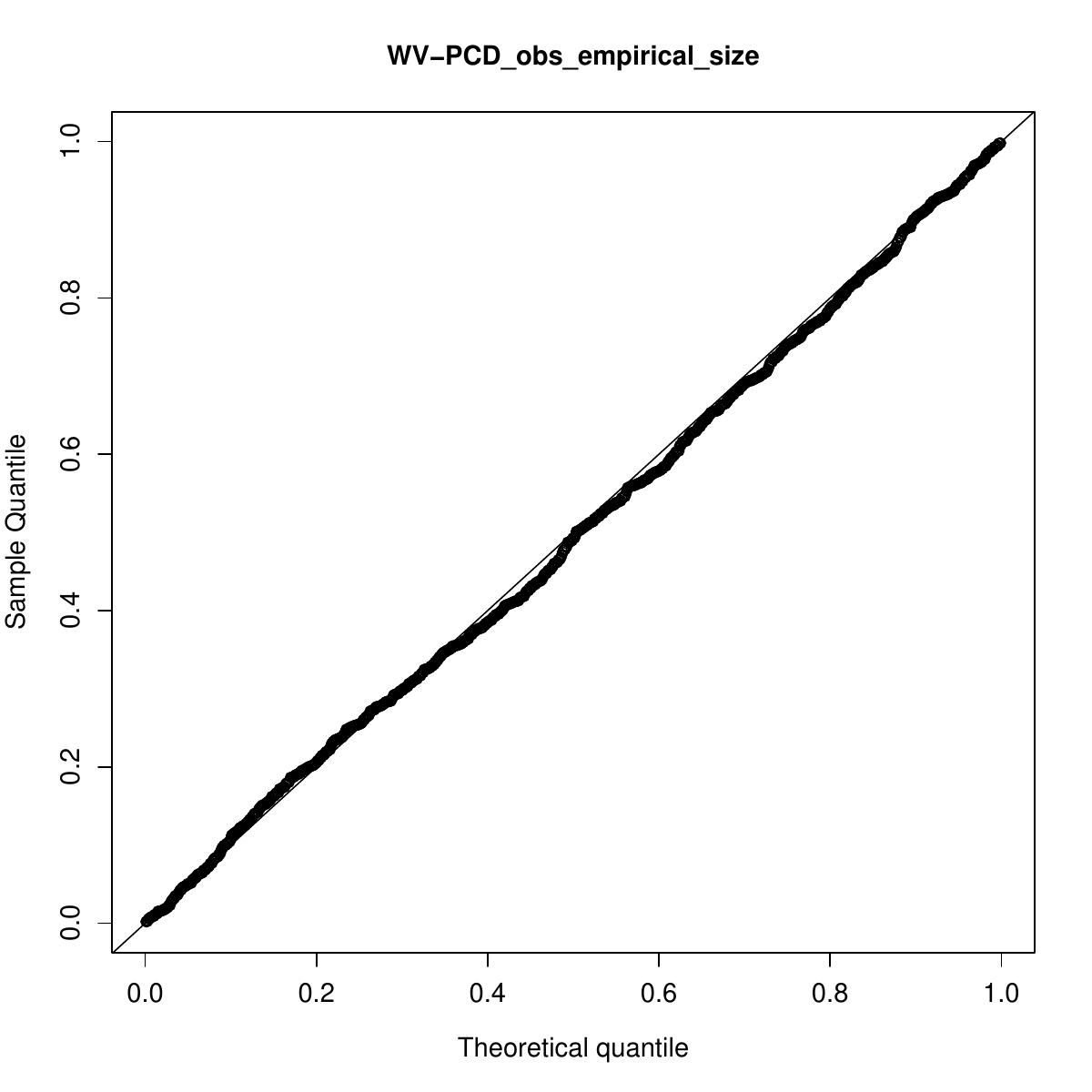}
\end{minipage}
\hfill
\begin{minipage}{0.47\textwidth}
\centering
\textbf{(D) } WV-PCD-SC, $n=150,\ p=25$\\
\includegraphics[width=0.92\linewidth]{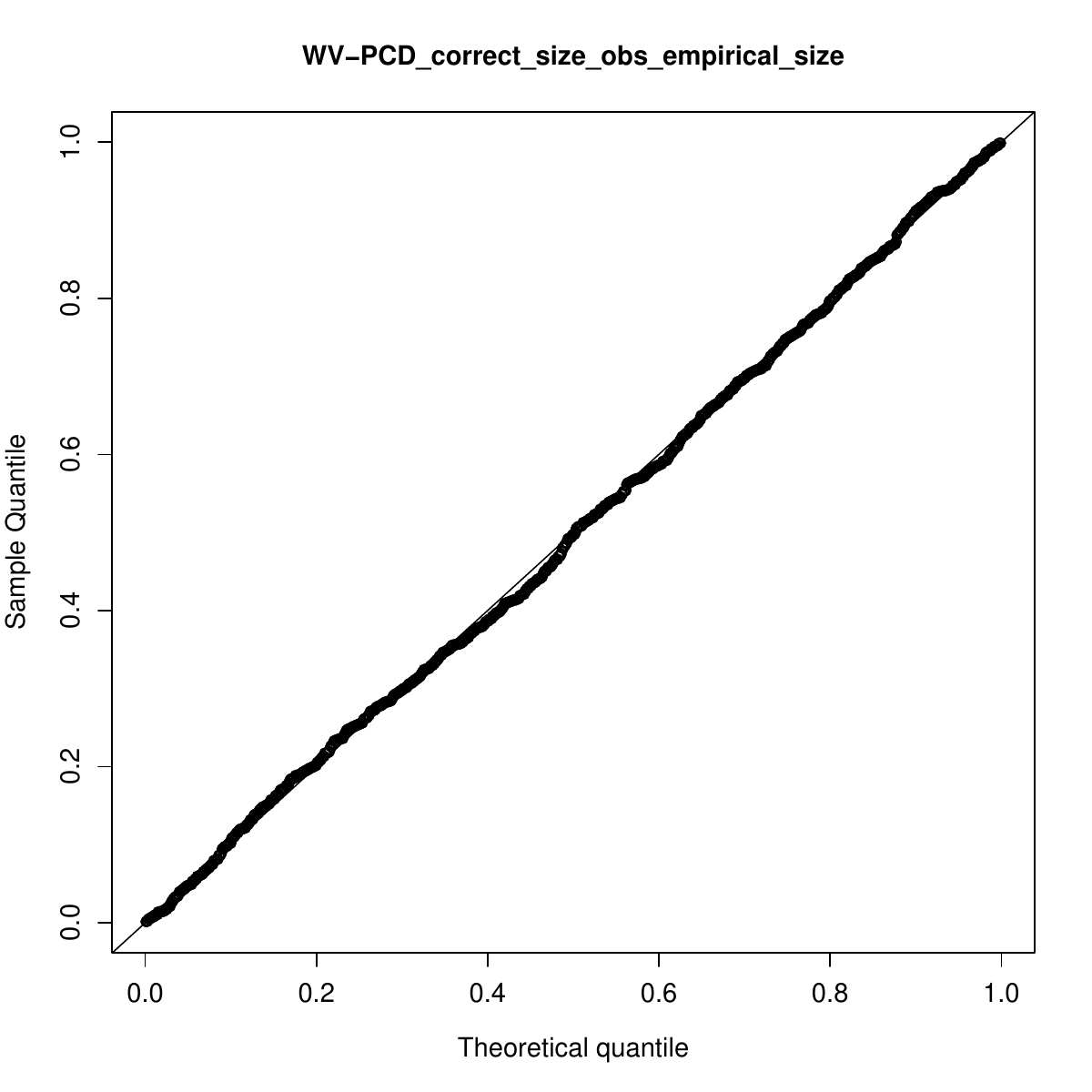}
\end{minipage}

\caption{Q--Q plots for the simulation setting with two observed subpopulations in the small-sample scenario with $n=150$ and $p=25$. Panels (A) and (B) correspond to HWV-PCD and its small-sample corrected version HWV-PCD-SC, respectively. Panels (C) and (D) correspond to WV-PCD and its small-sample corrected version WV-PCD-SC, respectively.}
\label{fig:qq-sc-2obs}
\end{figure}

% gene enrichment analysis
\clearpage
\onecolumn

\addcontentsline{toc}{section}{Appendix}

\subsection{Gene set enrichment analysis results}

Detailed significant Gene Ontology enrichment results from the full-model analysis are shown in Appendix Table~\ref{tab:appendix-gsea-go}. Only terms with Benjamini--Hochberg adjusted \(p<0.05\) are reported.

\begin{table}[H]
\centering
\caption{Significant Gene Ontology terms from gene set enrichment analysis based on the full model. Only terms with Benjamini--Hochberg adjusted \(p<0.05\) are shown. Full core enrichment genes are provided in Supplementary Table S1.}
\label{tab:appendix-gsea-go}
\scriptsize
\setlength{\tabcolsep}{3pt}
\renewcommand{\arraystretch}{1.12}
\begin{adjustbox}{max width=\textwidth}
\begin{tabular}{@{}p{0.9cm}p{8.0cm}r r p{1.5cm}p{1.6cm}@{}}
\toprule
Ontology & GO term & Size & NES & Raw \(p\) & Adjusted \(p\) \\
\midrule

\multicolumn{6}{@{}l}{\textbf{Genome-wide genetic background}}\\
MF & 5alpha-androstane-3beta,17beta-diol dehydrogenase activity & 6 & -2.036 & \(4.27\times10^{-5}\) & 0.040 \\
MF & U4 snRNA binding & 9 & 1.802 & \(3.28\times10^{-5}\) & 0.040 \\

\midrule
\multicolumn{6}{@{}l}{\textbf{Home water fluoride}}\\
BP & negative regulation of cellular glucuronidation & 8 & 1.954 & \(2.56\times10^{-7}\) & 0.002 \\
BP & regulation of cellular glucuronidation & 9 & 1.891 & \(9.11\times10^{-6}\) & 0.039 \\
BP & negative regulation of carbohydrate metabolic process & 64 & 1.529 & \(1.47\times10^{-5}\) & 0.042 \\
CC & keratin filament & 91 & 1.449 & \(4.13\times10^{-5}\) & 0.044 \\

\midrule
\multicolumn{6}{@{}l}{\textbf{None}}\\
MF & 5alpha-androstane-3beta,17beta-diol dehydrogenase activity & 6 & -2.040 & \(3.80\times10^{-5}\) & 0.035 \\
MF & U4 snRNA binding & 9 & 1.805 & \(3.34\times10^{-5}\) & 0.035 \\

\midrule
\multicolumn{6}{@{}l}{\textbf{SSBS}}\\
BP & negative regulation of cellular glucuronidation & 8 & 2.387 & \(9.53\times10^{-8}\) & \(8.21\times10^{-4}\) \\
BP & natural killer cell activation involved in immune response & 33 & 2.257 & \(2.15\times10^{-6}\) & 0.009 \\
BP & humoral immune response & 290 & 1.590 & \(1.20\times10^{-5}\) & 0.035 \\
BP & antimicrobial humoral response & 170 & 1.658 & \(1.80\times10^{-5}\) & 0.039 \\
BP & regulation of cellular glucuronidation & 9 & 2.206 & \(2.64\times10^{-5}\) & 0.045 \\
CC & caveola & 78 & 1.900 & \(2.72\times10^{-5}\) & 0.029 \\
MF & type I interferon receptor binding & 17 & 2.601 & \(1.09\times10^{-8}\) & \(2.03\times10^{-5}\) \\
MF & bitter taste receptor activity & 20 & 2.414 & \(3.67\times10^{-7}\) & \(3.42\times10^{-4}\) \\
MF & taste receptor activity & 26 & 2.297 & \(8.19\times10^{-6}\) & 0.005 \\
MF & GTPase regulator activity & 452 & 1.447 & \(1.66\times10^{-5}\) & 0.006 \\
MF & nucleoside-triphosphatase regulator activity & 452 & 1.447 & \(1.66\times10^{-5}\) & 0.006 \\
MF & N-acyltransferase activity & 82 & 1.847 & \(5.02\times10^{-5}\) & 0.016 \\
MF & vinculin binding & 9 & 2.136 & \(9.12\times10^{-5}\) & 0.024 \\

\midrule
\multicolumn{6}{@{}l}{\textbf{Sex}}\\
BP & negative regulation of cellular glucuronidation & 8 & 1.831 & \(4.59\times10^{-6}\) & 0.040 \\
MF & MHC class Ib receptor activity & 8 & 1.813 & \(1.19\times10^{-5}\) & 0.022 \\
MF & 5alpha-androstane-3beta,17beta-diol dehydrogenase activity & 6 & -2.049 & \(3.97\times10^{-5}\) & 0.025 \\
MF & U4 snRNA binding & 9 & 1.775 & \(4.03\times10^{-5}\) & 0.025 \\
MF & bile acid binding & 10 & -2.234 & \(7.81\times10^{-5}\) & 0.036 \\
MF & cysteine-type endopeptidase inhibitor activity involved in apoptotic process & 19 & 1.673 & \(1.22\times10^{-4}\) & 0.045 \\
MF & transforming growth factor beta binding & 21 & 1.643 & \(1.56\times10^{-4}\) & 0.048 \\

\bottomrule
\end{tabular}
\end{adjustbox}
\end{table}
\normalsize

\end{appendices}

\end{document}